\documentclass[aps,prd,twocolumn,showpacs,superscriptaddress,nofootinbib,groupedaddress]{revtex4} 
\usepackage{mathrsfs}
\usepackage{epsfig}
\usepackage{graphicx}
\usepackage{mathtools}
\usepackage{color}
\usepackage{url}
\usepackage{hyperref}
\usepackage{footnote}
\usepackage{cleveref}
\usepackage{amssymb}

\newcommand{\bn}{\textbf{n}}

\newcommand{\rr}{$\gamma$-}

\def\hp{ {\sc HEALPix}}

\begin{document}
\title{Planck Lensing and Cosmic Infrared Background Cross-Correlation with Fermi-LAT: Tracing Dark Matter Signals in the Gamma-Ray Background}

\author{Chang Feng\footnote{chang.feng@uci.edu}}
\affiliation{Department of Physics and Astronomy,
University of California, Irvine, CA 92697, USA }

\author{Asantha Cooray} 
\affiliation{Department of Physics and Astronomy,
University of California, Irvine, CA 92697, USA }

\author{Brian Keating}
\affiliation{Department of Physics,
University of California, San Diego, CA 92093, USA }

\begin{abstract}
The extragalactic $\gamma$-ray background and its spatial anisotropy could potentially contain a signature of dark matter (DM) annihilation or particle decay.
Astrophysical foregrounds, such as blazars and star-forming galaxies (SFGs), however, dominate the $\gamma$-ray background, precluding an easy detection of the
signal associated with the DM annihilation or decay in the background intensity spectrum. 
The DM imprint on the $\gamma$-ray background is expected to be correlated with large-scale
structure tracers. In some cases, such a cross-correlation is even expected to have a higher signal-to-noise ratio than the auto-correlation. One reliable tracer of the DM distribution in the
large-scale structure is lensing of the cosmic microwave background (CMB), and the cosmic infrared background (CIB) is a reliable tracer of SFGs.
We analyze Fermi-LAT data taken over 92 months and study the cross-correlation with Planck CMB lensing, Planck CIB, and Fermi-$\gamma$ maps. 
We put upper limits on the DM annihilation cross-section from the cross-power spectra with the $\gamma$-ray background anisotropies. 
The unbiased power spectrum estimation is validated with simulations that include cross-correlated signals. 
We also provide a set of systematic tests and show that no significant contaminations are found for the measurements presented here. Using $\gamma$-ray background map from data gathered over 92 months, we find the best constraint on the DM annihilation with a $1\sigma$ confidence level upper limit of $10^{-25}$-$10^{-24}$ cm$^{3}$ s$^{-1}$, when the mass of DM particles is between 20 and 100\,\rm{GeV}.

\end{abstract}

\maketitle

\newpage
\section{Introduction}

Dark matter (DM) constitutes 27\% of the energy density of the universe, relative to the critical density~\cite{planck_params}. The spatial distribution of DM in the large-scale structure can be mapped through gravitational distortions, such as lensing of the Cosmic Microwave Background (CMB) anisotropies and cosmic shear of galaxy shapes. DM halos emerged in spots where over-densities reached maxima and could host baryonic mass that later collapsed and cooled to form galaxies. Radiation coming out of DM halos spans a wide range of wavelengths in the electromagnetic spectrum. Inside DM halos, the dust produced by star formation absorbs ultraviolet radiation from hot, young stars, re-emitting it in the infrared wavelengths. Astrophysical sources like blazars and star-forming galaxies (SFGs), which also reside in DM halos, can emit \rr rays. Moreover, by self-annihilating or decaying into other particles, weakly interacting massive particles (WIMP),  which are thought of as the building block of DM halos, could produce \rr ray radiation as well. 

From the experimental side, measurements of all of these radiation signatures have been made with various all-sky or large area surveys. CMB lensing has been measured from CMB temperature and polarization anisotropies from both space and ground~\cite{2014AA571A17P,2015arXiv150201591P,wmapcross2007,hirata2004,hirata2008,cf2012,sptpol,act,pb,bicep}. These tiny distortions are well-explained by the linear perturbation theory of DM distribution. At Planck's high frequencies, nonlinear structures are resolved by the cosmic infrared background (CIB) from which the DM halo properties and star formation history can be understood. The $\gamma$-ray anisotropies are now mapped by the Fermi large area telescope (LAT). Anisotropies in the background have been detected, and evidence for extra-galactic \rr ray background has been claimed since 2012~\cite{fermi_auto_ps}.

It has been proposed that the cross-correlation between \rr ray and large-scale structure would be a better probe of the DM signals than the auto correlation because the cross-correlation can effectively isolate \rr ray contributions from other astrophysical sources~\cite{cross_letter,cross_long}, while also suppressing systematic effects. A component decomposition can be made by comparing models to data, and DM properties such as mass and cross-section can be constrained by the \rr ray angular correlation that is only responsible for DM signals. To date, DM properties have been constrained from data sets, such as the 2MASS and NVSS galaxy catalogs~\cite{pdm,cross_rg}, the weak lensing data from Canada-France-Hawaii Telescope Lensing Survey (CFHTLenS)~\cite{rs}, and Planck lensing~\cite{kr}. In this analysis, we use Planck lensing measurements, CIB and Fermi-LAT data to constrain the DM cross section for a variety of different masses. 

We organize this paper as follows. In Section II, we derive all the theoretical power spectra for all the components based on the halo-model approach. In Section III, we discuss different data sets. In Section IV, we describe the data analysis procedure. We conclude in Section V.\\

\section{Theoretical power spectrum with the halo model}

We assume a standard Navarro-Frenk-White (NFW) profile to establish the halo model. The NFW profile is
\begin{equation}
\rho_{\rm{NFW}}=\rho_s\Big(\frac{r}{r_s}\Big)^{-1}\Big(1+\frac{r}{r_s}\Big)^{-2},
\end{equation}
where $\rho_s=\rho_c(z)\Delta_c(z)c^3/[3(\ln(1+c)-c/(1+c))]$, $\Delta_c(z)=18\pi^2+82x-39x^2$, $x=\Omega_m(1+z)^3/E^2(z)-1$, $E(z)=H(z)/H_0$, $r_s$ is the characteristic radius of the halo, $\rho_c$ is the critical density, and $c$ is the concentration factor.

The halo mass function is~\cite{tau21_2013} 
\begin{equation}
\frac{dn}{dM}=\frac{\bar\rho_m(0)}{M^2}f(\tilde\nu)\frac{d\tilde\nu}{d\ln M},\label{massfunc}
\end{equation}
where 
\begin{equation}
\tilde\nu f(\tilde\nu)=A\sqrt{\frac{2}{\pi}a\tilde\nu^2}e^{-\frac{1}{2}a\tilde\nu^2}[1+(a\tilde\nu^2)^{-p}],
\end{equation}
and $A=0.322$, $a=0.707$ and $p=0.3$. The quantity $\tilde\nu$ is defined as $\tilde\nu=\delta_{\rm{sc}}/\sigma(M,z)$. Here, $\delta_{\rm{sc}}$ is the critical density contrast for the collapse and is almost redshift-independent, $\bar\rho_m$ is the comoving matter density, and $\sigma(M,z)$ is the variance for all the halos with mass $M$ at redshift $z$. The concentration factor is determined from $9/(1+z)(M/M^{\ast})^{-0.13}$~\cite{concentration} and the critical mass $M^{\ast}$ is the solution when $\tilde\nu(M^{\ast},z)=1$.

The Fourier transform of the NFW profile $\rho_{\rm{NFW}}$ is
\begin{equation}
u(k,M,z)=\frac{1}{M}\int_0^{r_{\rm{vir}}}dr\,4\pi r^2\,\frac{\sin{kr}}{kr}\rho_{\rm{NFW}},
\end{equation}

and the Fourier transform of $\rho^2_{\rm{NFW}}$ is
\begin{equation}
\tilde u_2(k,M,z)=\frac{1}{\bar\rho^2_{\rm{m}}}\int^{r_{\rm{vir}}}_0 dr\,4\pi r^2\,\frac{\sin(kr)}{kr}\rho^2_{\rm{NFW}},
\end{equation}
which includes the $\mathcal{J}$ factor automatically when $k$ approaches 0~\cite{jfac}.

To describe halo bias, we make use of the ``GIF" model given in Ref.\cite{halomodel}, 
\begin{equation}
b(M,z)=1+\frac{a\tilde\nu^2-1}{\delta_{\rm{sc}}}+\frac{2p}{\delta_{\rm{sc}}[1+(a\tilde\nu^2)^p]}.
\end{equation}

The profile of a halo with mass $M$ at redshift $z$ that produces a CMB lensing field ($\kappa=-\nabla^2\phi/2$) is derived from the NFW profile. Its Fourier transform is
\begin{equation}
\kappa(k,M,z)=\frac{W_{\kappa}M}{a\bar\rho_m}u(k,M,z),
\end{equation}
where
\begin{equation}
W^{\kappa}(z)=\frac{3\Omega_m}{2}\Big(\frac{H_0}{c_{0}}\Big)^2\frac{1}{a}\chi\Big(\frac{\chi_{\ast}-\chi}{\chi_{\ast}}\Big).
\end{equation}
The viral radius $r_{\rm{vir}}$ is $cr_s$, $c_0$ is the speed of light, $\chi$ is the comoving distance, and $\chi_{\ast}$ is the comoving distance at the last scattering surface.

We follow the details of the CIB modeling in Ref.~\cite{planckCIB2013} with parameter set $\{\alpha,T_0,\beta,\gamma,$$\,s_z,\sigma^2_{L/M}\} = \{0.36,24.4,1.75,1.7,3.6,0.5\}$. The CIB luminosity function is given as
\begin{equation}
L_{(1+z)\nu}(M,z)=L_0(1+z)^{s_z}\Sigma(M)\Theta[(1+z)\nu].
\end{equation}
We use the mean level of CIB  given in Ref.~\cite{planckCIB2013} to determine the parameter $L_0$ as Ref.~\cite{cibg} indicates. The conditional mass distribution adopts a logarithmic form
\begin{equation}
\Sigma(M)=M\frac{1}{\sqrt{2\pi\sigma^2_{L/M}}}e^{-\frac{1}{2}\Big(\frac{\log_{10}M-\log_{10}M_{\rm{eff}}}{\sigma_{L/M}}\Big)^2},
\end{equation}
and the SED is
\begin{eqnarray}
\Theta (\nu) \propto
\left\{\begin{array}{ccc}
\nu^{\beta}B_{\nu}(T_d)& \nu<\nu_0\\
\nu^{-\gamma}&  \nu\geqslant \nu_0 
\end{array}\right.,
\label{eqn:thetanu}
\end{eqnarray}
where the frequency $\nu_0$ is determined by smoothing the gradient and the dust temperature $T_d=T_0(1+z)^{\alpha}$. The effective halo mass is $10^{12.6}M_{\odot}$.
The occupation number of the central galaxy is

\begin{eqnarray}
  N_{\rm{cen}} =
\left\{\begin{array}{ccc}
0& M<M_{\rm{min}}\\
1&  M\geqslant M_{\rm{min}} 
\end{array}\right.,
\label{eqn:ncen}
\end{eqnarray} where the minimum mass is $10^{10}M_{\odot}$ as used in Ref.~\cite{xCIBg}.

The conditional luminosity functions that determine the fractional CIB emission from the central and satellite galaxies are described as
\begin{eqnarray}
f_{\nu}^{\rm{cen}}(M,z) = N_{\rm{cen}}\frac{L_{(1+z)\nu}(M,z)}{4\pi}
\label{eqn:fcen}
\end{eqnarray}
and
\begin{eqnarray}
f_{\nu}^{\rm{sat}}(M,z) = \int_{M_{\rm{min}}}^{M}dm\frac{dn}{dm}\frac{L_{(1+z)\nu}(m,z)}{4\pi},
\label{eqn:fsat}
\end{eqnarray} where $dn/dm$ is the mass function of the sub-halo and is given in Refs.~\cite{subhalo,subhalo1}. The total CIB emissivity is
\begin{equation}
j_{\nu}(z)=\int dM \frac{dn}{dM}(z)[f_{\nu}^{\rm{cen}}(M,z)+f_{\nu}^{\rm{sat}}(M,z)].
\end{equation}

The emission from DM annihilation traces the squared density, i.e., $\rho^2_{\rm{DM}}$ which is weighted by

\begin{eqnarray}
W^{\gamma,\rm{ann}}(E,\chi; m_{\rm{DM}},\langle\sigma\nu\rangle)&=&\frac{(\Omega_m\rho_c)^2}{4\pi}\frac{\langle\sigma\nu\rangle}{2m^2_{\rm{DM}}}(1+z)^3\nonumber\\
&&\Delta^2(\chi)\frac{dN_{\rm{ann}}}{dE}e^{-\tau(\chi,E(\chi))}.\nonumber\\
&&
\end{eqnarray}

Here, we consider specific DM candidates in order to constrain $\langle\sigma\nu\rangle$\mbox{--}$m_{\rm{DM}}$ relation. We begin with $b\bar b$ annihilation channel and consider it as a representative channel for our case. The DM energy spectrum $dN_{\rm{ann}}/dE$ for the $b\bar b$ channel is provided by PPPC 4 DM ID (A Poor Particle Physicist Cookbook for Dark Matter Indirect Detection)~\cite{PPPC4DMID}. The $\gamma$-ray attenuation function $\tau(E,z)$ is tabulated between $0.01<z<9.0$ and 1\,\rm{GeV} $< E <$ $10^5$\,\rm{GeV}~\cite{tau_table}. The DM mass $m_{\rm{DM}}$ and its annihilation cross-section $\langle\sigma\nu\rangle$ are two free parameters in this analysis. The clumping factor is

\begin{equation}
\Delta^2(z)=\int^{M_{\rm{max}}}_{M_{\rm{min}}}dM\frac{dn}{dM}\int d^3x\frac{\rho_h^2(x|M,z)}{\bar\rho^2_{m}}. 
\end{equation}
Here, $\rho_h$ is the NFW profile. For our case, we take a conservative approach and do not include subhalos which can boost \rr ray emission~\cite{boost_sub,gao_sim}.

Unlike the annihilation, the DM decay signal traces the DM density $\rho_{\rm{DM}}$, which is weighted by
\begin{equation}
W^{\gamma,\rm{dec}}(E,\chi;m_{\rm{DM}},\Gamma_d)=\frac{\Omega_m\rho_c}{4\pi}\frac{\Gamma_d}{m_{\rm{DM}}}\frac{dN_{\rm{dec}}}{dE}e^{-\tau(\chi,E(\chi))}.
\end{equation}
Here, $\Gamma_d$ is the decay time and the energy spectrum of the decay $dN_{\rm{dec}}/dE(E)=dN_{\rm{ann}}/dE(2E)$. We set the minimum halo mass to $10^{-6}M_{\odot}$ for the DM signals~\cite{pdm}.

For Planck lensing, CIB and DM annihilation and decay, the angular correlation functions are determined by the biased NFW profiles in Fourier space. The equations for the 1-halo and 2-halo terms are
\begin{eqnarray}
C_l^{1h,XY}&=&\int dz\frac{d\chi}{dz}\Big(\frac{a}{\chi}\Big)^2\int dM n(M,z)\nonumber\\
&&X_l(k,M,z)Y_l(k,M,z),
\end{eqnarray}
and 
\begin{eqnarray}
C_l^{2h,XY}&=&\int dz\frac{d\chi}{dz}\Big(\frac{a}{\chi}\Big)^2P_{\rm{lin}}(k,z)\nonumber\\
&&\Big[\int dMb(M,z)n(M,z)\tilde X_l(k,M,z)\Big]\nonumber\\
&&\Big[\int dMb(M,z)n(M,z)\tilde Y_l(k,M,z)\Big],
\end{eqnarray}
here $X_l(k,M,z)$ or $Y_l(k,M,z)$ = $W^{\kappa}u_1(k,M,z)/a$ for Planck lensing, $W^{a,\rm{CIB}}_{\nu}u(k,M,z)/a$ for Planck CIB, $W^{\gamma,\rm{ann}}/\Delta^2\tilde u_2(k,M,z)/a$ for DM annihilation and $W^{\gamma,\rm{dec}}u_1(k,M,z)/a$ for DM decay. $u_1=M/\bar\rho_mu$. For $\tilde X_l(M,z)$ or $\tilde Y_l(M,z)$, only Planck CIB takes a different form, $W^{b,\rm{CIB}}_{\nu}u(k,M,z)/a$. $W^{a,\rm{CIB}}_{\nu}=\sqrt{2f_{\nu}^{\rm{cen}}(M,z)f_{\nu}^{\rm{sat}}(M,z)+f_{\nu}^{\rm{sat}}(M,z)f_{\nu}^{\rm{sat}}(M,z)}$ where $W^{b,\rm{CIB}}_{\nu}=(f_{\nu}^{\rm{cen}}(M,z)+f_{\nu}^{\rm{sat}}(M,z))$. The linear matter power spectrum $P_{\rm{lin}}(k,z)$ is calculated from $z=0$ to 8 using \rm{CAMB}. The function $n(M,z)$ is the halo mass function defined in Eq. (\ref{massfunc}).

Astrophysical sources such as blazars, SFGs, flat spectrum radio quasars (FSRQs), and misaligned active galactic nuclei (mAGN), are also significant \rr ray emitters. They are point-like sources, so the NFW profile does not apply, but the power spectrum can be calculated from the luminosity function alternatively. In the following, we model each emitter separately. 

For blazars and FSRQs, the $\gamma$ ray luminosity function (GLF) is taken from Ref.~\cite{XiaGamma,blazarLF}
\begin{equation}
\Phi(L_{\gamma},z,\Gamma)=\tilde\Phi(L_{\gamma},z=0,\Gamma)\times e(L_{\gamma},z),
\end{equation} where

\begin{equation}
\tilde\Phi(L_{\gamma},z=0,\Gamma)=\frac{A}{\ln(10)L_{\gamma}}\Big[\Big(\frac{L_{\gamma}}{L_{\ast}}\Big)^{\gamma_1}+\Big(\frac{L_{\gamma}}{L_{\ast}}\Big)^{\gamma_2}\Big]^{-1}
\end{equation} and $e(L_{\gamma},z)$ is 
\begin{equation}
e(L_{\gamma},z)=\Big[\Big(\frac{1+z}{1+z_c(L_{\gamma})}\Big)^{p_1(L_{\gamma})}+\Big(\frac{1+z}{1+z_c(L_{\gamma})}\Big)^{p_2}\Big]^{-1}.
\end{equation}
Here, $z_c(L_{\gamma})=z_c^{\ast}(L_{\gamma}/10^{48})^{\alpha}$, $p_1(L_{\gamma})=p_1^{\ast}+\tau\times[\log_{10}(L_{\gamma})-46]$. The second luminosity dependent density evolution function (LDDE2) is taken from ~\cite{blazarLF} for blazars and the parameter set $\{A, \gamma_1,\gamma_2,L_{\ast},z^{\ast}_c,p^{\ast}_1,\tau,p_2,\alpha\}$ is given in Table 3. For FSRQs, we use the LDDE parameters in Table 3 in Ref.~\cite{fsrq}.

The mean luminosity produced by unresolved sources such as blazars and SFGs can be generally expressed as
\begin{equation} g_s(z)=\int^{L_{\gamma,\rm{max}}(F_{\rm{max}},z)}_{L_{\gamma,\rm{min}}} dL_{\gamma}\Phi(L_{\gamma},z)L_{\gamma}\label{gs}.
\end{equation}
The maximum luminosity is determined by the threshold flux above which the detector can resolve the sources, and $F=1\times10^{-10}\rm{cm}^{-2}s^{-1}$ for $E>E_{\ast}$ and $E_{\ast}=1\,\rm{GeV}$~\cite{2016PhR...636....1C}.

\begin{figure*}
\rotatebox{0}{\includegraphics[width=8cm, height=7.2cm]{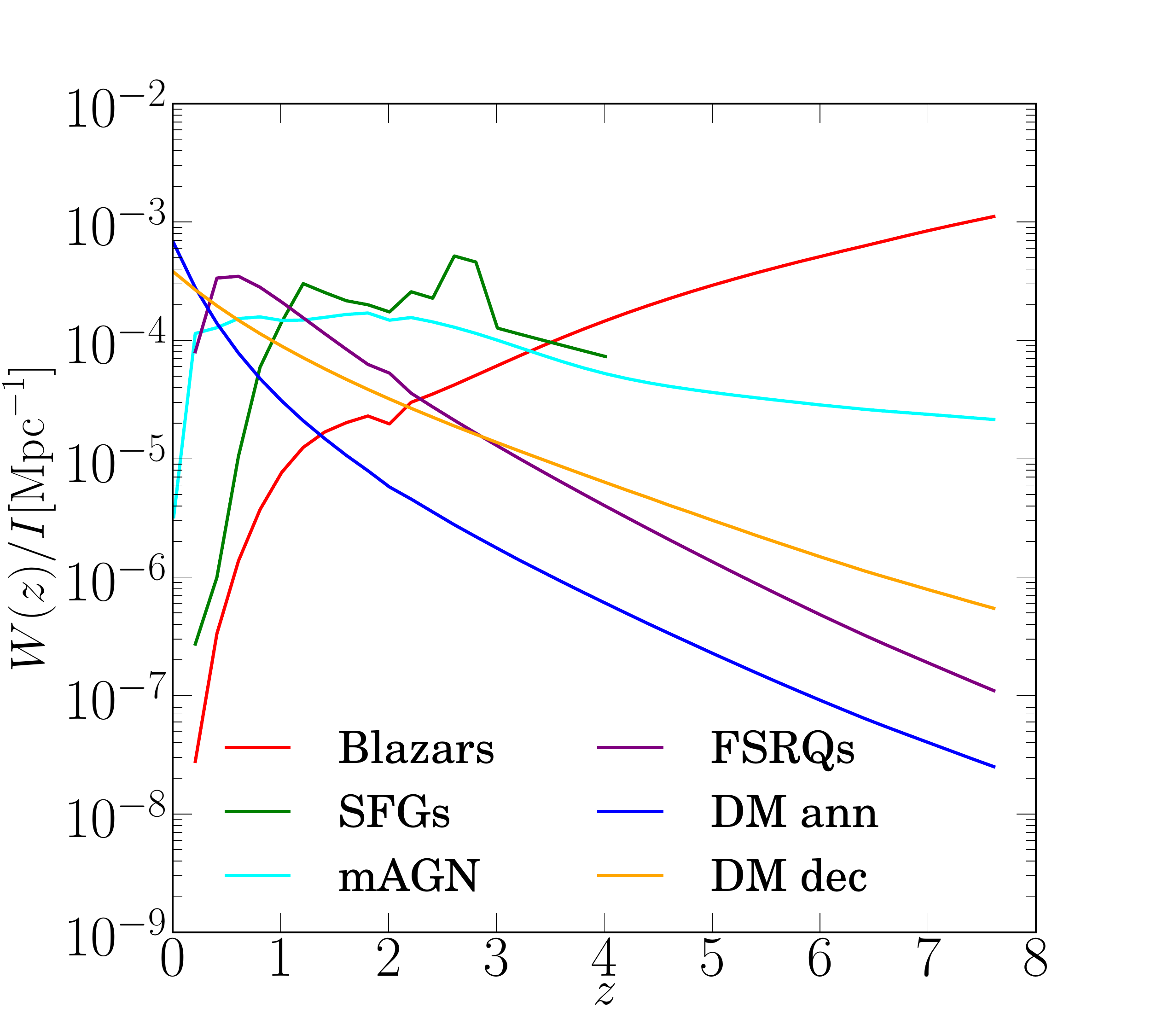}}
\rotatebox{0}{\includegraphics[width=8cm, height=7.2cm]{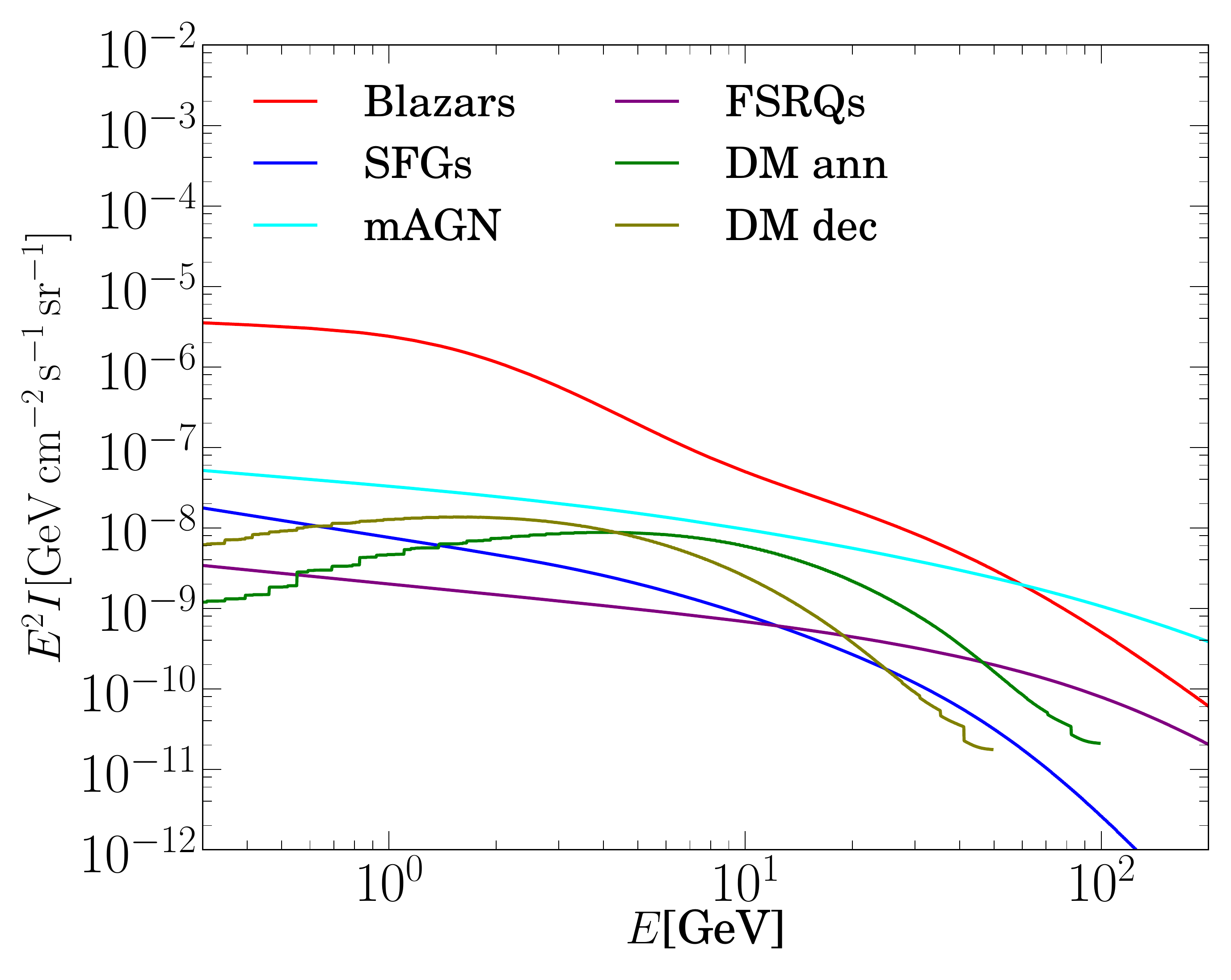}}
\caption{The representative plots for the redshift windows (left) at the energy band 1-5 \rm{GeV} for $m_{\rm{DM}}=100\,\rm{GeV}$ and the energy spectra (right). }
\label{fermi-theory-es}
\end{figure*}

For a single energy bin, the weighting function of the point-like sources is
\begin{equation}
W(E,z)=\frac{A_s(z)g_s(z)}{4\pi E_0^2}\Big(\frac{E}{E_0}\Big)^{-\alpha_s}e^{-\tau(E(1+z),z)}.\label{wt}
\end{equation}
The energy spectrum $dN/dE$ of blazar (FSRQs) is assumed to be a simple power law $A_s(E/E_0)^{-\alpha_s}$ 
with $\alpha_s$=2.2 (2.44), $A_s=(1+z)^{-\alpha_s}$, $E_0=100\,\rm{MeV}$ and the FSRQ spectra index is calculated from 57 FSRQs~\cite{2009ApJ...700..597A}. The luminosity and halo mass is related by~\cite{sando2007}
\begin{equation}
M=10^{11.3}M_{\odot}\Big(\frac{L_{\gamma}}{10^{44.7}\rm{erg/s}}\Big)^{1.7}.
\end{equation}

For SFGs, the GLF is
\begin{equation}
\Phi(L_{\gamma},z)=\frac{\Phi^{\ast}}{\alpha_{\Phi}\ln(10)L_{\gamma}}\Big(\frac{L_{\rm{IR}}}{L_{\ast}}\Big)^{1-\alpha}e^{-\frac{1}{2}\Big[\frac{\log_{10}(1+\frac{L_{\rm{IR}}}{L_{\ast}})}{\sigma}\Big]^2}.
\end{equation}
The factor $\alpha_{\Phi}$ converts the infrared luminosity (IR) to the \rr ray. The piece-wise GLF is estimated for SFGs so the parameter set $\{\Phi^{\ast},L^{\ast},\alpha,\sigma\}$ is determined at each redshift band in Ref.~\cite{sfg}.
We can also calculate the mean luminosity and weighting function for SFGs using Eqs. (\ref{gs}, \ref{wt}). The parameters for SFGs in Eq. (\ref{wt}) are $\alpha_s=2.7$ and $A_s=(\alpha_s-2)/(1+z)^2$. The choice of such a spectral index is due to the fact that the interaction between cosmic-rays and interstellar gas leads to gamma rays mostly from pion decay in flight and the gamma-ray spectrum has the same spectral index as the cosmic-ray spectrum~\cite{2010ApJ...722L.199F}. The IR luminosity required by the SFGs' GLF should be converted from the \rr ray luminosity first following the power law
\begin{equation}
\log_{10}\Big(\frac{L_{\gamma}}{\rm{erg/s}}\Big) = \alpha_{\Phi} \log_{10}\Big(\frac{L_{\rm{IR}}}{10^{10}L_{\odot}}\Big) + \beta_{\Phi},
\end{equation}
where $\alpha_{\Phi}=1.09$ and $\beta_{\Phi}=39.19$~\cite{Ackermann12}. One has to relate the halo mass to the luminosity by
\begin{equation}
M=10^{12}M_{\odot}\Big(\frac{L_{\gamma}}{10^{39}\rm{erg/s}}\Big)^{0.5},
\end{equation}
when the redshift-dependent bias is taken into account.

The GLF of mAGN is derived from the radio GLF at 151 \rm{MHz}~\cite{radioGLF}. We convert the $\gamma$-ray luminosity $L_{\gamma}$ to total radio luminosity $L_{5 \rm{GHz}}$ using the best-fit $L_{\gamma}$\mbox{--}$L_{\rm{core}}^{5\rm{GHz}}$ and $L_{\rm{core}}^{5\rm{GHz}}$\mbox{--}$L_{\rm{tot}}^{1.4\rm{GHz}}$ correlation functions fitted from 12 mAGN~\cite{Lr_L}. We then shift the total luminosity from $L_{151 \rm{MHz}}$ to $L_{5 \rm{GHz}}$ with a power-law and adopt Model C with ($\Omega_M=0$) in~\cite{radioGLF}. The comoving volume is also corrected following the procedure in~\cite{Lr_L}. The mean spectral index is averaged from 12 mAGN samples listed in Table 1 of~\cite{Lr_L} and the parameters for mAGN in Eq. (\ref{wt}) are $\alpha_s=2.37$ and $A_s=(\alpha_s-2)/(1+z)^2$. A $L\mbox{--}M$ relation for mAGN is built from a chain of steps. First, the host galaxy mass $M$ is converted to the black hole mass $M_{\rm{bh}}$ by the best fit correlation function given in~\cite{Mbh_M}. Next, the radio luminosity is interpolated from a table that gives black hole mass and radio luminosity properties in a few galaxies~\cite{Mbh_Lr}, and the $\gamma$-ray luminosity is calculated from the radio luminosity according to the best-fit model in Ref.~\cite{Lr_L}. Last, we use the samples to make a line fit to $\log_{10}{(M/M_{\odot})}$ and $\log_{10}{[L_{\gamma}/(\rm{erg\,s^{-1}})]}$ and derive a $L\mbox{--}M$ relation for mAGN.

Built on all the individual window functions, the mean window function and mean intensity are simply integrated over either energy or redshift, such as
\begin{equation}
W(\chi)=\int^{E_{\rm{max}}}_{E_{\rm{min}}}dEW(E,\chi),
\end{equation}
and
\begin{equation}
I(E)=\int d\chi W(E,\chi).
\end{equation}
All the energy spectra and window functions shown in Fig. \ref{fermi-theory-es} have different amplitudes and shapes that are determined by a few parameters and functions, such as the DM annihilation/decay channel, the photon index of the astrophysical source, the luminosity range for the source, the $\gamma$-ray luminosity function, and the mass-luminosity relation. The step-like feature in the DM annihilation/decay energy spectra is due to the discrete mass step in the PPPC 4 DM ID data.

\begin{figure}
\rotatebox{0}{\includegraphics[width=8cm, height=5.1cm]{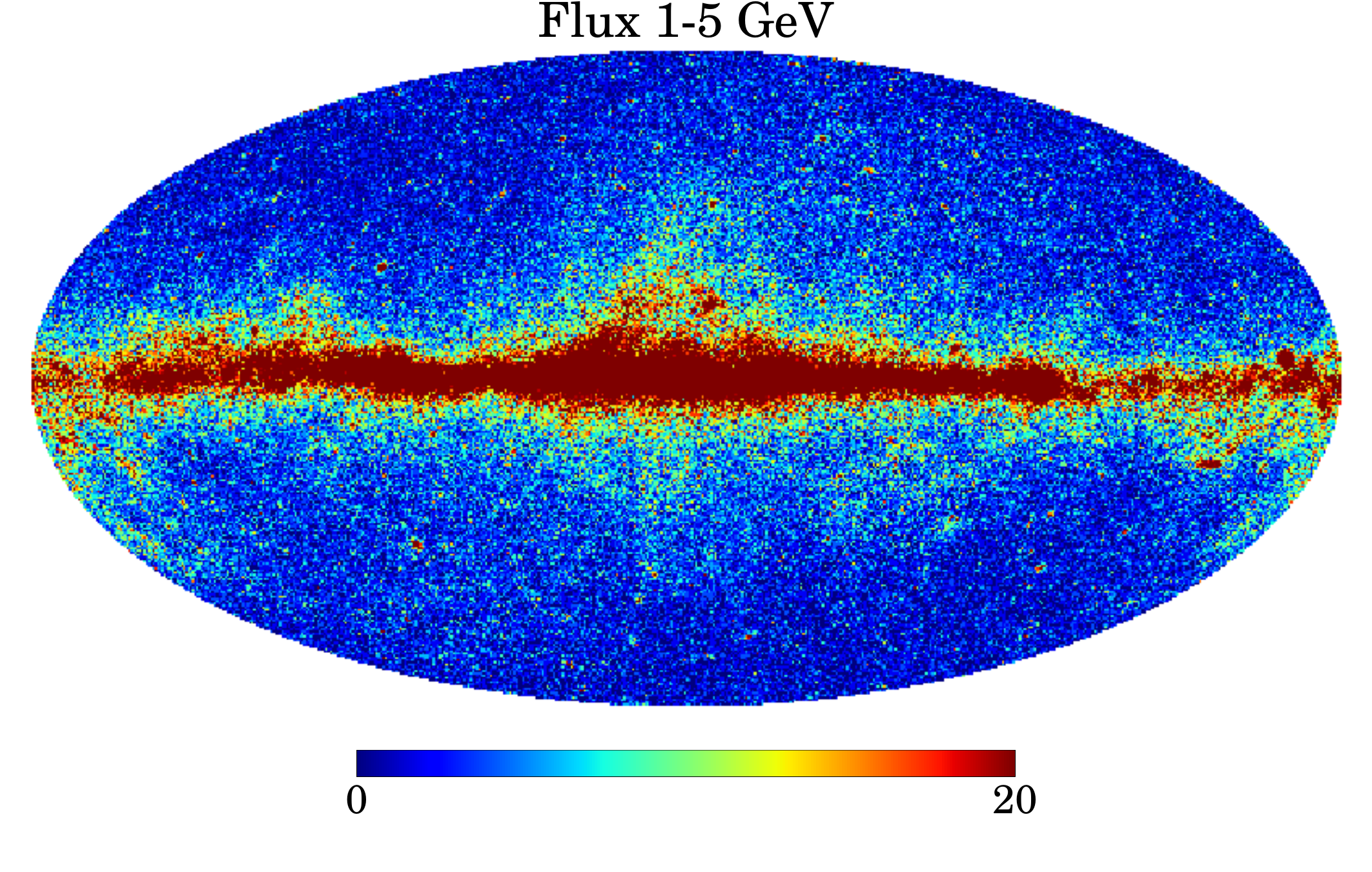}}
\rotatebox{0}{\includegraphics[width=8cm, height=5.1cm]{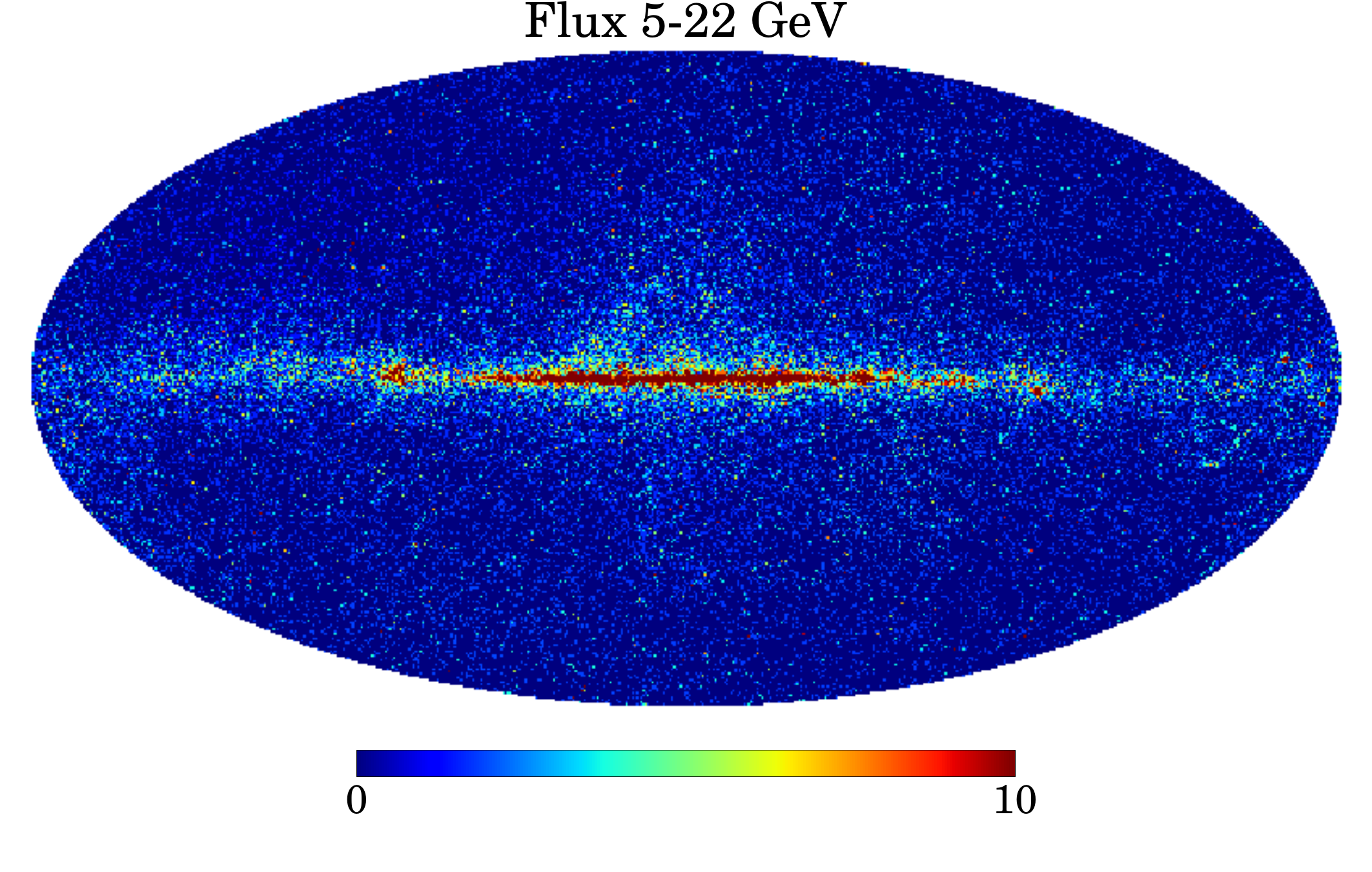}}
\rotatebox{0}{\includegraphics[width=8cm, height=5.1cm]{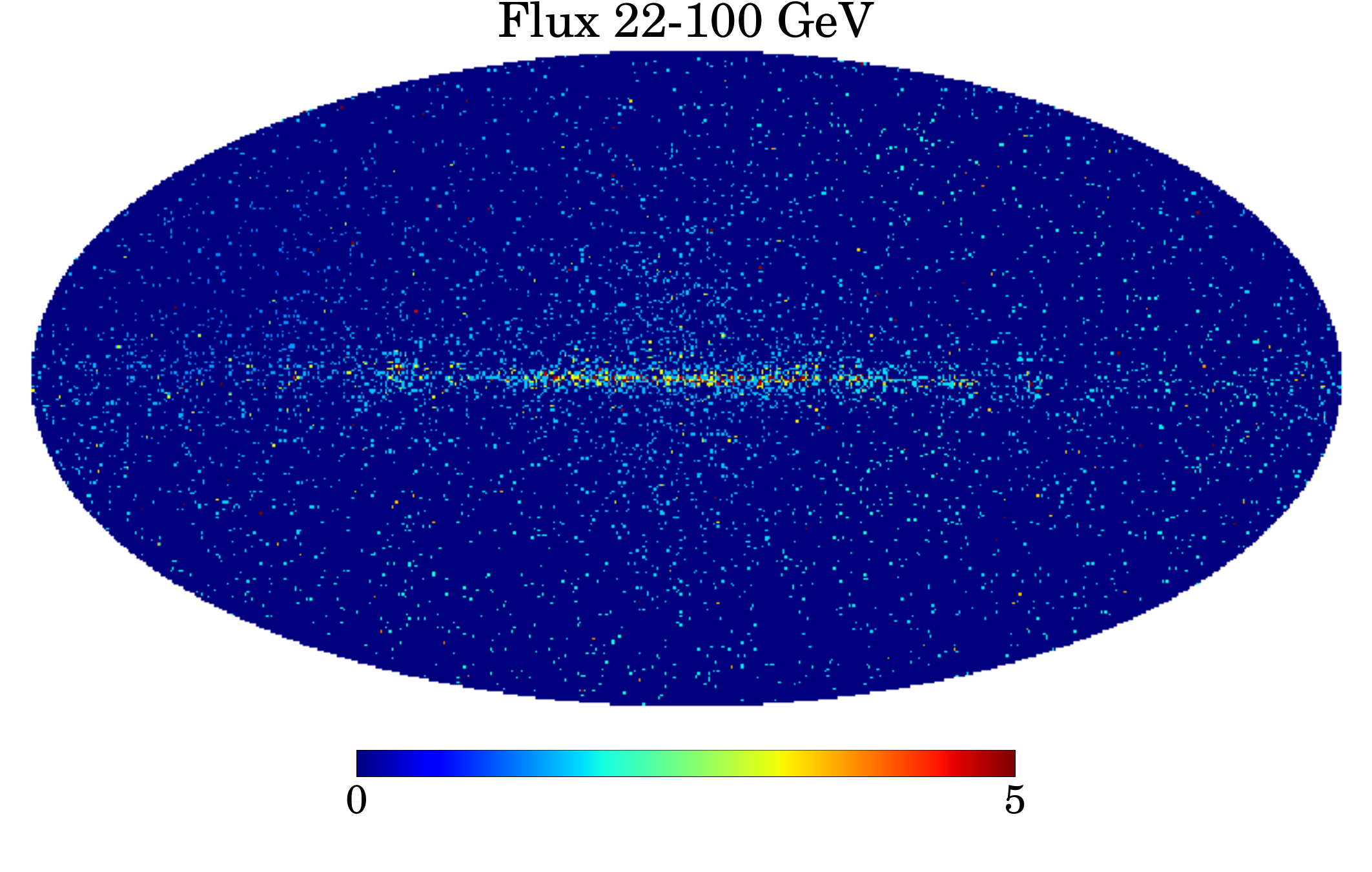}}
\caption{The full-sky Fermi flux maps (in unit of $10^{-6}\rm{photon}\,\rm{cm}^{-2}\,\rm{s}^{-1}\,\rm{sr}^{-1}$) made from different photon energies over 92 months. From top to bottom: 1-5 GeV, 5-22 GeV, 22-100 GeV.}
\label{fermiflux}
\end{figure}

\begin{figure}
\rotatebox{0}{\includegraphics[width=8cm, height=7.2cm]{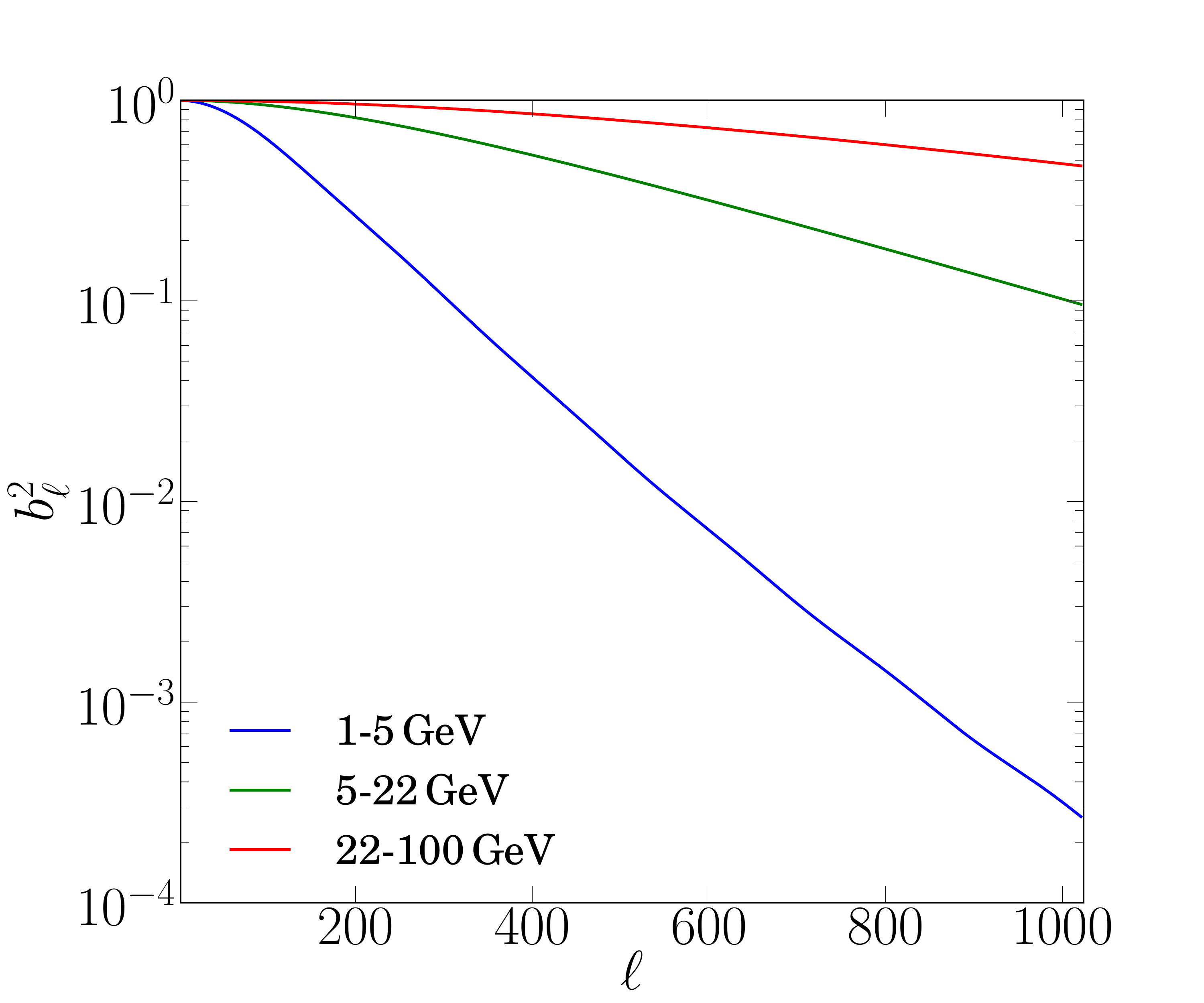}}
\caption{The beam transfer functions for the Fermi-LAT data at energy bands 1-5 \rm{GeV}, 5-22 \rm{GeV} and 22-100 \rm{GeV}. }
\label{fermi-beam}
\end{figure}
\begin{figure}
\rotatebox{0}{\includegraphics[width=8cm, height=5.1cm]{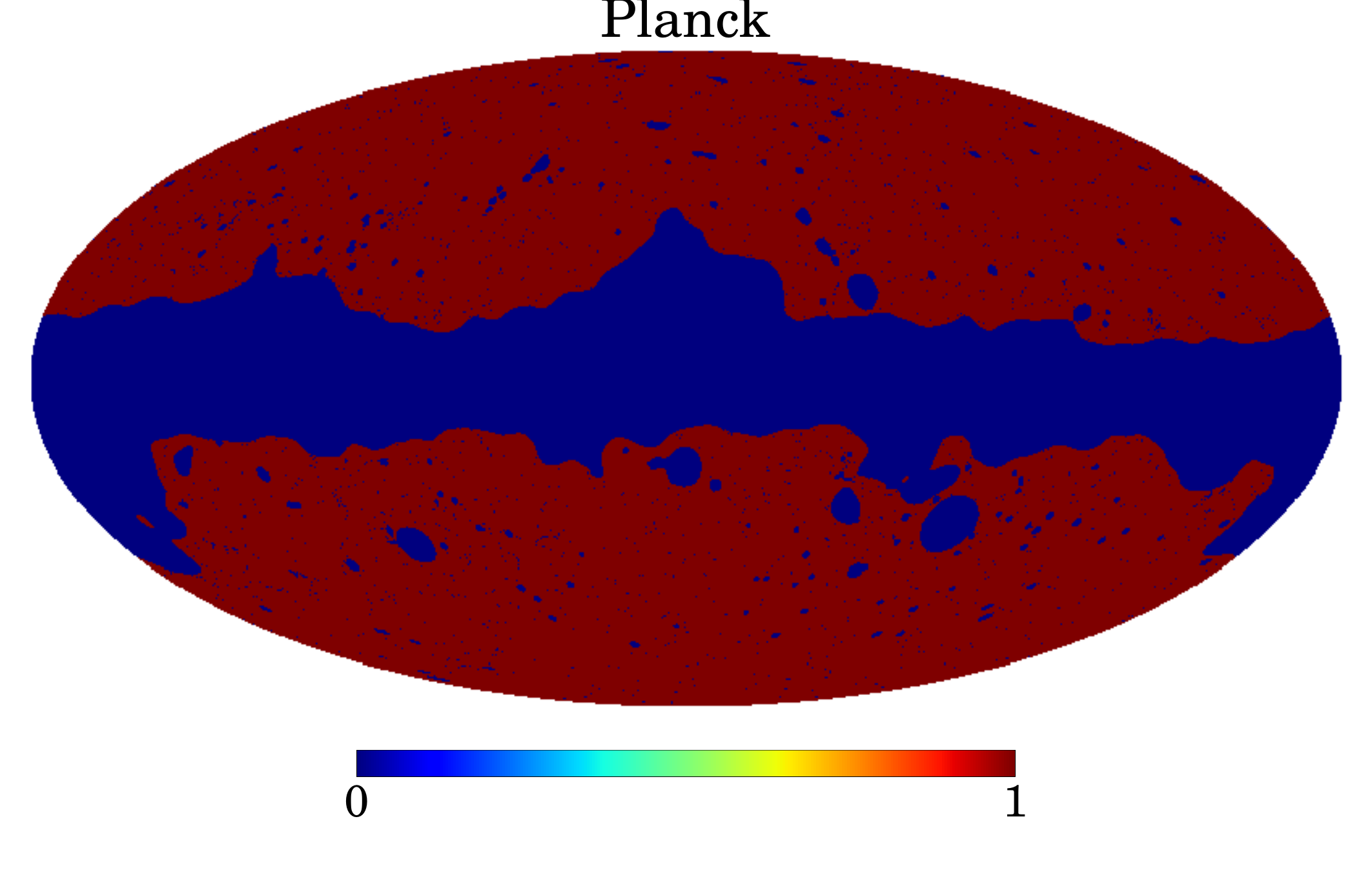}}
\rotatebox{0}{\includegraphics[width=8cm, height=5.1cm]{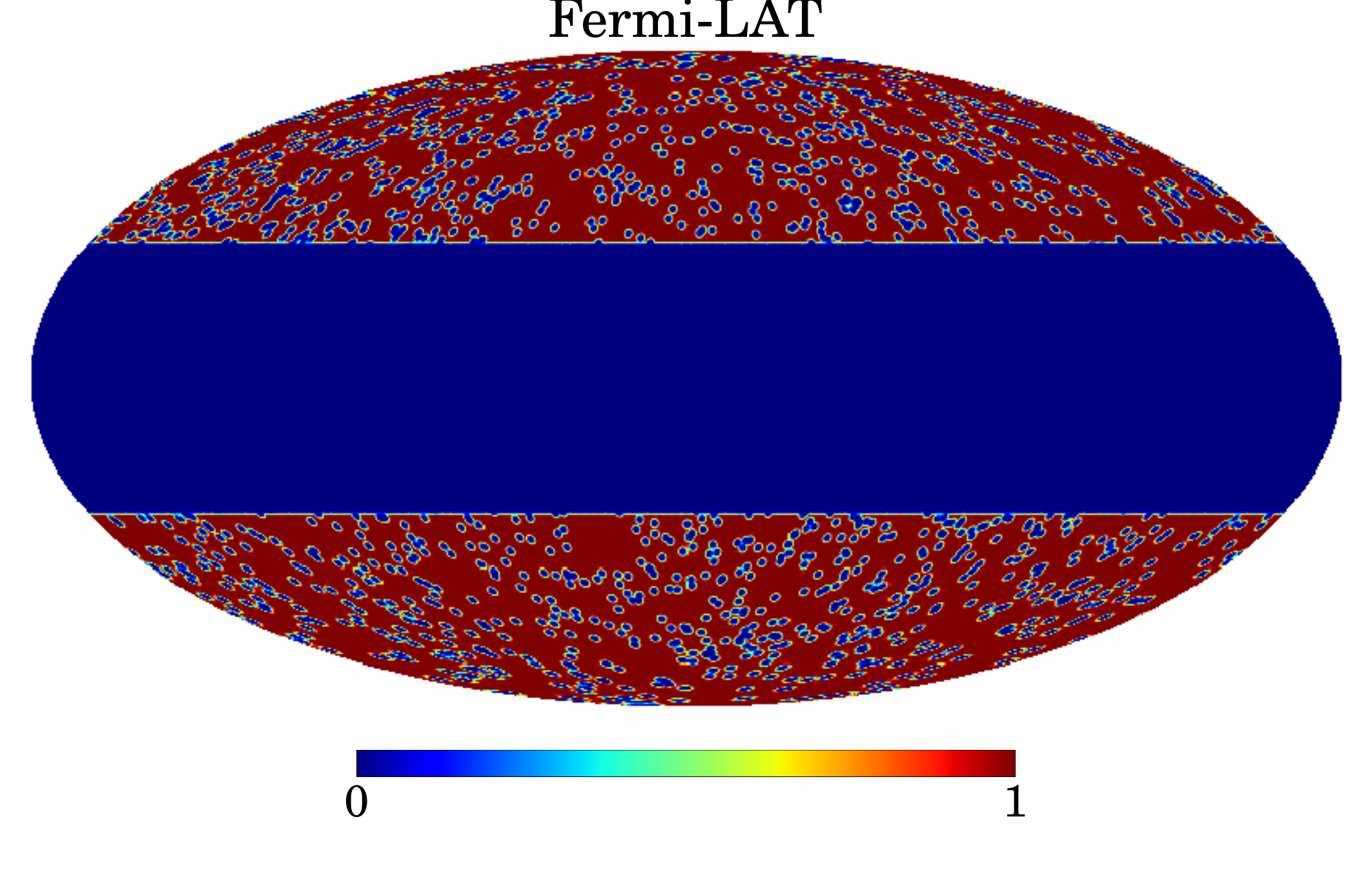}}
\caption{Masks applied to Planck (top) and Fermi-LAT (bottom) data.}
\label{masks}
\end{figure}

For the point-like sources ($X$ or $Y$=\,\rm{Blazars},  \rm{SFGs}, \rm{FSRQs}, \rm{mAGN}), the power spectra are
\begin{eqnarray}
C_l^{1h,XX/YY}&=&\int dz\frac{d\chi}{dz}\Big(\frac{a}{\chi}\Big)^2\int^{L_{\rm{max}}(z)}_{L_{\rm{min}}(z)} dL \Phi_X(L,z)\nonumber\\
&&\Big[\frac{W^{(X)}(z)}{a}\frac{L}{\langle g_X\rangle}\Big]\Big[\frac{W^{(X)}(z)}{a}\frac{L}{\langle g_X\rangle}\Big]
\end{eqnarray}
and

\begin{eqnarray}
C_l^{2h,XY}&=&\int dz\frac{d\chi}{dz}\Big(\frac{a}{\chi}\Big)^2P_{\rm{lin}}(k,z)\nonumber\\
&&\Big[\int^{L_{\rm{max}}(z)}_{L_{\rm{min}}(z)} dL \Phi_X(L,z)b(L(m),z)\frac{W^{(X)}(z)}{a}\frac{L}{\langle g_X\rangle}\Big]\nonumber\\
&&\Big[\int^{L_{\rm{max}}(z)}_{L_{\rm{min}}(z)} dL \Phi_Y(L,z)b(L(m),z)\frac{W^{(Y)}(z)}{a}\frac{L}{\langle g_Y\rangle}\Big].\nonumber\\
\end{eqnarray}

The 1-halo term $C_l^{1h,XY} (X\neq Y)$ is negligibly small because we assume they are both point-like sources, and it is Poisson noise, which is independent of spatial clustering.

The point-like sources should also trace the underlying DM distribution; thus, the correlation between the matter distribution and the point-like sources is non-vanishing. Their angular correlation can be estimated from the two categories discussed above and expressed as
\begin{eqnarray}
C_l^{1h,XY}&=&\int dz\frac{d\chi}{dz}\Big(\frac{a}{\chi}\Big)^2\int^{L_{\rm{max}}(z)}_{L_{\rm{min}}(z)} dL \Phi_Y(L,z)\nonumber\\
&&X_l(k,M(L),z)\Big[\frac{W^{(Y)}(z)}{a}\frac{L}{\langle g_Y\rangle}\Big]
\end{eqnarray}
and 
\begin{eqnarray}
C_l^{2h,XY}&=&\int dz\frac{d\chi}{dz}\Big(\frac{a}{\chi}\Big)^2P_{\rm{lin}}(k,z)\Big[\int dMb(M,z)n(M,z)\nonumber\\
&&\tilde X_l(k,M,z)\Big]\Big[\int^{L_{\rm{max}}(z)}_{L_{\rm{min}}(z)} dL \Phi_Y(L,z)b(L(m),z)\nonumber\\
&&\frac{W^{(Y)}(z)}{a}\frac{L}{\langle g_Y\rangle}\Big].
\end{eqnarray}

In these equations, $X$=\{\rm{lensing}, \rm{CIB}, \rm{DM\, annihilation}, \rm{DM\, decay}\} and $Y$=\{\rm{Blazars}, \rm{SFGs}, \rm{FSRQs}, \rm{mAGN}\}. In our analysis, we will focus on this type of cross-correlation.

\begin{figure}
\rotatebox{0}{\includegraphics[width=9cm, height=7.5cm]{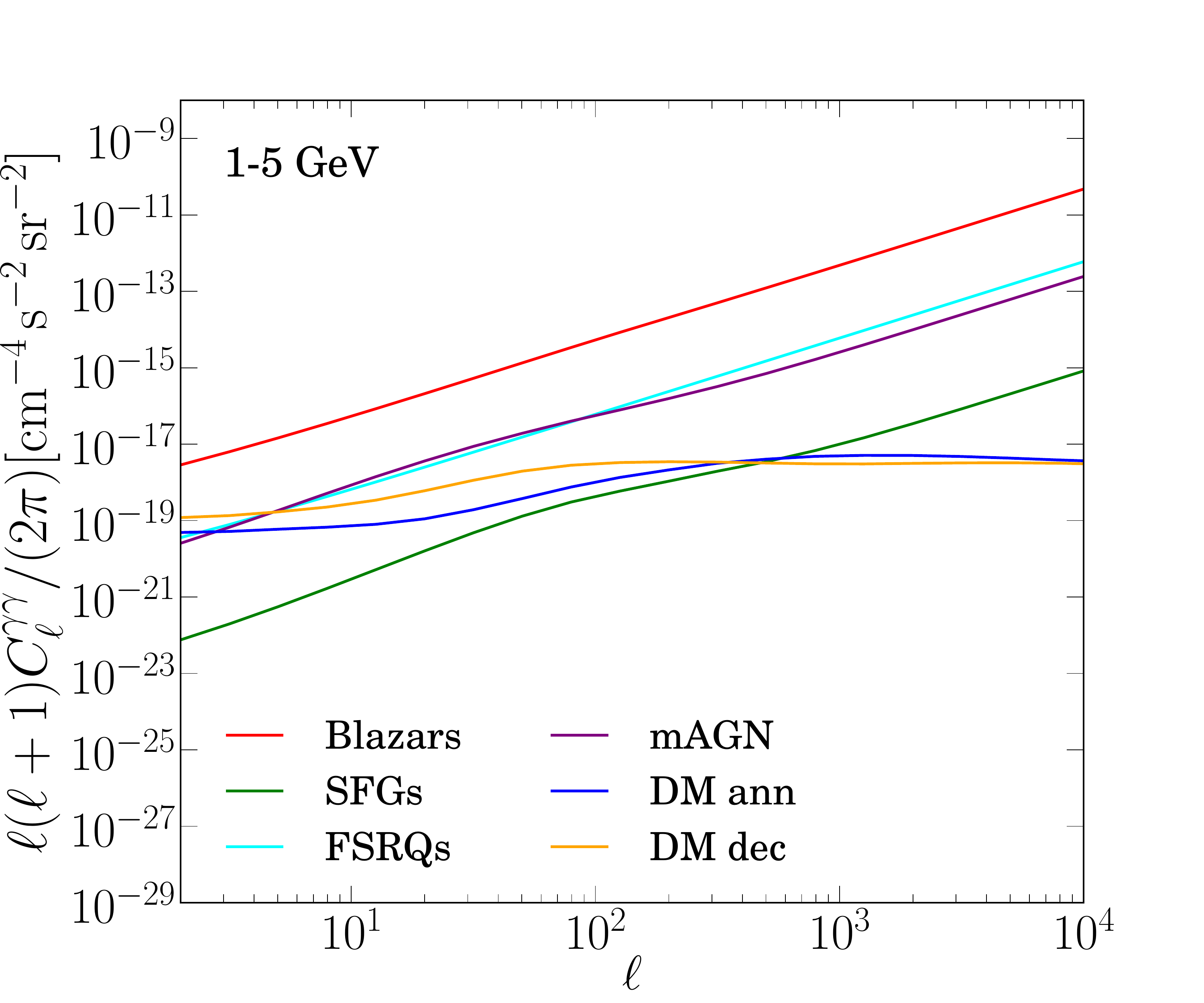}}
\rotatebox{0}{\includegraphics[width=9cm, height=7.5cm]{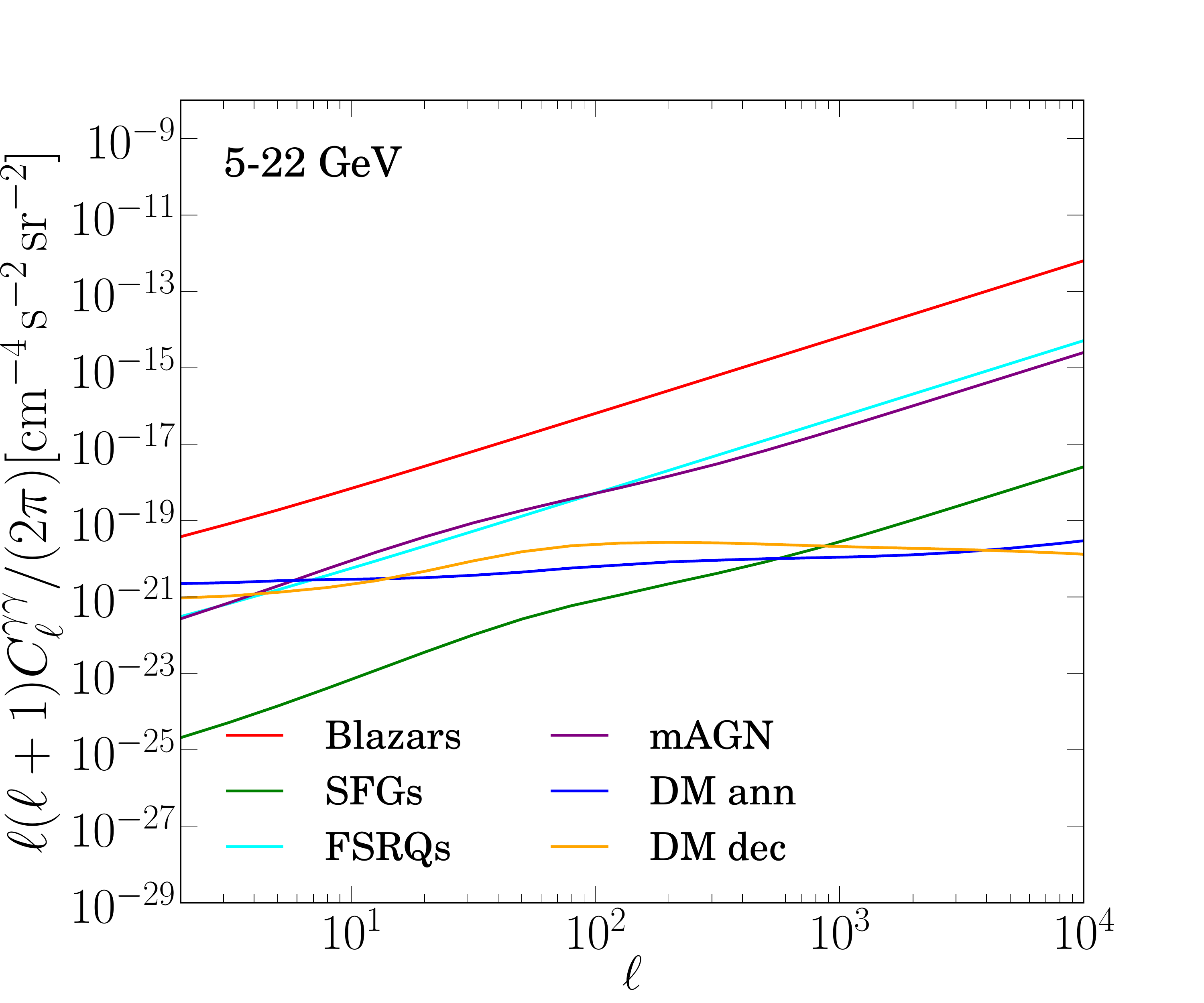}}
\rotatebox{0}{\includegraphics[width=9cm, height=7.5cm]{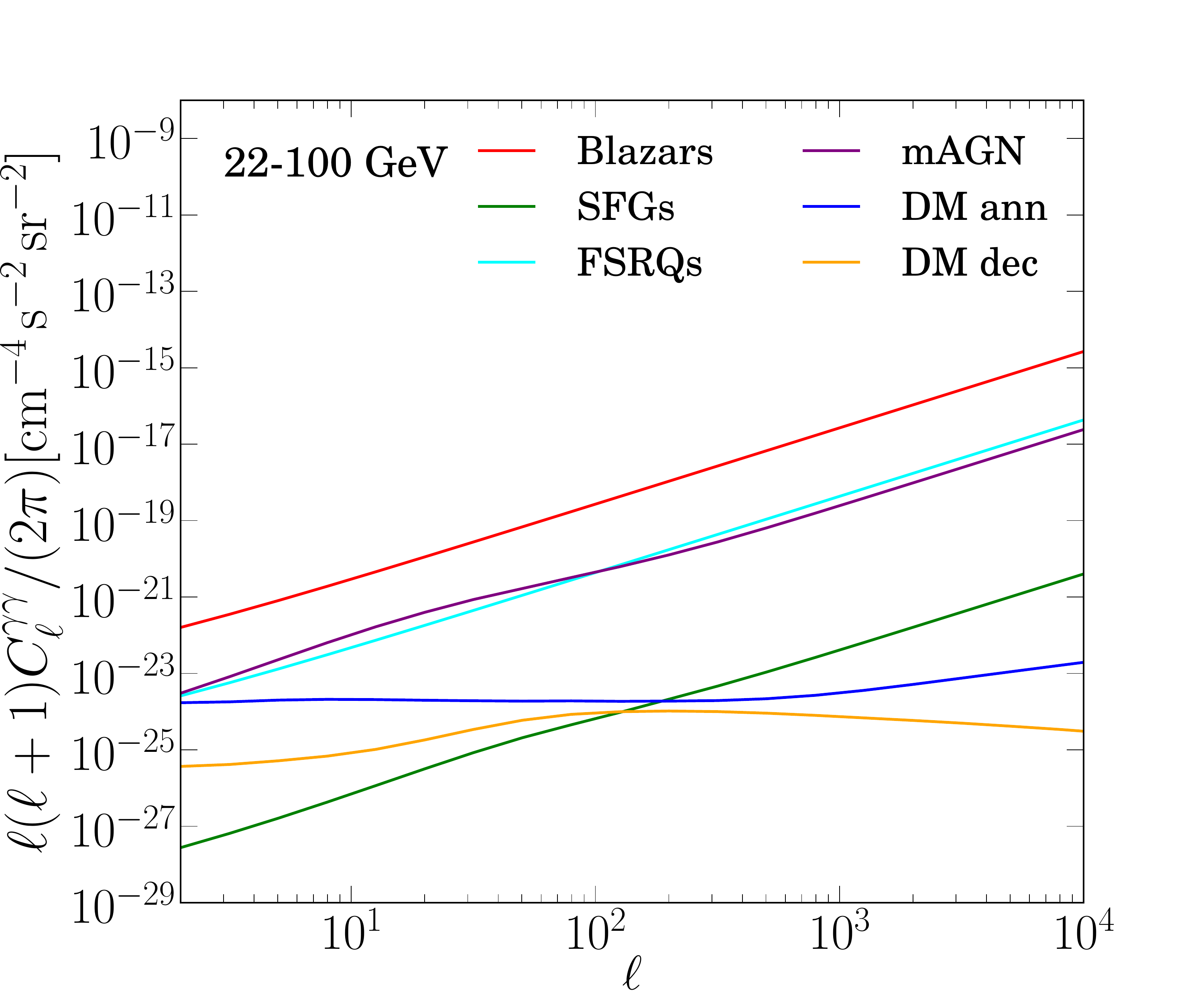}}
\caption{The theoretical auto-power spectra of the \rr ray anisotropies. The $\gamma$-ray anisotropy is mainly dominated by Blazars. }
\label{fermi-theory-auto}
\end{figure}

\begin{figure}
\rotatebox{0}{\includegraphics[width=9cm, height=7.5cm]{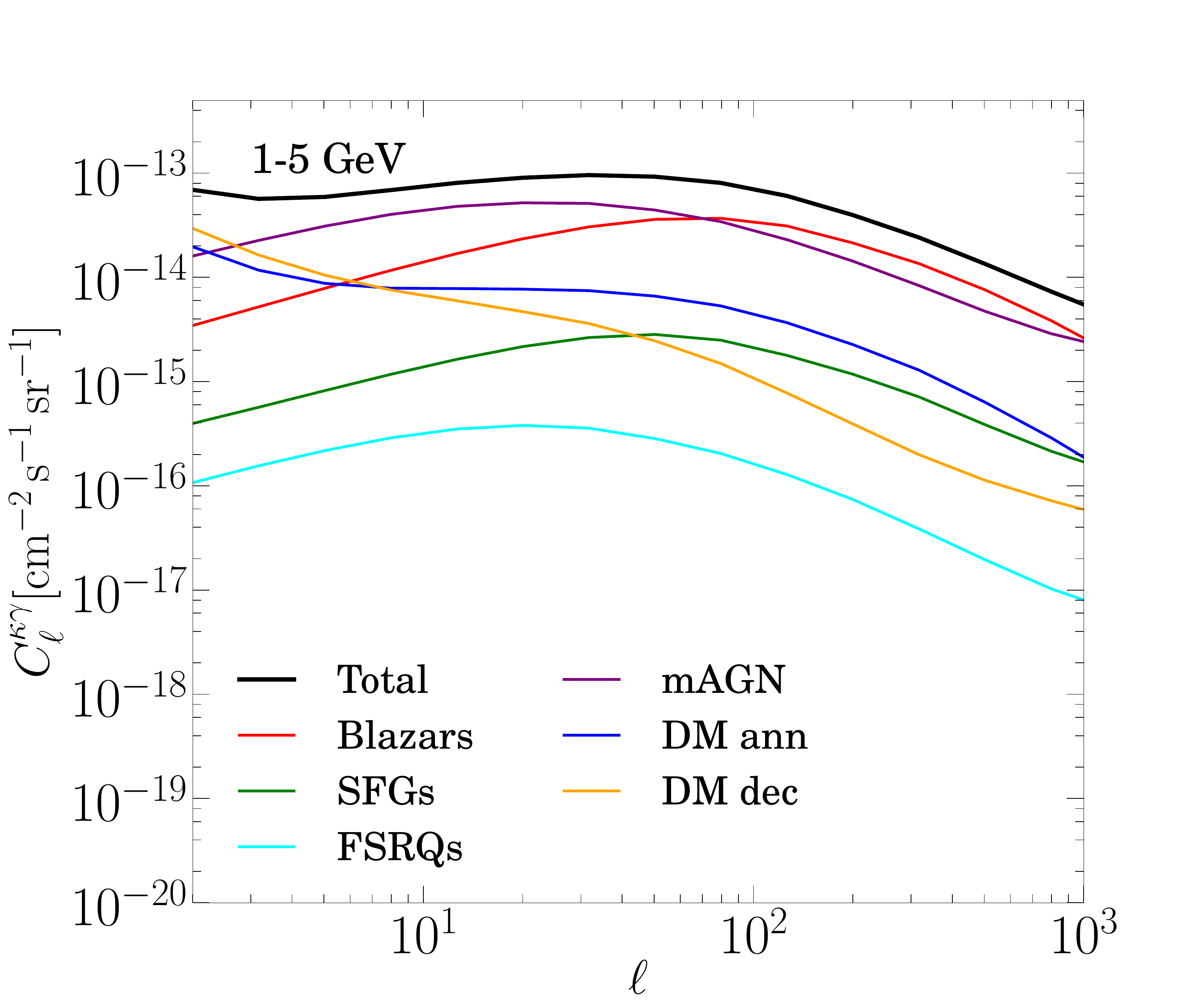}}
\rotatebox{0}{\includegraphics[width=9cm, height=7.5cm]{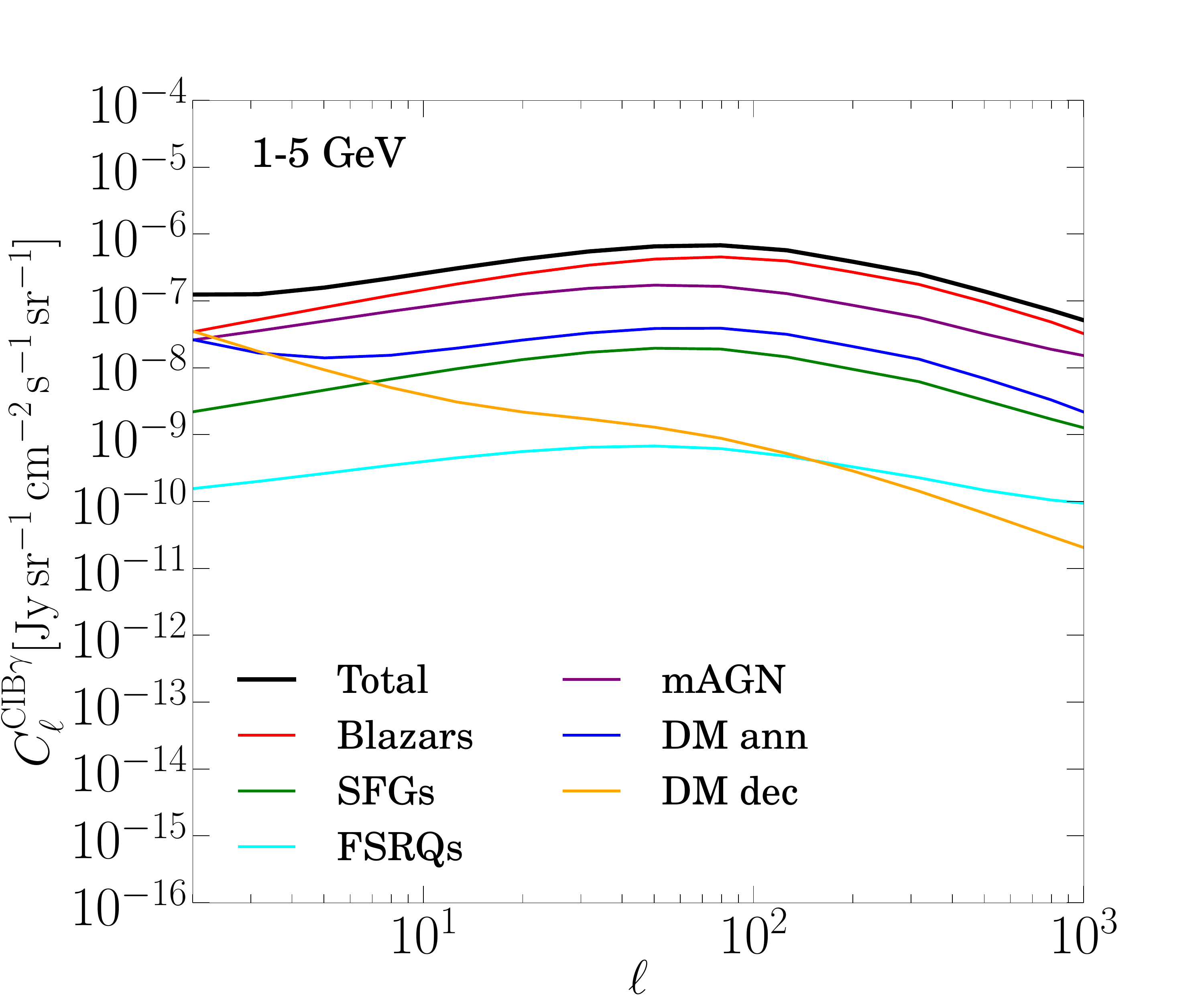}}
\caption{The DM signals that are probed in this cross-correlation analysis. (Top) The predicted cross-correlation between Planck lensing and the $\gamma$-ray. (Bottom) The predicted cross-correlation between Planck CIB and the $\gamma$-ray.}
\label{fermi-theory-cross}
\end{figure}

\begin{figure}
\rotatebox{0}{\includegraphics[width=9cm, height=7.5cm]{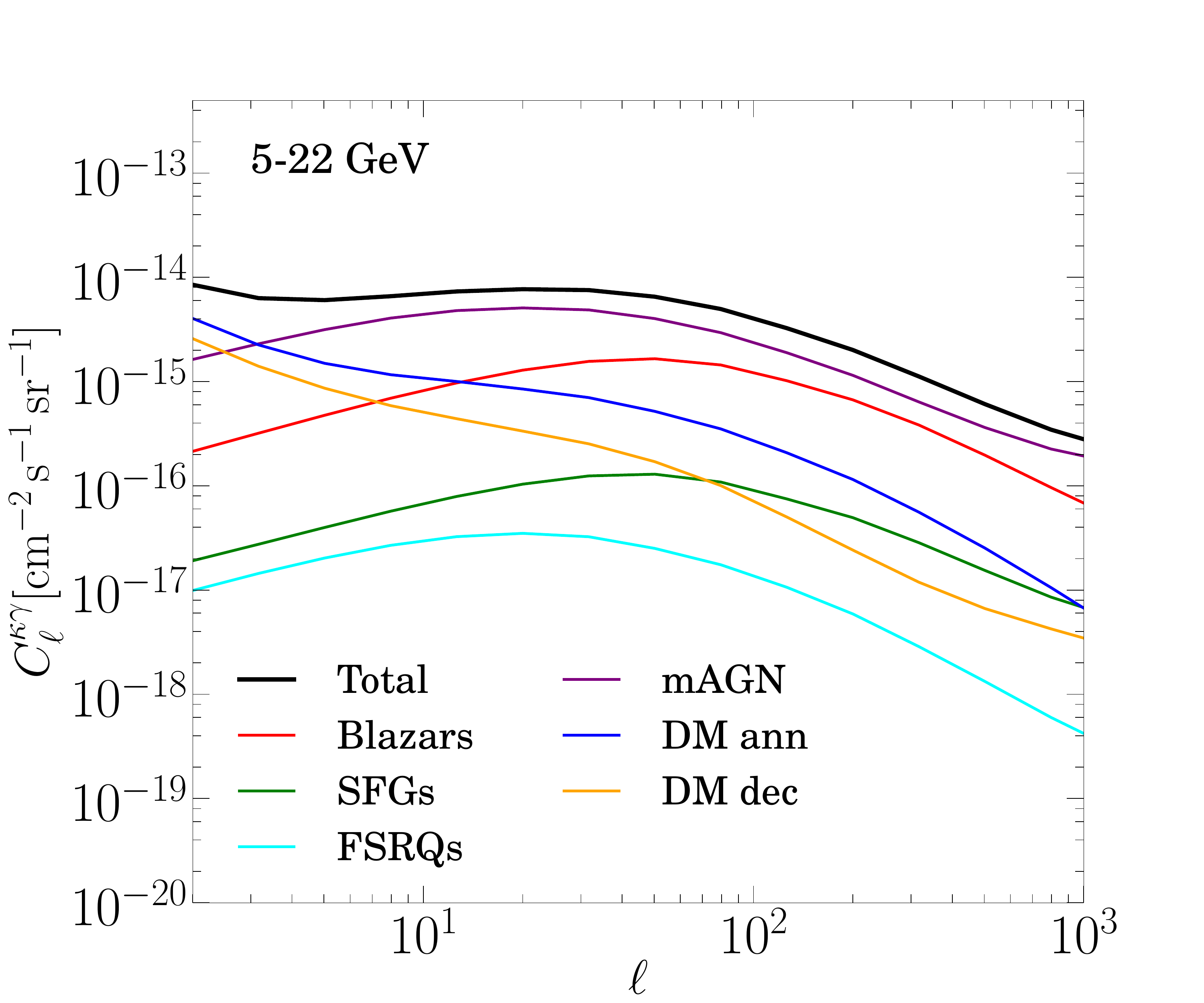}}
\rotatebox{0}{\includegraphics[width=9cm, height=7.5cm]{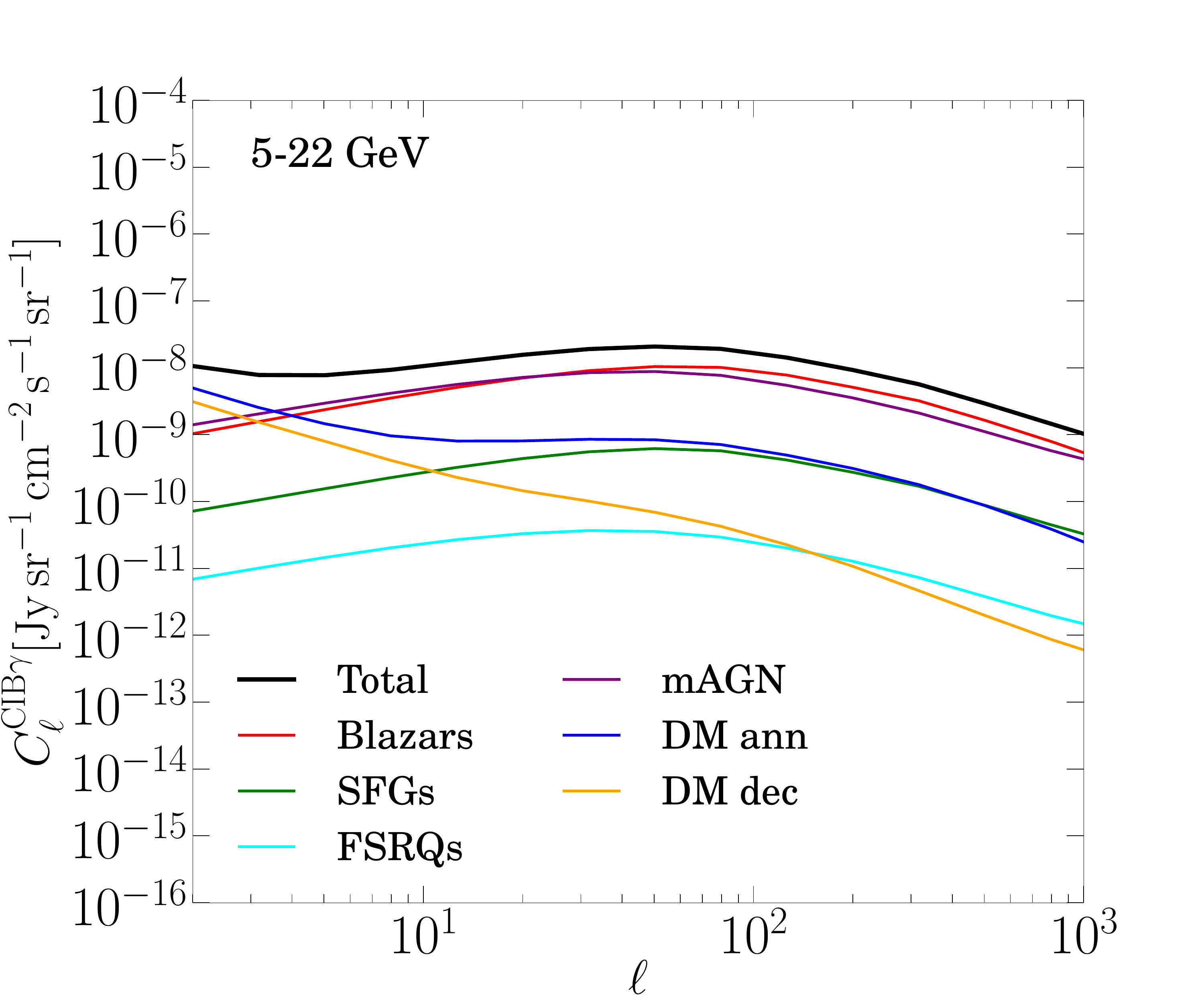}}
\caption{The DM signals that are probed in this cross-correlation analysis. (Top) The predicted cross-correlation between Planck lensing and the $\gamma$-ray. (Bottom) The predicted cross-correlation between Planck CIB and the $\gamma$-ray.}
\label{fermi-theory-cross_1}
\end{figure}

\begin{figure}
\rotatebox{0}{\includegraphics[width=9cm, height=7.5cm]{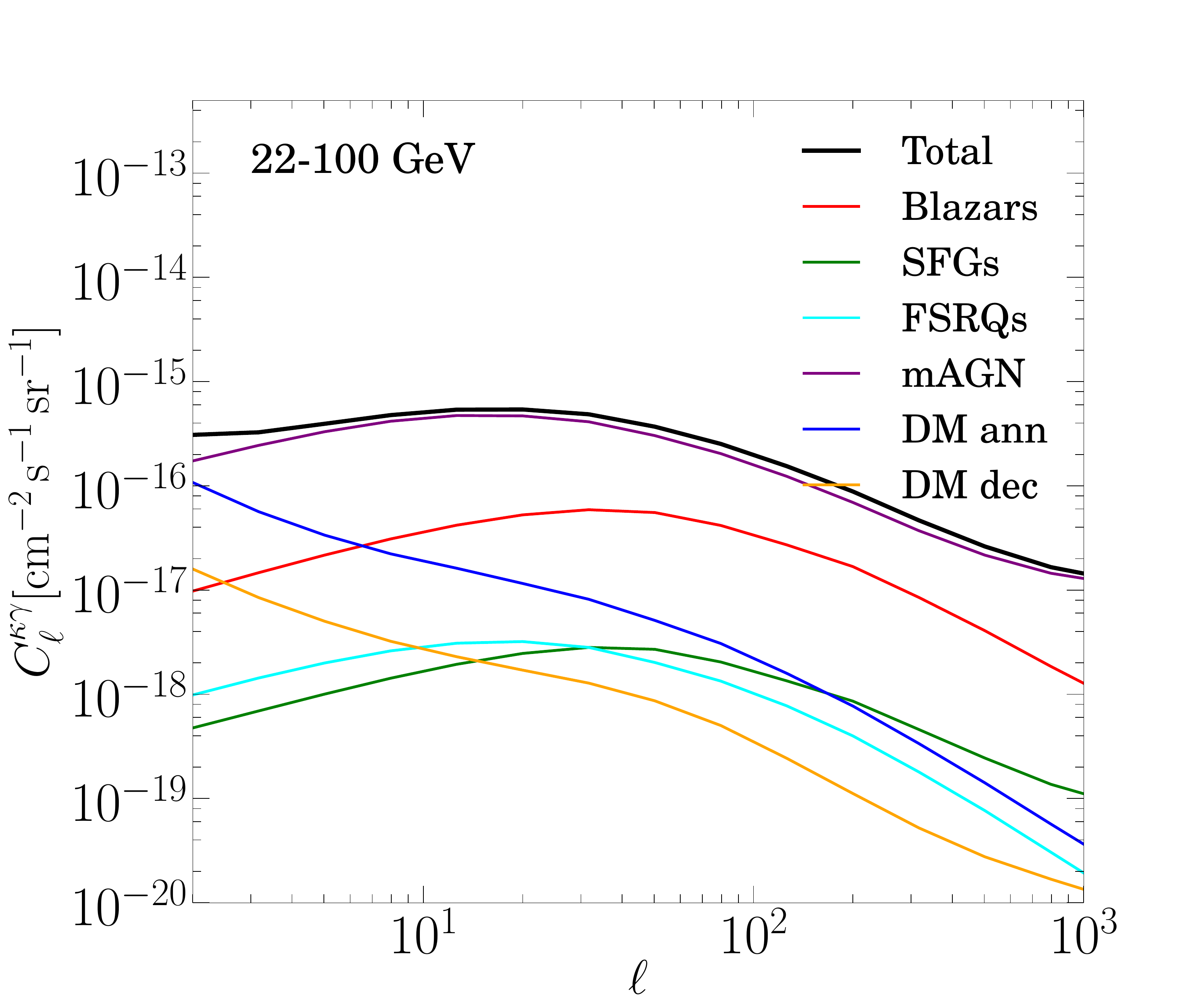}}
\rotatebox{0}{\includegraphics[width=9cm, height=7.5cm]{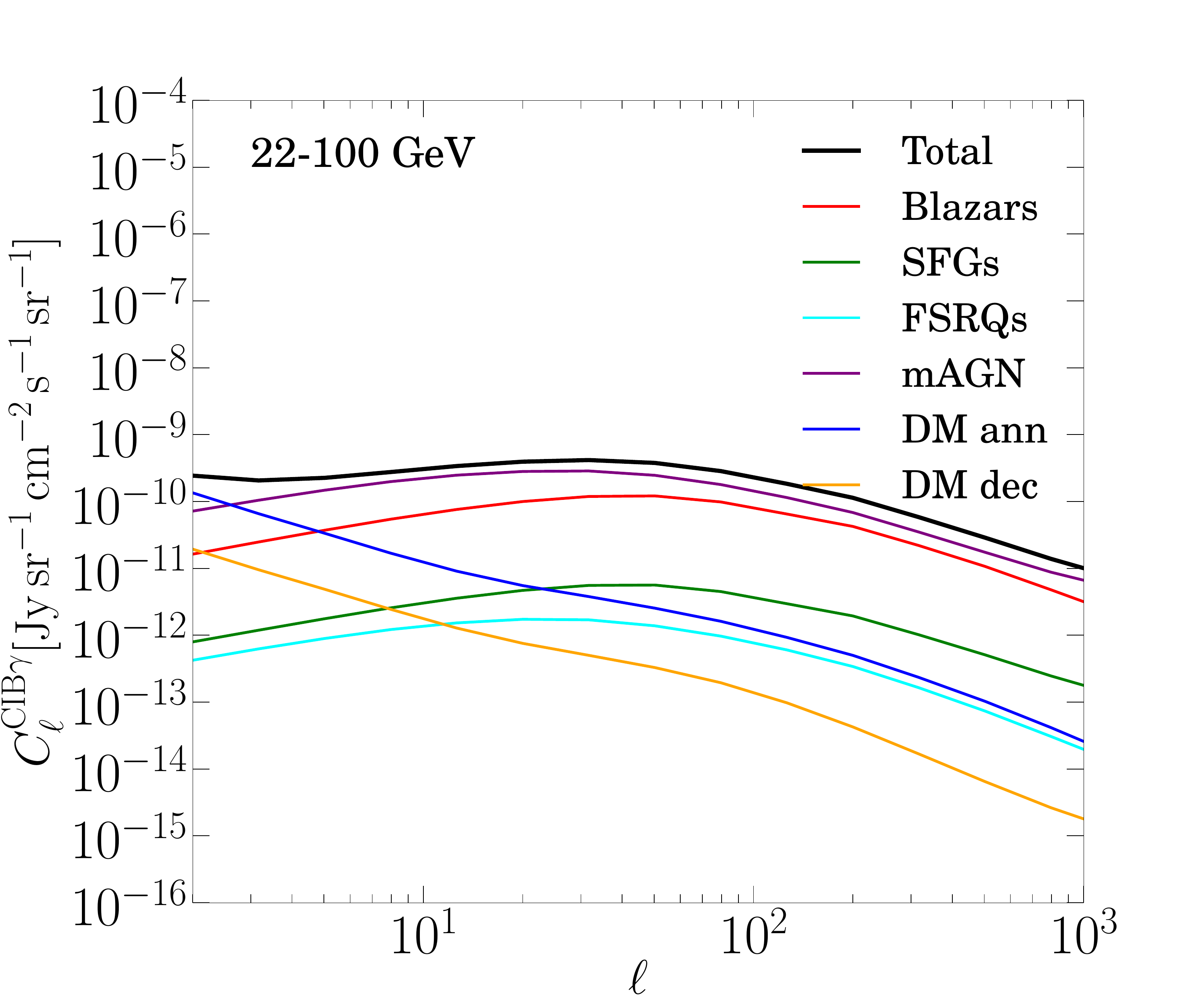}}
\caption{The DM signals that are probed in this cross-correlation analysis. (Top) The predicted cross-correlation between Planck lensing and the $\gamma$-ray. (Bottom) The predicted cross-correlation between Planck CIB and the $\gamma$-ray.}
\label{fermi-theory-cross_2}
\end{figure}

The auto-power spectra for all the species are shown in Fig. \ref{fermi-theory-auto}. It is easy to find that they are dominated by the astrophysical components at different energy levels. Thus, it is difficult to detect the DM signals directly from the auto correlations. However, the cross-correlation with large scale structure (LSS) tracers can significantly boost the DM signals as Figs. (\ref{fermi-theory-cross}, \ref{fermi-theory-cross_1}, \ref{fermi-theory-cross_2}) indicate. A benchmark particle physics model with $m_{\rm{DM}}=100\,\rm{GeV}$, $\langle\sigma\nu\rangle=3\times10^{-26}\rm{cm}^3\,\rm{s}^{-1}$ and $\Gamma_d=1/(6\times10^{27})\,\rm{s}^{-1}$ is used to make these plots~\cite{cross_rg}.

\begin{figure}
\rotatebox{0}{\includegraphics[width=8.5cm, height=7.2cm]{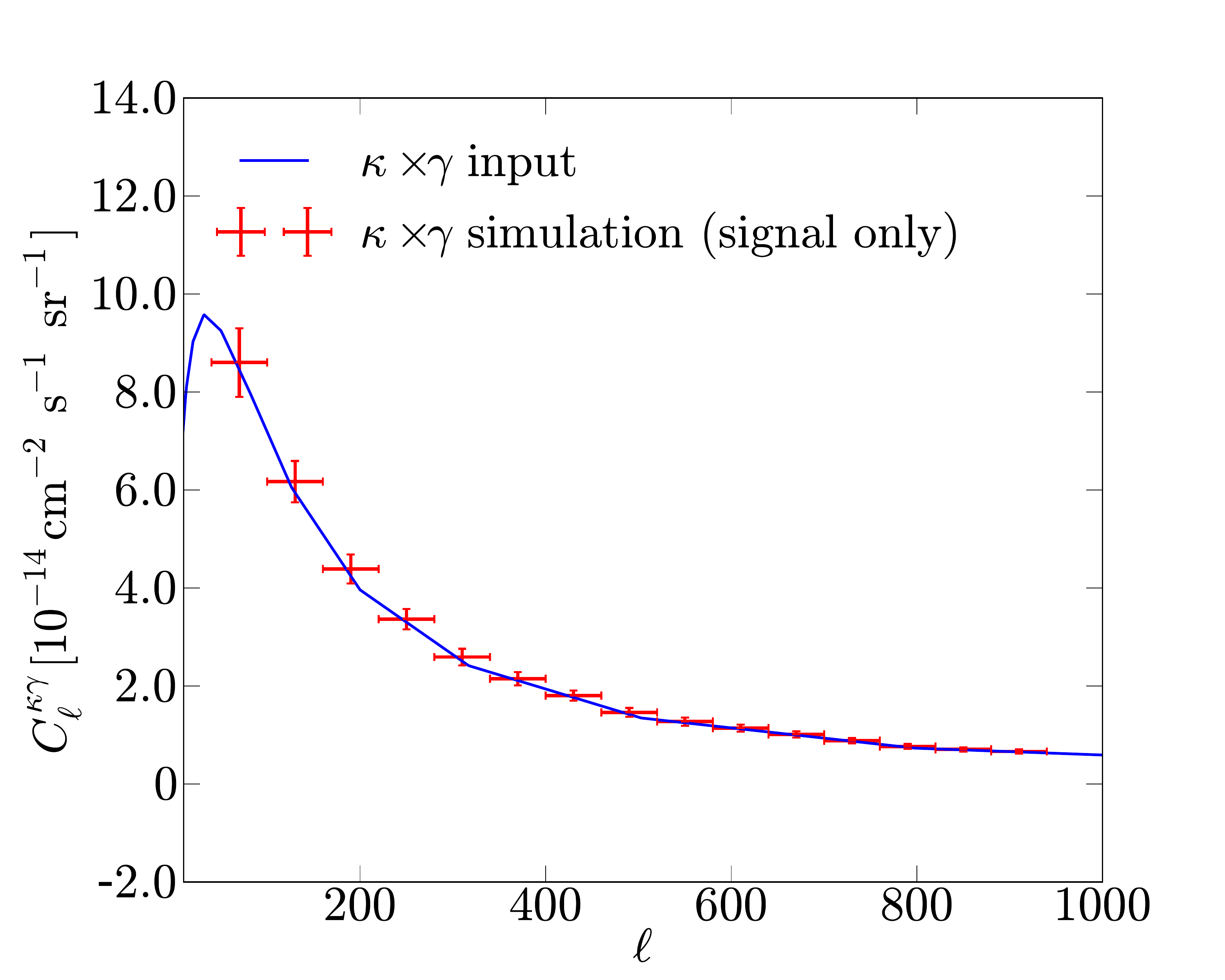}}
\caption{The cross-power spectrum spectrum recovered from correlated simulations.}
\label{cross_sim}
\end{figure}

\begin{figure}
\rotatebox{0}{\includegraphics[width=8cm, height=7.2cm]{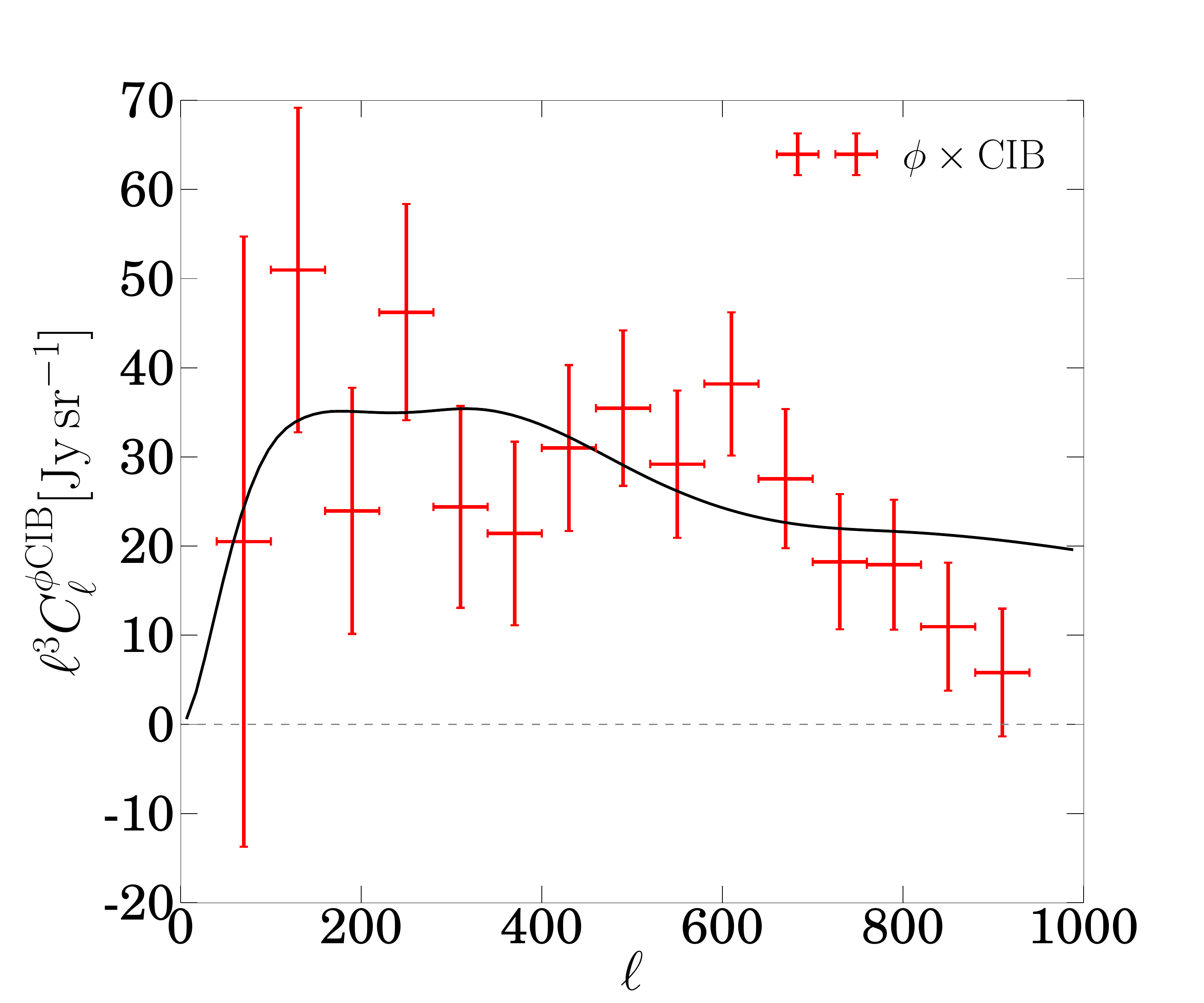}}
\rotatebox{0}{\includegraphics[width=8cm, height=7.2cm]{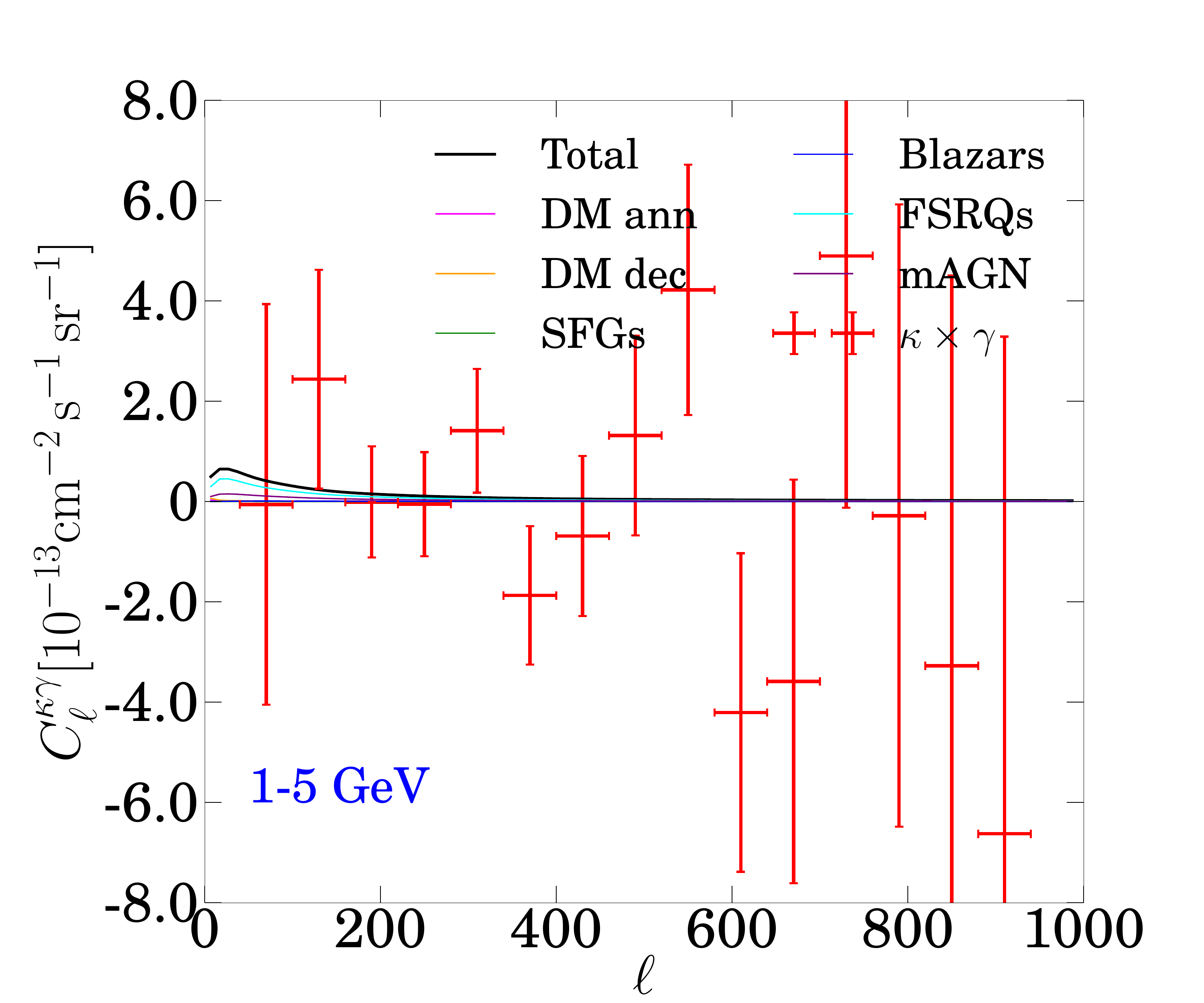}}
\rotatebox{0}{\includegraphics[width=8cm, height=7.2cm]{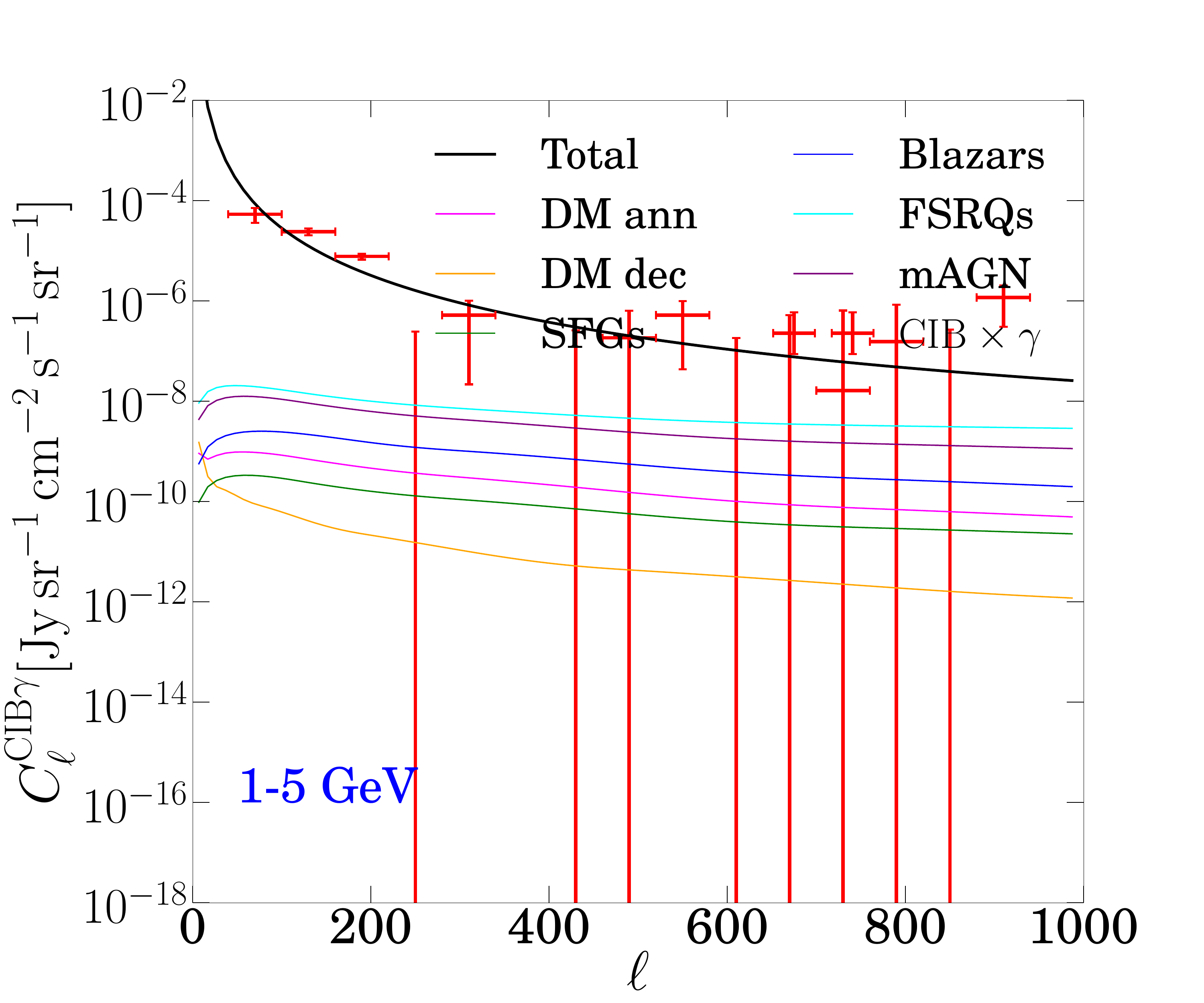}}
\caption{The measured cross-correlation power spectra from Planck lensing, Planck CIB and Fermi-LAT $\gamma$-ray maps at 1-5 GeV.}
\label{ps1}
\end{figure}
\begin{figure}
\rotatebox{0}{\includegraphics[width=8cm, height=7.2cm]{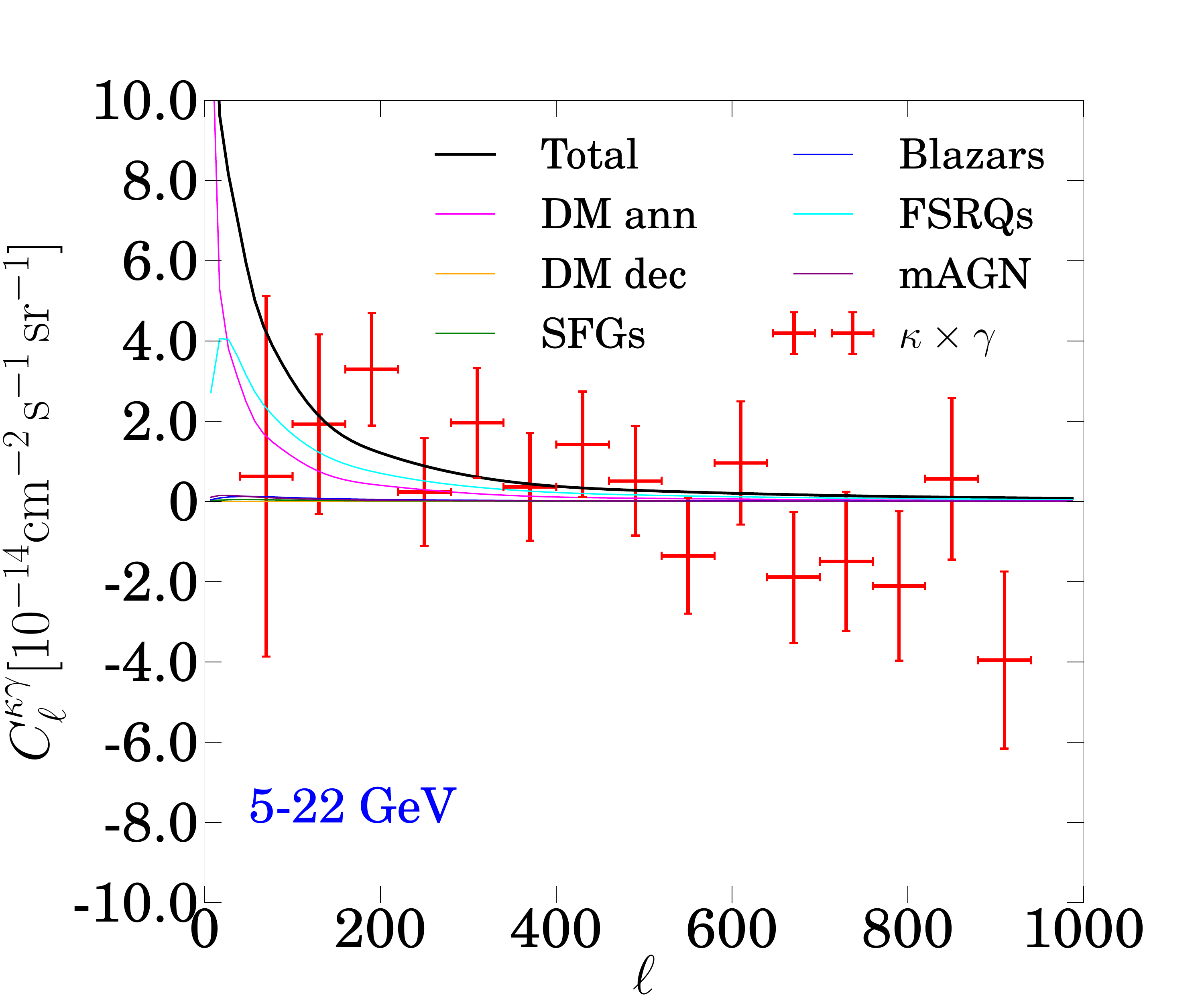}}
\rotatebox{0}{\includegraphics[width=8cm, height=7.2cm]{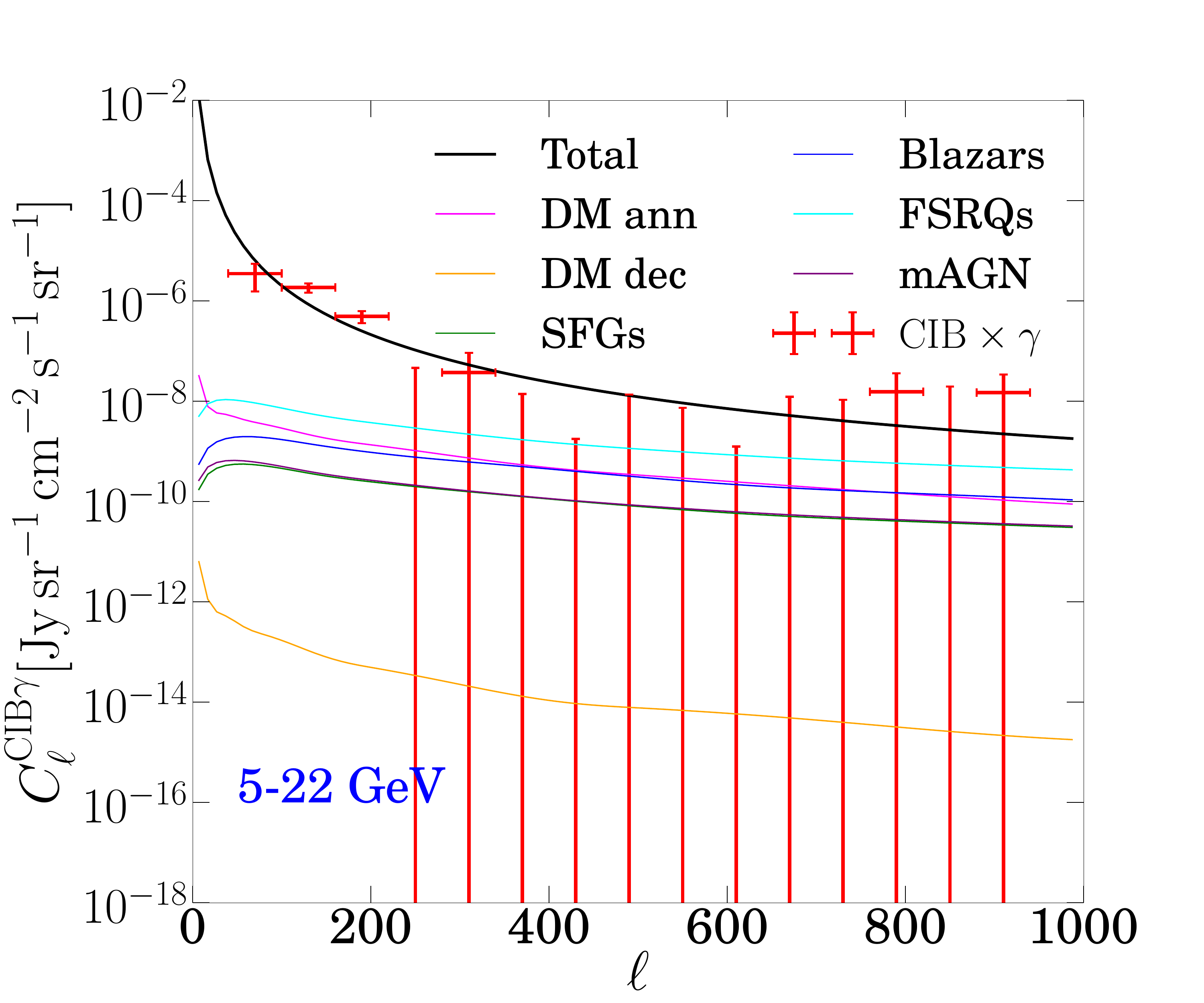}}
\caption{The measured cross-correlation power spectra from Planck lensing, Planck CIB and Fermi-LAT $\gamma$-ray maps at 5-22 GeV.}
\label{ps2}
\end{figure}
\begin{figure}
\rotatebox{0}{\includegraphics[width=8cm, height=7.2cm]{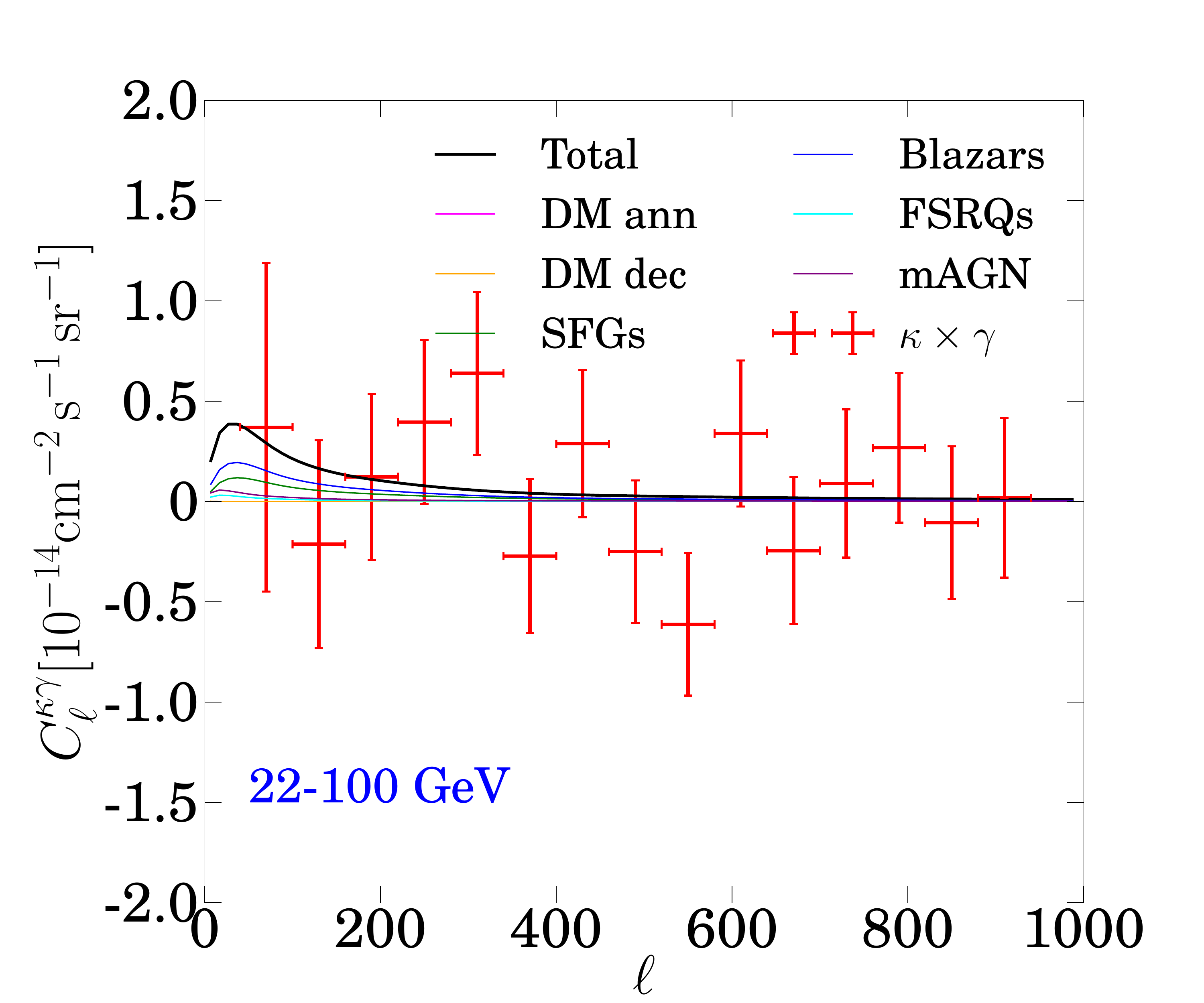}}
\rotatebox{0}{\includegraphics[width=8cm, height=7.2cm]{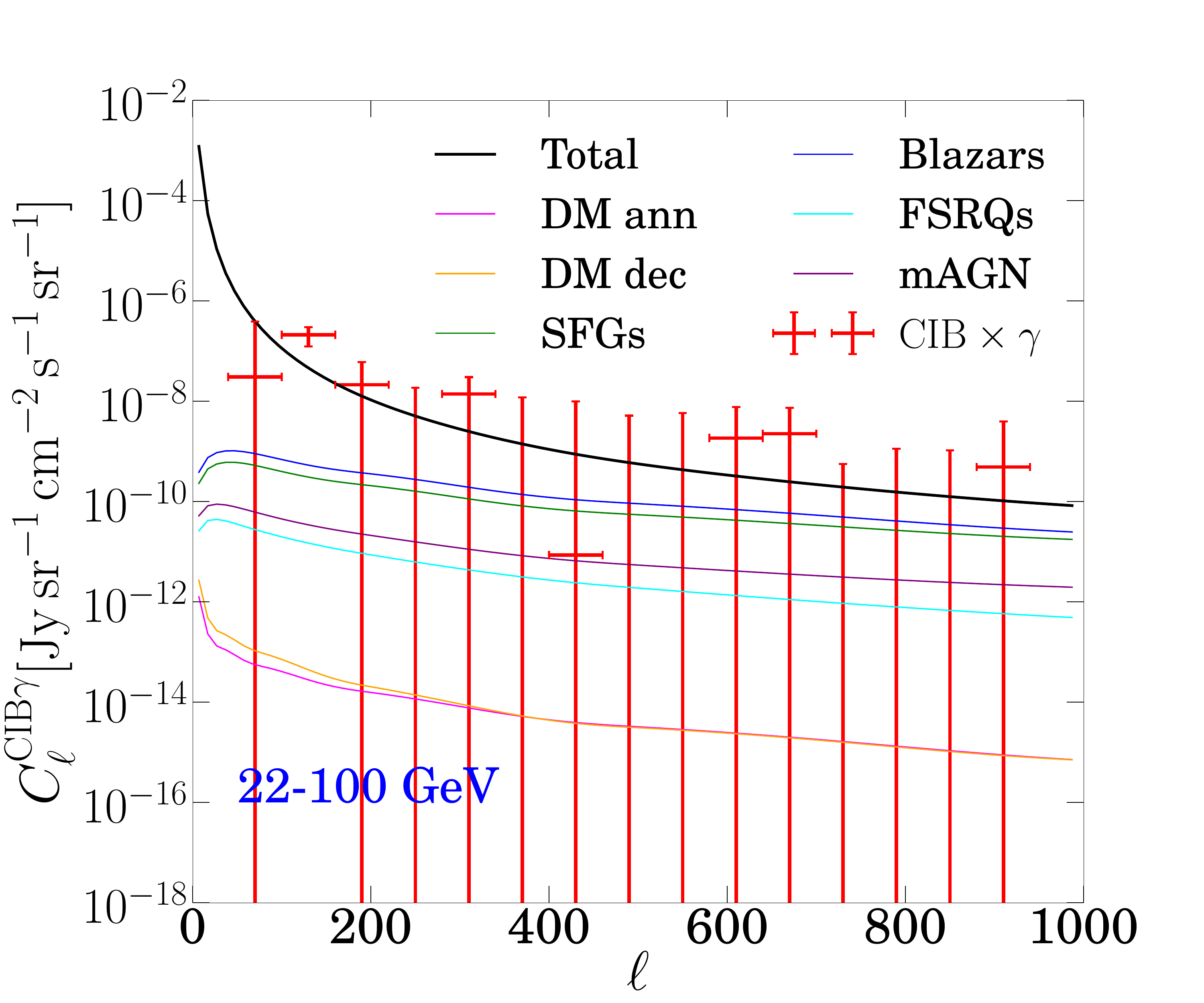}}
\caption{The measured cross-correlation power spectra from Planck lensing, Planck CIB and Fermi-LAT $\gamma$-ray maps at 22-100 GeV.}
\label{ps3}
\end{figure}

\begin{figure}
\rotatebox{0}{\includegraphics[width=8cm, height=7.2cm]{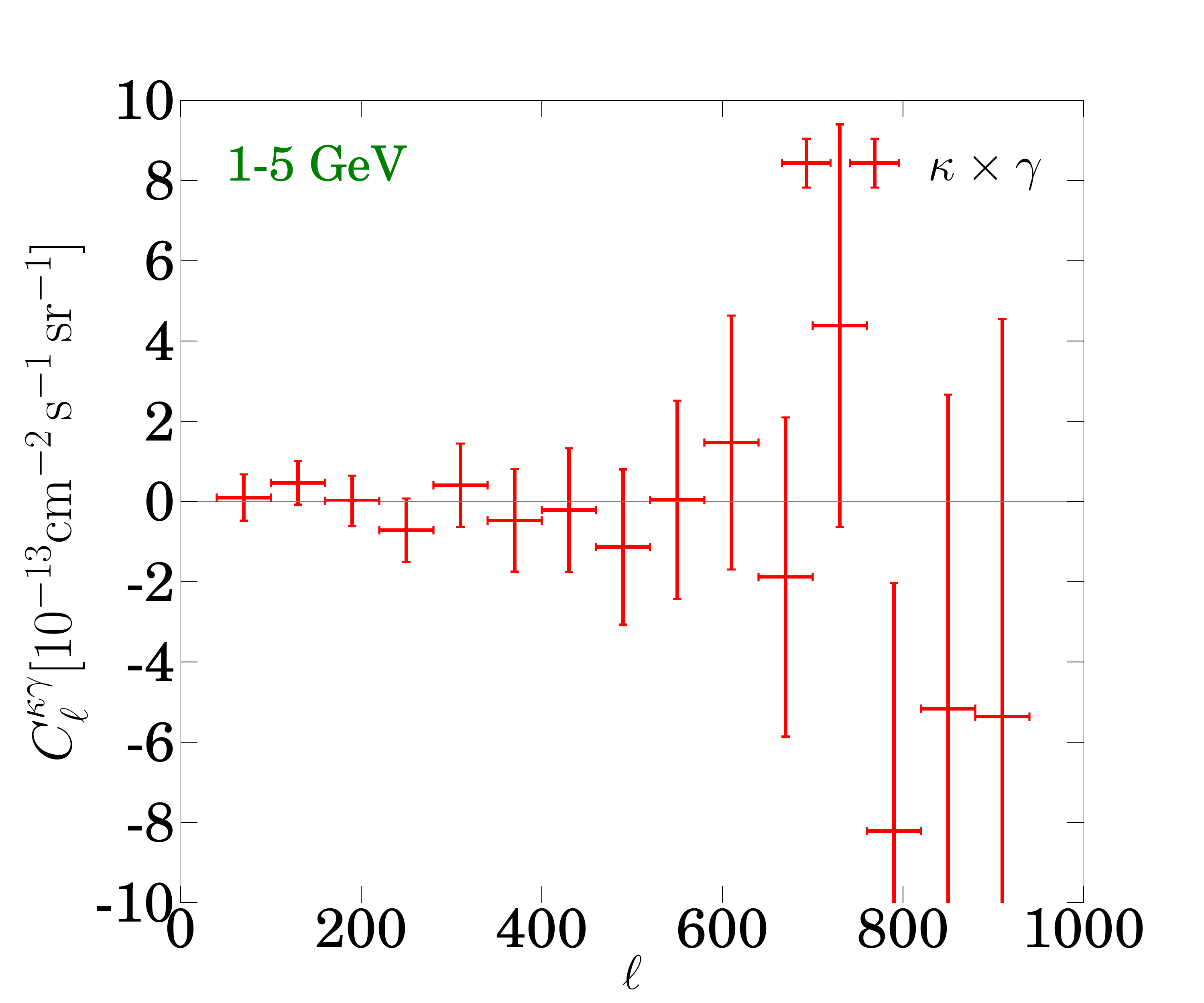}}
\rotatebox{0}{\includegraphics[width=8cm, height=7.2cm]{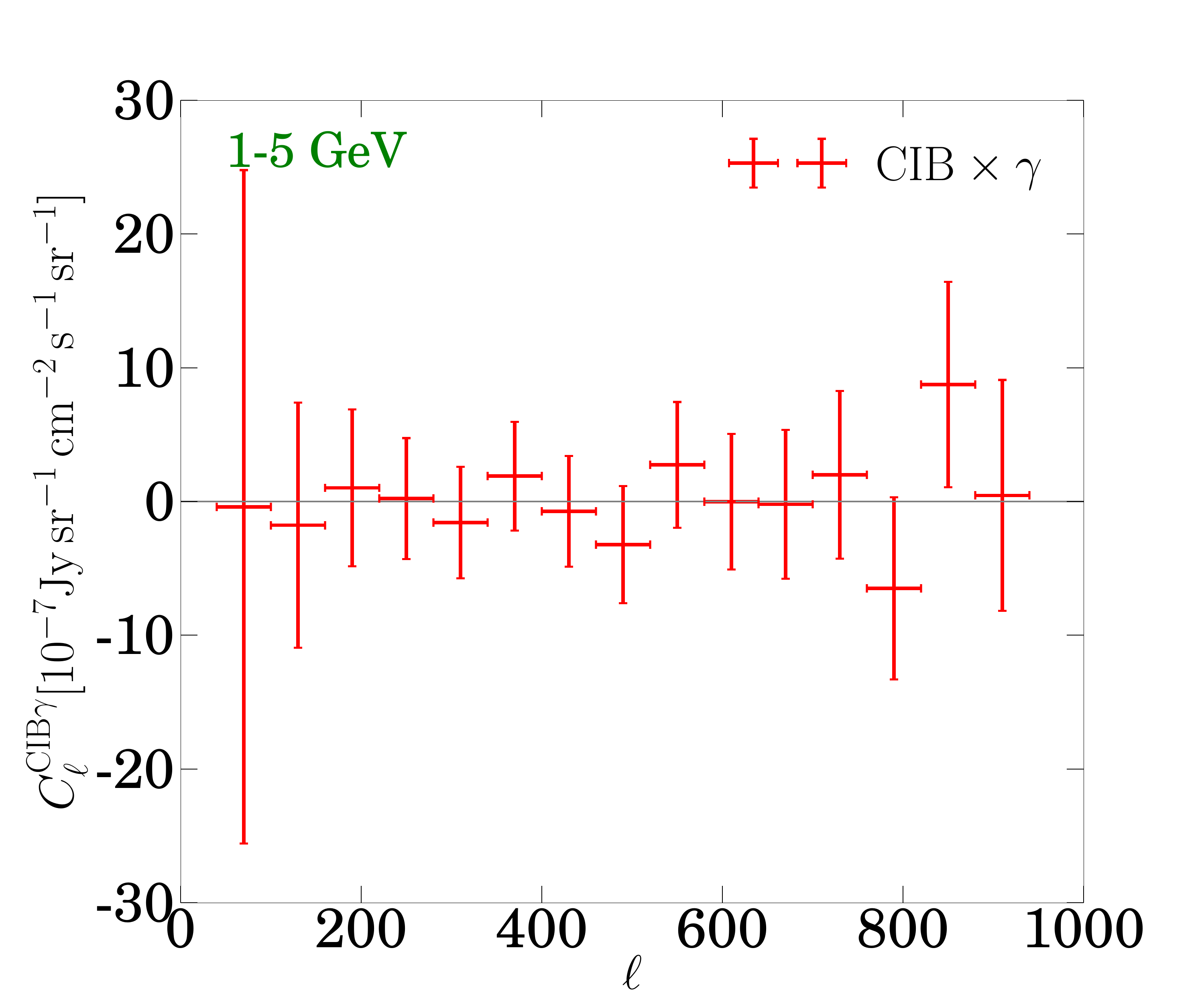}}
\caption{The measured cross-correlation power spectra from Planck lensing, Planck CIB and Fermi-LAT $\gamma$-ray jackknife maps at 1-5 GeV.}
\label{jkk1}
\end{figure}
\begin{figure}
\rotatebox{0}{\includegraphics[width=8cm, height=7.2cm]{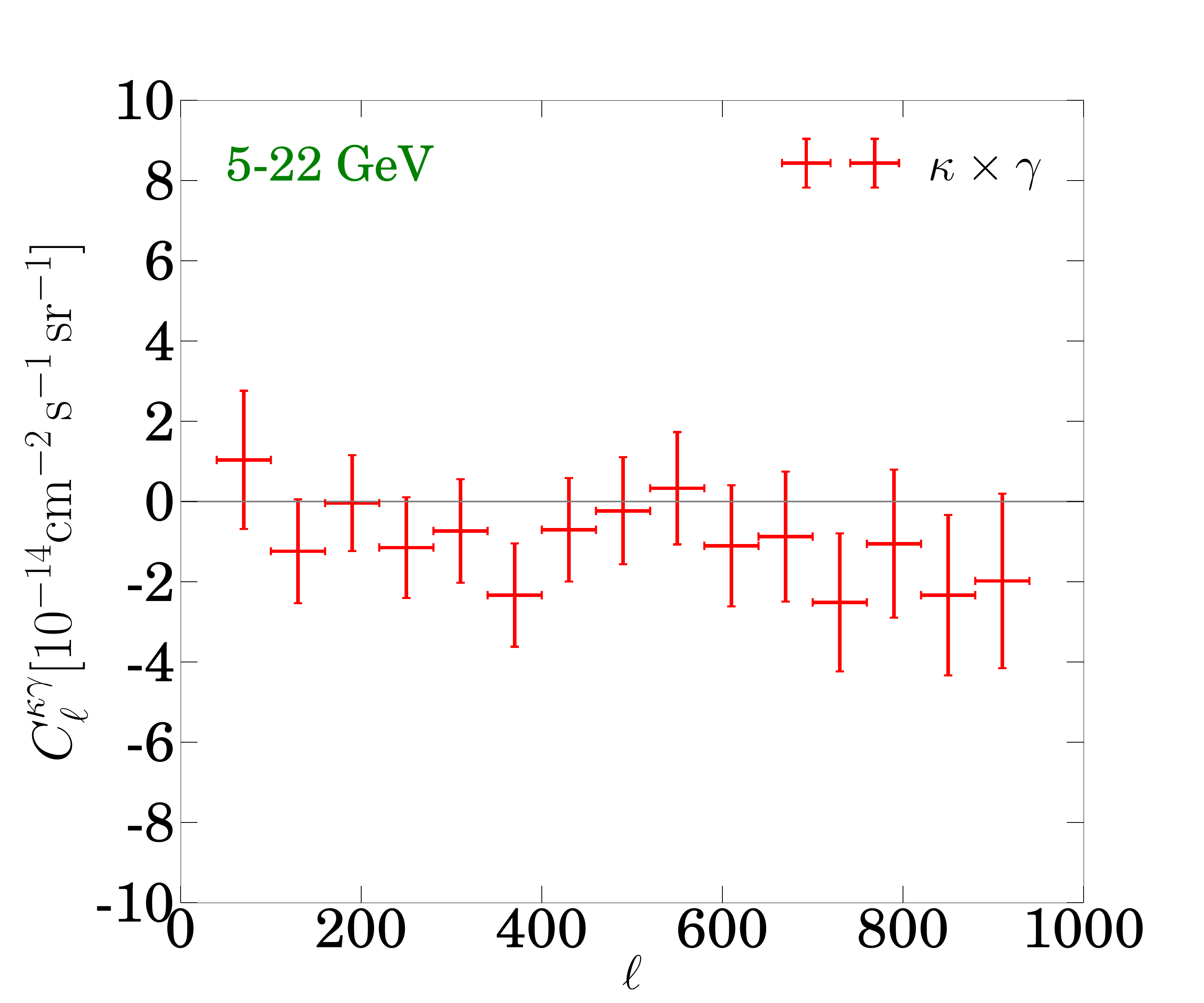}}
\rotatebox{0}{\includegraphics[width=8cm, height=7.2cm]{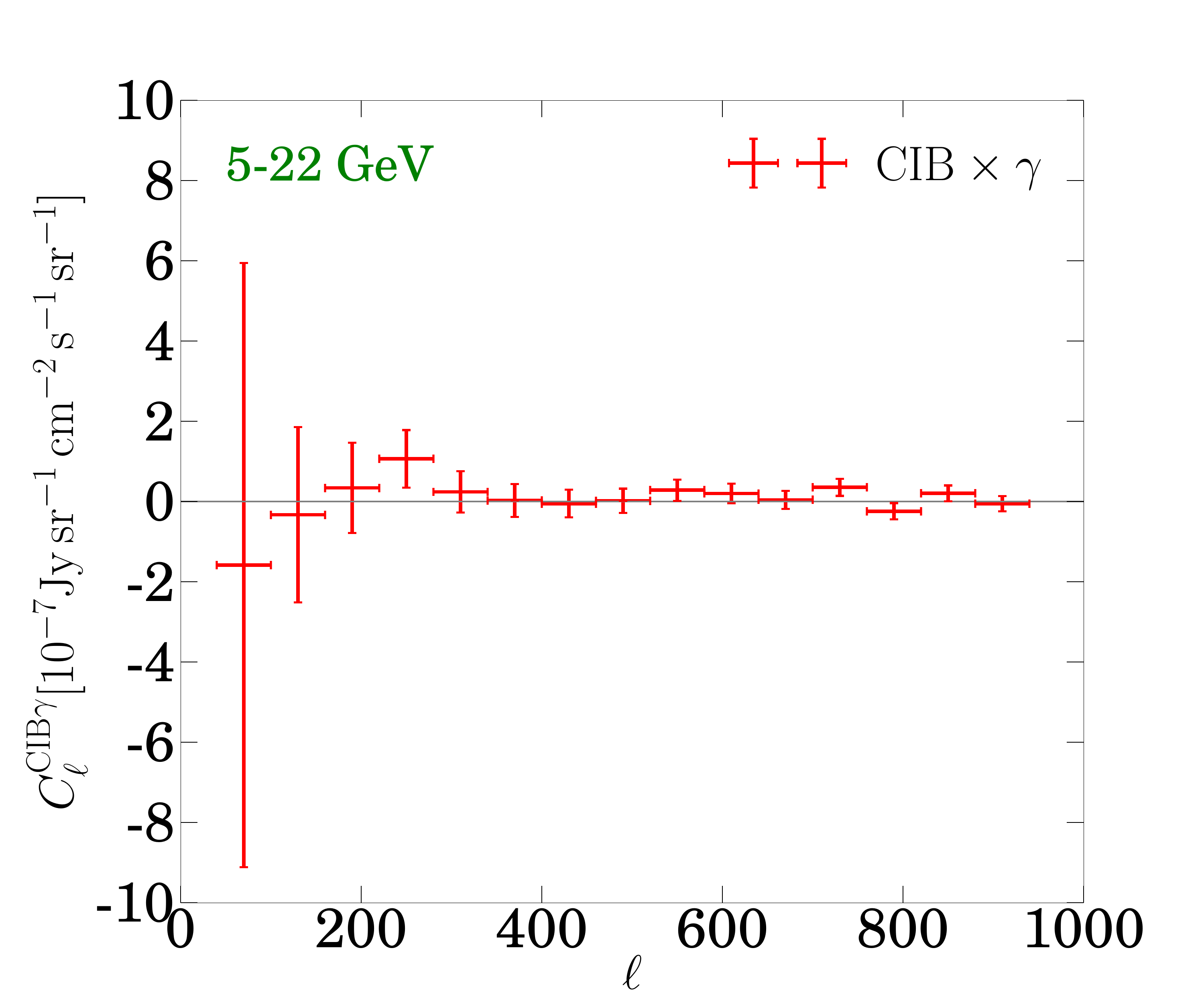}}
\caption{The measured cross-correlation power spectra from Planck lensing, Planck CIB and Fermi-LAT $\gamma$-ray jackknife maps at 5-22 GeV.}
\label{jkk2}
\end{figure}
\begin{figure}
\rotatebox{0}{\includegraphics[width=8cm, height=7.2cm]{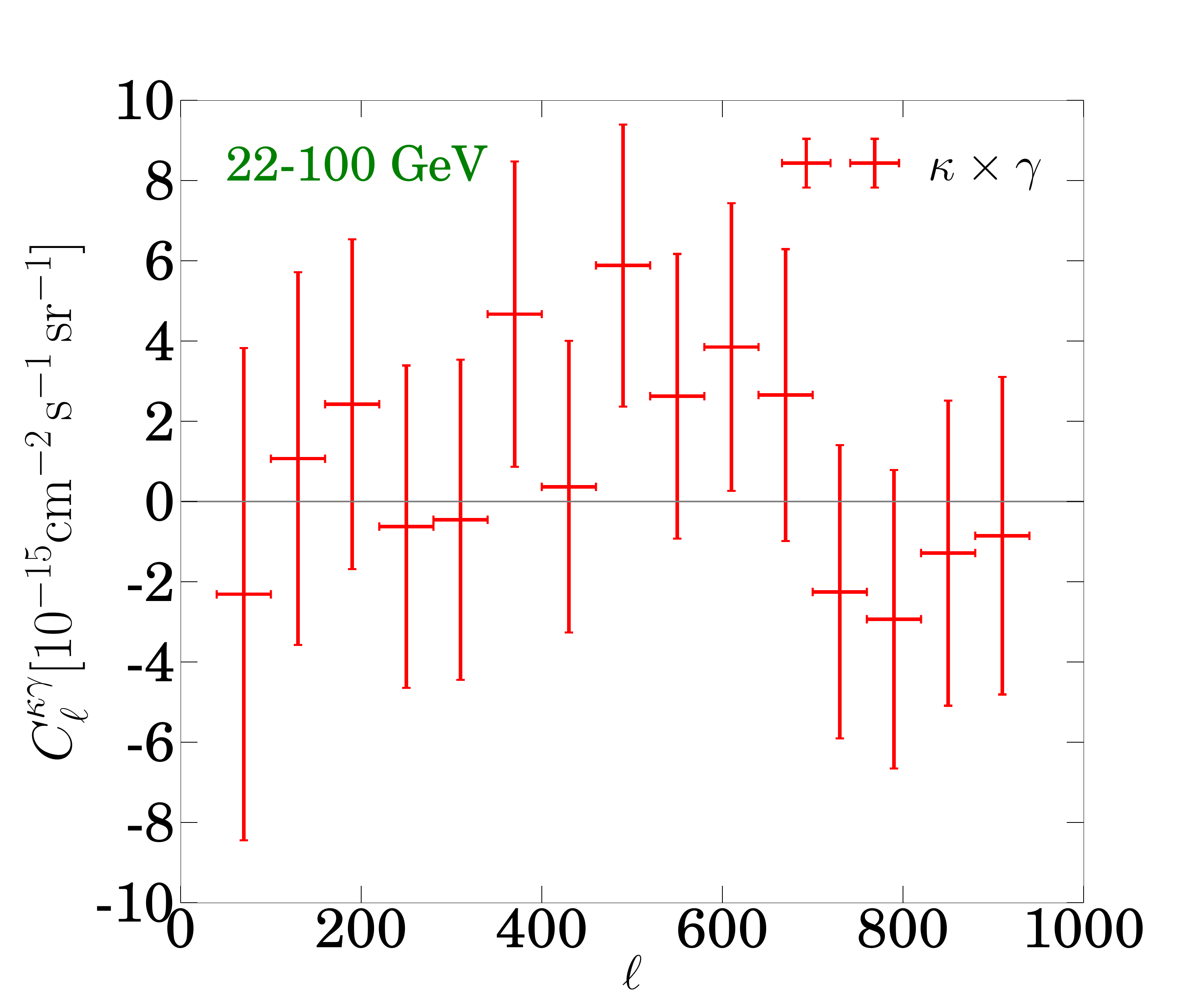}}
\rotatebox{0}{\includegraphics[width=8cm, height=7.2cm]{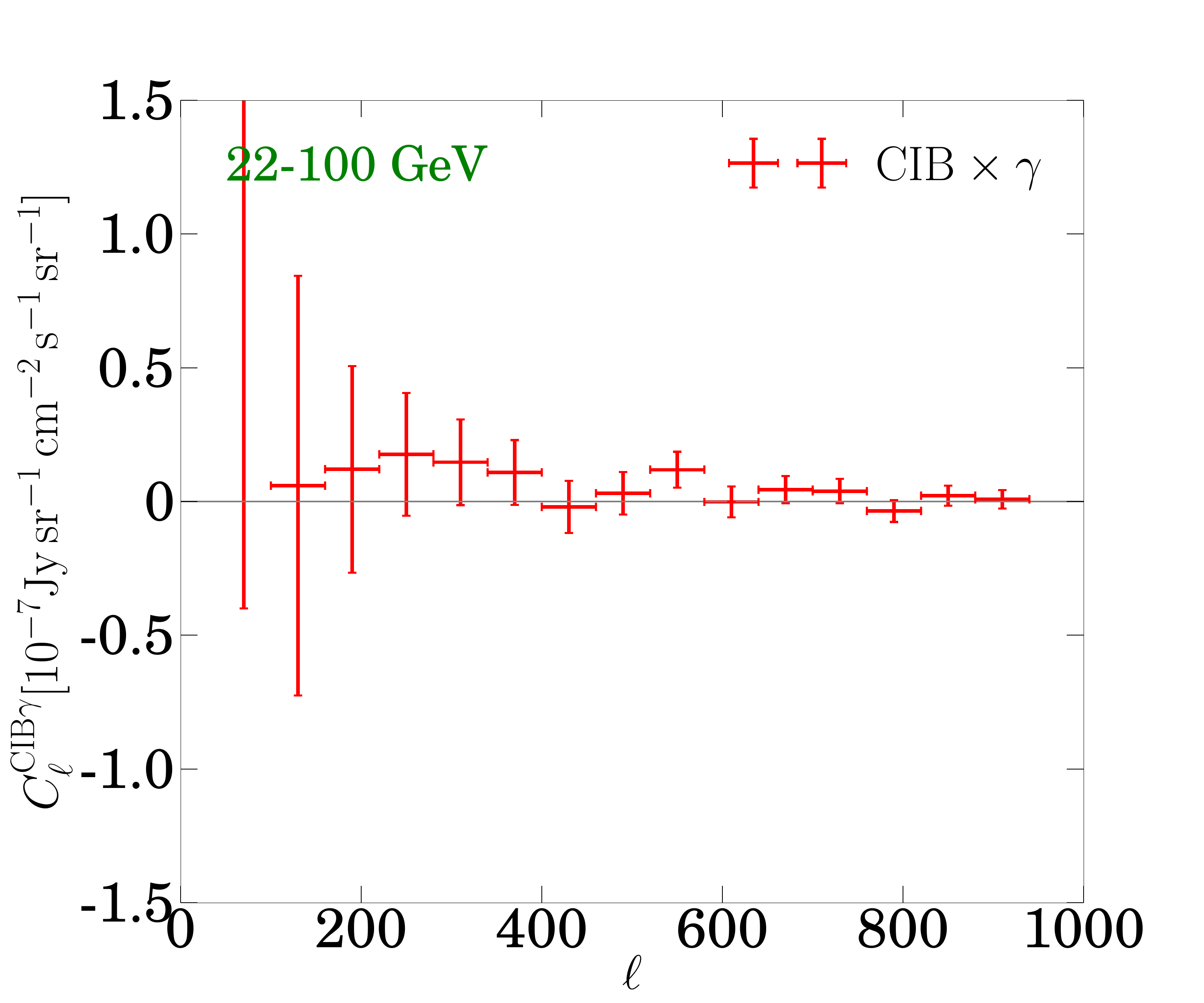}}
\caption{The measured cross-correlation power spectra from Planck lensing, Planck CIB and Fermi-LAT $\gamma$-ray jackknife maps at 22-100 GeV.}
\label{jkk3}
\end{figure}

\section{Data sets}
We use the Planck lensing map derived from its temperature to trace the DM distribution. We only use 857 \rm{GHz} data because the CIB is the strongest at this frequency. A Gaussian beam for 857 \rm{GHz} is assumed and the full-width-at-half-maximum (FWHM) is $4'.63$. 

The $\gamma$-ray data are taken from the Fermi-LAT satellite's 92-month observations from 2008 August to 2016 April. The weekly photon data are processed by the science tools v10r0p5 provided by the Fermi-LAT collaboration\footnote[1]{\url{http://fermi.gsfc.nasa.gov/ssc/data/analysis/software/}}. We split the \rr ray data into three energy bins which are uniform in $\log$ space: 1-5 GeV, 5-22 GeV and 22-100 GeV. The photon counts and coverage maps made separately by the science tools are repixelized by \hp\footnotemark[2] \footnotetext[2]{\url{http://healpix.sourceforge.net}} at resolution $n_{\rm{side}}=512$. To reduce the background and filter out poorly reconstructed events, we use the Pass7-reprocessed instrument response functions for the ULTRACLEAN event class P7REP\_ULTRACLEAN\_V15 and the step size $\cos(\theta)=0.025$ for the exposure. The flux map is finally generated from the ratio of the photon counts to coverage map. We split the photon count data into two halves and made the flux maps at three energy bands for them. In Fig.\ref{fermiflux}, the flux maps for all the 92 months are shown.

For the sanity check of our \rr ray map-maker, we applied it to 22 months of data and made maps within the energy bands discussed in Ref.\cite{fermi_auto_ps}. The exposure map is made with the good time intervals, \textit{gtis}, which are recorded in the event files and updated by the science tool \textit{gtmktime}. The choice of a rocking angle negligibly changes the maps, so we use the default value of the science tool \textit{gtbin}. We chose 1FGL point source catalog and masked $2^{\circ}$ circle as in Ref.\cite{fermi_auto_ps}. The resulting fluxes and raw auto-power spectra from our maps agree with the 22-month analysis. 

The Fermi-LAT point spread function (PSF) can be approximated by a Gaussian beam at $l<500$ but quickly deviates from Gaussian beyond that, and the resulting PSF is dependent on the energy bands. To get the correct PSF, the $\gamma$-ray PSF $\xi_{\rm{PSF}}(E,\theta)$ is modeled at any energy $E$ and inclination angle $\theta$ by functional forms. We take the parameters provided by the Fermi collaboration to build the energy-dependent PSF~\cite{xCFHLS}. The $\gamma$-ray beam transfer function from energy $E_1$ to $E_2$ is averaged from each PSF, i.e.,
\begin{equation}
W_l(E_1<E<E_2)\sim\int_{E_1}^{E_2}dEW_l(E)\frac{dN}{dE},
\end{equation}
where $W_l(E)$ is the Legendre transform of the PSF profile at energy $E$. Specifically, it is
\begin{equation}
W_l(E)=2\pi\int_{-1}^{+1}d\cos(\theta)P_l(\cos(\theta))\xi_{\rm{PSF}}(\theta,E).
\end{equation}
and $dN/dE\sim E^{-2.4}$~\cite{fermi_auto_ps}. The beam transfer functions used in this analysis are shown in Fig.\ref{fermi-beam}. 

We use the mask released by the Planck lensing products to mask point sources and galactic plane. For the $\gamma$-ray data, we first remove the bright Galactic emission within the latitude $|b|<30^{\circ}$. We then identify the point sources from the Three-year Point Source Catalog (3FGL) and remove them with a disk of $1.5^{\circ}$ angular radius. The sharp edges in the \rr ray mask are also apodized with a $\cos$-like taper. Both masks are shown in Fig.\ref{masks}.

The power spectrum defined as 
\begin{equation}C_l^{XY}=\frac{1}{2l+1}\sum_m\langle a^X_{lm}a^{\ast Y}_{lm}\rangle
\end{equation} with maps $X$ and $Y$, can be easily calculated from the standard MASTER method~\cite{master}, i.e., 
\begin{equation}
\tilde C^{XY}_l=\displaystyle\sum_{l'} M_{ll'}b^{2,XY}_{l'}C^{XY}_{l'}+N^{XY}_l.
\end{equation}
Here, $X$ and $Y$ refer to Planck lensing $\phi(\bn)$, Planck CIB $T(\bn)$ and $\gamma$-ray maps $\gamma(\bn)$. The hybrid beam transfer function is formed through $b^{XY}_l=\sqrt{b^{X}_lb^{Y}_l}$. We made all the beam transfer functions for CIB and \rr ray maps. We also calculated the exact mode-coupling matrix $M_{ll'}$ using simulations and validated that it can be well-approximated by a simple $f_{\rm{sky}}$ scaling. For the cross-correlations, the noise bias $N^{XY}_l$ is negligible.

To test if there is any bias in the cross-power spectrum estimation, we use Cholesky decomposition of the covariance matrix between lensing and \rr ray to make correlated simulations. We then feed all of these simulations into the pipeline, which incorporates different masks and beam transfer functions, and estimate the power spectra. We find that the averaged power spectra from 200 realizations agree with the theory as shown by Fig.\ref{cross_sim}. Also the statistical uncertainties for all the bands calculated from these simulations agree with the Knox-formula predication and the overall difference between them is less than 5\%.

\section{Data analysis}
For Planck lensing, CIB, and Fermi-LAT \rr  ray maps, we consider different components
\begin{eqnarray}
\tilde\phi(\bn)&=&A^{\phi}\phi(\bn)+n^{\phi}(\bn),
\nonumber\\
\tilde T(\bn)&=&A^{\rm{CIB}}T^{\rm{CIB}}(\bn)+A^{\rm{dust}}T^{\rm{dust}}(\bn),\nonumber\\
\tilde \gamma(\bn)&=&A^{\rm{DM,ann}}\gamma^{\rm{DM,ann}}(\bn)+A^{\rm{DM,dec}}\gamma^{\rm{DM,dec}}(\bn)\nonumber\\
&+&A^{\rm{SFG}}\gamma^{\rm{SFG}}(\bn)+A^{\rm{Blazar}}\gamma^{\rm{Blazar}}(\bn)\nonumber\\
&+&A^{\rm{FSRQ}}\gamma^{\rm{FSRQ}}(\textbf{n})+A^{\rm{mAGN}}\gamma^{\rm{mAGN}} (\textbf{n})\nonumber\\
&+&n^{\gamma}(\bn).
\end{eqnarray}
Here, $\tilde X$ means it is the observed map and $n(\bn)$ is the noise contribution. 

From the raw power spectra of CIB and \rr ray maps, the dust contribution is seen from the first three bins. In principle, one could devise a template in map space, such as the thermal dust template (``$\rm{COM\_CompMap\_dust\mbox{-}commrul\_2048\_R}$") for Planck CIB and the diffuse galactic emission (``$\rm{gll\_iem\_v06}$") from Fermi-LAT, then subtract it from the map. To simplify the discussion, we instead equivalently introduce a power spectrum template to capture this contribution, which is only significant for the CIB-$\gamma$ correlation. The template is simply chosen as a power law, i.e., $C_l=Al^{-n}$. Our goal is to marginalize over this component and estimate the remaining DM signals. Our parameter set is defined as $\{A^{\phi},A^{\rm{CIB}},m_{\rm{DM}},\langle\sigma v\rangle, \Gamma_d, A^{\rm{SFG}},A^{\rm{Blazar}}, A^{\rm{FSRQ}},A^{\rm{mAGN}},$ $A,n\}$.

From Ref.~\cite{fermi_auto_ps}, it argues that the low multipoles are contaminated by the large-scale-features-introduced signals in the data. Also, the Planck CIB auto correlation is highly dominated by the dust from large to moderate angular scales. Therefore, we do not include the Planck CIB and $\gamma$-ray auto-power spectra in our model fitting, and only focus on the cross-correlations which are less affected by systematic effects. The $\chi^2$ is thus defined as
\begin{eqnarray}
\chi^2(\textbf{P})&=&\displaystyle\sum_{XY=\{\kappa\kappa, \kappa T, \kappa\gamma,T\gamma\}}\nonumber\\
&&\displaystyle\sum_{bb'}(\tilde C^{XY}_b-\hat C^{XY}_b)\textbf{C}^{-1}_{bb'}(\tilde C^{XY}_{b'}-\hat C^{XY}_{b'}).\label{like}\nonumber\\
\end{eqnarray}
The covariance matrix can be approximated by $\textbf{C}_{bb'}=(\Delta C^{XY}_b)^2\delta_{bb'}$ and $b$ is the index of the band power $\hat C_b$ which is combination of the power spectra $C_l^{\kappa\kappa}$, $C_l^{\kappa\rm{CIB}}$, $C_l^{\kappa\rm{DM,ann}}$, $C_l^{\kappa\rm{DM,dec}}$, $C_l^{\kappa\rm{SFGs}}$, $C_l^{\kappa\rm{Blazars}}$, $C_l^{\rm{CIB}\mbox{-}\rm{DM,ann}}$, $C_l^{\rm{CIB}\mbox{-}\rm{DM,dec}}$, $C_l^{\rm{CIB}\mbox{-}\rm{SFGs}}$, $C_l^{\rm{CIB}\mbox{-}\rm{Blazars}}$, etc.

The cosmological parameters we use are \{$A_s$, $n_s$, $k_{\rm{pivot}}$, $H_0$, $\Omega_b$, $\Omega_c$\}=\{$2.1\times10^{-9}$,0.96,0.05,70,0.0461,0.222857\}. The power spectrum with DM signals, i.e., $C_l^{\kappa\rm{DM,ann}}$, $C_l^{\kappa\rm{DM,dec}}$, $C_l^{\rm{CIB}\mbox{-}\rm{DM,ann}}$, and $C_l^{\rm{CIB}\mbox{-}\rm{DM,dec}}$, are interpolated from templates for any given mass $m_{\rm{DM}}$. The $L_{\rm{min}}$ for Blazars, SFGs, FSRQs and mAGN are $7\times 10^{42}$, $10^{42}$, $4\times 10^{43}$, and  $10^{41}$$\,\rm{erg}\,\rm{s}^{-1}$.

The error bars $\Delta C^{XY}_b$ of the power spectra are given by the Knox formula
\begin{eqnarray}
(\Delta C^{XY}_b)^2&=&\frac{1}{(2l+1)\Delta_b f_{\rm{sky}}}[\tilde C^X_b\tilde C^Y_b+(\tilde C^{XY}_b)^2].
\end{eqnarray}
Here, $\tilde C$ is the raw power spectrum with noise, and it automatically reduces to the error bars for the auto-power spectrum when $X=Y$. The error bars calculated from MCMC simulations agree with this analytical calculation.

\begin{figure*}
\rotatebox{0}{\includegraphics[width=8cm, height=7.2cm]{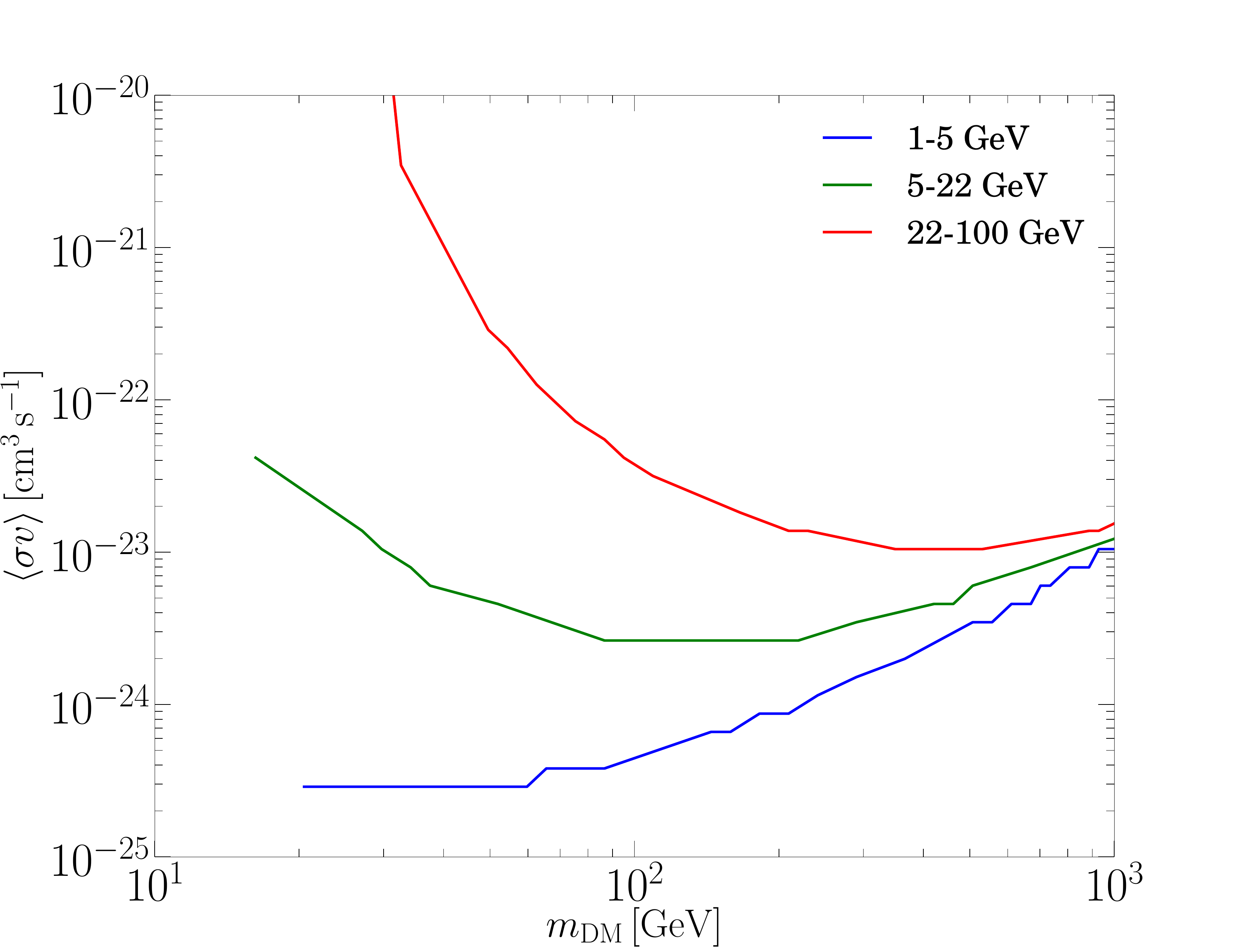}}
\rotatebox{0}{\includegraphics[width=8cm, height=7.2cm]{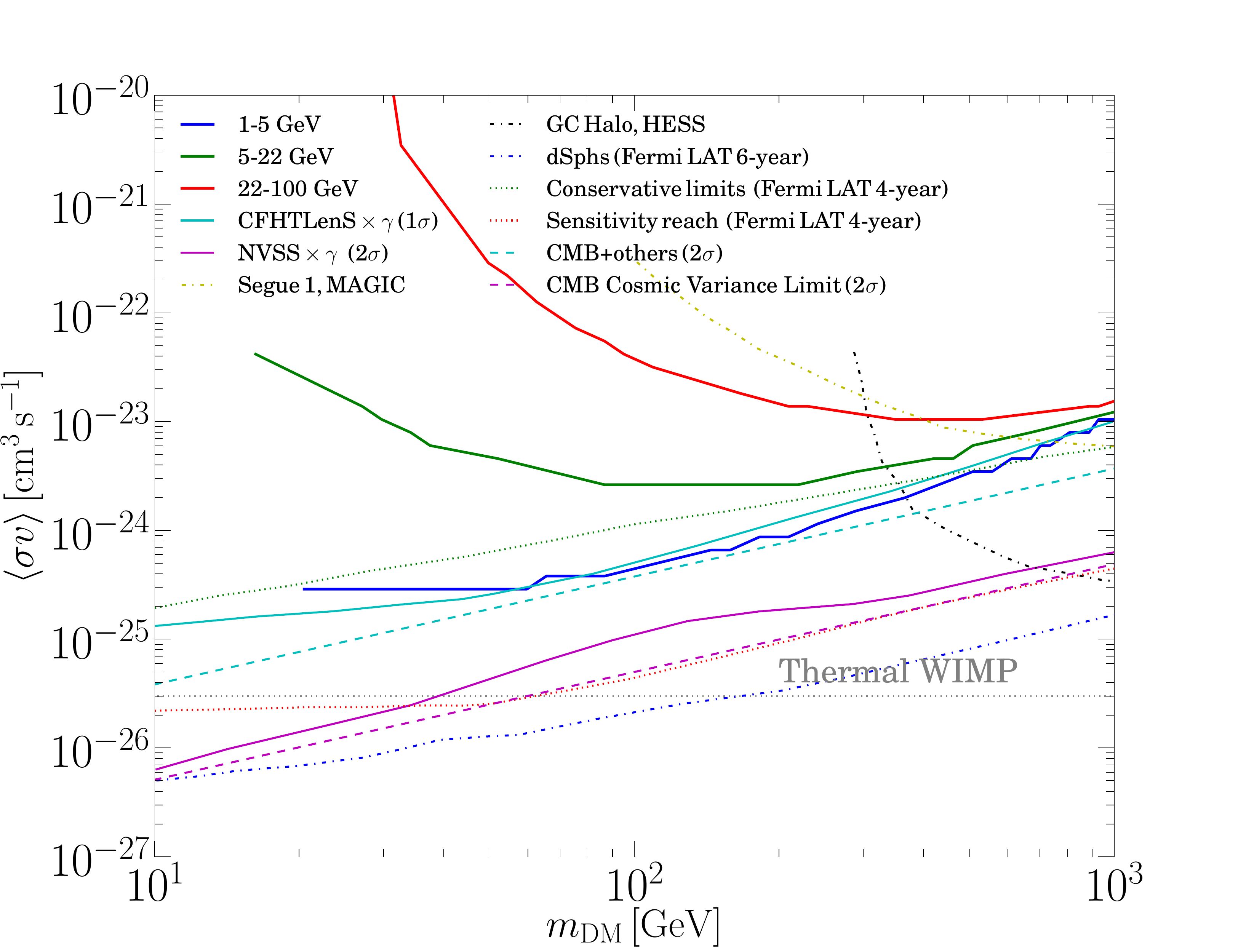}}
\caption{(Left) The cross-sections vs. DM masses at three energy bands. It is seen that the 1-5\,\rm{GeV} band favors smaller cross sections than other two. (Right) The previous upper bounds reported in the literature are shown for comparison. These limits are derived from observations with CMB~\cite{cmbandothers}, cosmic shear~\cite{rs}, NVSS~\cite{cross_rg}, the Galactic Center~\cite{gchalo}, the dwarf spheroidal satellite galaxies (dSphs) of the Milky Way~\cite{dsph}, the satellite galaxy Segue 1~\cite{magic}, and the isotropic \rr background~\cite{fermi2}.}
\label{cross}
\end{figure*}

We show all the measured power spectra in Figs. (\ref{ps1},\ref{ps2},\ref{ps3}), as well as the best-fit components.  We split the photon event data into two halves and make two sets of flux maps. The jackknife maps for the \rr ray are cross-correlated with Planck lensing and CIB and the results are shown in Figs. (\ref{jkk1},\ref{jkk2},\ref{jkk3}). All of these results indicate that there are no significant systematic issues. We use MCMC procedure to sample O($10^7$) parameter sets from the likelihood function Eq. (\ref{like}). We make two-dimensional posterior distribution functions for parameter pair $m_{\rm{DM}}$\mbox{--}$\langle\sigma v\rangle$ from these samples and calculate $\Delta\chi^2(\textbf{P})$ to determine the confidence contours. The 1$\sigma$ contours for $m_{\rm{DM}}$\mbox{--}$\langle\sigma v\rangle$ combination at different energy bands are calculated from this procedure and we show the upper bands on the DM annihilation cross-section in Fig. \ref{cross}.

The comparison between our results and previous limits is made in Fig.\ref{cross} where the DM properties are constrained by different data sets, such as the weak lensing~\cite{rs}, the radio galaxy NVSS~\cite{cross_rg}, the latest CMB measurements (WMAP9+Planck+ACT+SPT) with BAO+HST+SN~\cite{cmbandothers}, Fermi-LAT 4 year isotropic \rr background~\cite{fermi2}, the Galactic Center~\cite{gchalo}, the dwarf spheroidal satellite galaxies (dSphs) of the Milky Way~\cite{dsph}, and the satellite galaxy Segue 1~\cite{magic}. Our limits are complementary and comparable to those with similar approaches; moreover, this is the first time the CIB fluctuation is cross-correlated with anisotropic \rr ray background. The overall signal-to-noise ratio is moderately improved because fluctuations span slightly different redshift ranges and the theoretical uncertainty introduced by the modeling of astrophysical \rr ray emitters is so large that the DM constraint is weakened. However, our approach is important for the understanding of the \rr ray composition and would be potentially enhanced with more data coming from either Fermi-LAT satellite or other LSS tracers.

\section{Conclusions}
We use all the 92-month weekly photon data from Fermi-LAT to make count and exposure maps at different energy bands. The data are split into two halves and the jackknife maps of the \rr ray are cross-correlated with Planck lensing and CIB. No significant systematic effects are found. We further measure the cross-power spectra between Planck lensing, CIB and Fermi-LAT \rr ray maps at energy bands 1-5, 5-22 and 22-100 \rm{GeV}. Based on halo-model approach, the composition of the \rr ray background is studied. The contributions of different \rr ray emitters, such as DM annihilation, DM decay, SFGs, Blazars, FSRQs and mAGN are estimated from the measurements. We finally place the upper bounds on the DM annihilation cross-section with respect to the masses from the \rr ray signals, with the astrophysical sources excluded.

\section{Acknowledgements}
We are grateful for helpful discussions with Tim Linden, Jennifer Gaskins, and Kevork Abazajian. C.F. acknowledges support from NASA grants NASA NNX16AJ69G and NASA NNX16AF39G. We also acknowledge the use of the \hp~\cite{hp} package.

\bibliography{Fermi_LAT}

\begin{thebibliography}{52}
\expandafter\ifx\csname natexlab\endcsname\relax\def\natexlab#1{#1}\fi
\expandafter\ifx\csname bibnamefont\endcsname\relax
  \def\bibnamefont#1{#1}\fi
\expandafter\ifx\csname bibfnamefont\endcsname\relax
  \def\bibfnamefont#1{#1}\fi
\expandafter\ifx\csname citenamefont\endcsname\relax
  \def\citenamefont#1{#1}\fi
\expandafter\ifx\csname url\endcsname\relax
  \def\url#1{\texttt{#1}}\fi
\expandafter\ifx\csname urlprefix\endcsname\relax\def\urlprefix{URL }\fi
\providecommand{\bibinfo}[2]{#2}
\providecommand{\eprint}[2][]{\url{#2}}

\bibitem[{\citenamefont{{Planck Collaboration}
  et~al.}(2015{\natexlab{a}})\citenamefont{{Planck Collaboration}, {Ade},
  {Aghanim}, {Arnaud}, {Ashdown}, {Aumont}, {Baccigalupi}, {Banday},
  {Barreiro}, {Bartlett} et~al.}}]{planck_params}
\bibinfo{author}{\bibnamefont{{Planck Collaboration}}},
  \bibinfo{author}{\bibfnamefont{P.~A.~R.} \bibnamefont{{Ade}}},
  \bibinfo{author}{\bibfnamefont{N.}~\bibnamefont{{Aghanim}}},
  \bibinfo{author}{\bibfnamefont{M.}~\bibnamefont{{Arnaud}}},
  \bibinfo{author}{\bibfnamefont{M.}~\bibnamefont{{Ashdown}}},
  \bibinfo{author}{\bibfnamefont{J.}~\bibnamefont{{Aumont}}},
  \bibinfo{author}{\bibfnamefont{C.}~\bibnamefont{{Baccigalupi}}},
  \bibinfo{author}{\bibfnamefont{A.~J.} \bibnamefont{{Banday}}},
  \bibinfo{author}{\bibfnamefont{R.~B.} \bibnamefont{{Barreiro}}},
  \bibinfo{author}{\bibfnamefont{J.~G.} \bibnamefont{{Bartlett}}},
  \bibnamefont{et~al.}, \bibinfo{journal}{ArXiv e-prints}
  (\bibinfo{year}{2015}{\natexlab{a}}), \eprint{1502.01589}.

\bibitem[{\citenamefont{{Planck Collaboration}
  et~al.}(2014{\natexlab{a}})\citenamefont{{Planck Collaboration}, {Ade},
  {Aghanim}, {Armitage-Caplan}, {Arnaud}, {Ashdown}, {Atrio-Barandela},
  {Aumont}, {Baccigalupi}, {Banday} et~al.}}]{2014AA571A17P}
\bibinfo{author}{\bibnamefont{{Planck Collaboration}}},
  \bibinfo{author}{\bibfnamefont{P.~A.~R.} \bibnamefont{{Ade}}},
  \bibinfo{author}{\bibfnamefont{N.}~\bibnamefont{{Aghanim}}},
  \bibinfo{author}{\bibfnamefont{C.}~\bibnamefont{{Armitage-Caplan}}},
  \bibinfo{author}{\bibfnamefont{M.}~\bibnamefont{{Arnaud}}},
  \bibinfo{author}{\bibfnamefont{M.}~\bibnamefont{{Ashdown}}},
  \bibinfo{author}{\bibfnamefont{F.}~\bibnamefont{{Atrio-Barandela}}},
  \bibinfo{author}{\bibfnamefont{J.}~\bibnamefont{{Aumont}}},
  \bibinfo{author}{\bibfnamefont{C.}~\bibnamefont{{Baccigalupi}}},
  \bibinfo{author}{\bibfnamefont{A.~J.} \bibnamefont{{Banday}}},
  \bibnamefont{et~al.}, \bibinfo{journal}{\aap} \textbf{\bibinfo{volume}{571}},
  \bibinfo{eid}{A17} (\bibinfo{year}{2014}{\natexlab{a}}),
  \eprint{arXiv:1303.5077}.

\bibitem[{\citenamefont{{Planck Collaboration}
  et~al.}(2015{\natexlab{b}})\citenamefont{{Planck Collaboration}, {Ade},
  {Aghanim}, {Arnaud}, {Ashdown}, {Aumont}, {Baccigalupi}, {Banday},
  {Barreiro}, {Bartlett} et~al.}}]{2015arXiv150201591P}
\bibinfo{author}{\bibnamefont{{Planck Collaboration}}},
  \bibinfo{author}{\bibfnamefont{P.~A.~R.} \bibnamefont{{Ade}}},
  \bibinfo{author}{\bibfnamefont{N.}~\bibnamefont{{Aghanim}}},
  \bibinfo{author}{\bibfnamefont{M.}~\bibnamefont{{Arnaud}}},
  \bibinfo{author}{\bibfnamefont{M.}~\bibnamefont{{Ashdown}}},
  \bibinfo{author}{\bibfnamefont{J.}~\bibnamefont{{Aumont}}},
  \bibinfo{author}{\bibfnamefont{C.}~\bibnamefont{{Baccigalupi}}},
  \bibinfo{author}{\bibfnamefont{A.~J.} \bibnamefont{{Banday}}},
  \bibinfo{author}{\bibfnamefont{R.~B.} \bibnamefont{{Barreiro}}},
  \bibinfo{author}{\bibfnamefont{J.~G.} \bibnamefont{{Bartlett}}},
  \bibnamefont{et~al.}, \bibinfo{journal}{ArXiv e-prints}
  (\bibinfo{year}{2015}{\natexlab{b}}), \eprint{1502.01591}.

\bibitem[{\citenamefont{{Smith} et~al.}(2007)\citenamefont{{Smith}, {Zahn}, and
  {Dor{\'e}}}}]{wmapcross2007}
\bibinfo{author}{\bibfnamefont{K.~M.} \bibnamefont{{Smith}}},
  \bibinfo{author}{\bibfnamefont{O.}~\bibnamefont{{Zahn}}}, \bibnamefont{and}
  \bibinfo{author}{\bibfnamefont{O.}~\bibnamefont{{Dor{\'e}}}},
  \bibinfo{journal}{\prd} \textbf{\bibinfo{volume}{76}}, \bibinfo{eid}{043510}
  (\bibinfo{year}{2007}), \eprint{0705.3980}.

\bibitem[{\citenamefont{{Hirata} et~al.}(2004)\citenamefont{{Hirata},
  {Padmanabhan}, {Seljak}, {Schlegel}, and {Brinkmann}}}]{hirata2004}
\bibinfo{author}{\bibfnamefont{C.~M.} \bibnamefont{{Hirata}}},
  \bibinfo{author}{\bibfnamefont{N.}~\bibnamefont{{Padmanabhan}}},
  \bibinfo{author}{\bibfnamefont{U.}~\bibnamefont{{Seljak}}},
  \bibinfo{author}{\bibfnamefont{D.}~\bibnamefont{{Schlegel}}},
  \bibnamefont{and}
  \bibinfo{author}{\bibfnamefont{J.}~\bibnamefont{{Brinkmann}}},
  \bibinfo{journal}{\prd} \textbf{\bibinfo{volume}{70}}, \bibinfo{eid}{103501}
  (\bibinfo{year}{2004}), \eprint{astro-ph/0406004}.

\bibitem[{\citenamefont{{Hirata} et~al.}(2008)\citenamefont{{Hirata}, {Ho},
  {Padmanabhan}, {Seljak}, and {Bahcall}}}]{hirata2008}
\bibinfo{author}{\bibfnamefont{C.~M.} \bibnamefont{{Hirata}}},
  \bibinfo{author}{\bibfnamefont{S.}~\bibnamefont{{Ho}}},
  \bibinfo{author}{\bibfnamefont{N.}~\bibnamefont{{Padmanabhan}}},
  \bibinfo{author}{\bibfnamefont{U.}~\bibnamefont{{Seljak}}}, \bibnamefont{and}
  \bibinfo{author}{\bibfnamefont{N.~A.} \bibnamefont{{Bahcall}}},
  \bibinfo{journal}{\prd} \textbf{\bibinfo{volume}{78}}, \bibinfo{eid}{043520}
  (\bibinfo{year}{2008}), \eprint{0801.0644}.

\bibitem[{\citenamefont{{Feng} et~al.}(2012)\citenamefont{{Feng}, {Aslanyan},
  {Manohar}, {Keating}, {Paar}, and {Zahn}}}]{cf2012}
\bibinfo{author}{\bibfnamefont{C.}~\bibnamefont{{Feng}}},
  \bibinfo{author}{\bibfnamefont{G.}~\bibnamefont{{Aslanyan}}},
  \bibinfo{author}{\bibfnamefont{A.~V.} \bibnamefont{{Manohar}}},
  \bibinfo{author}{\bibfnamefont{B.}~\bibnamefont{{Keating}}},
  \bibinfo{author}{\bibfnamefont{H.~P.} \bibnamefont{{Paar}}},
  \bibnamefont{and} \bibinfo{author}{\bibfnamefont{O.}~\bibnamefont{{Zahn}}},
  \bibinfo{journal}{\prd} \textbf{\bibinfo{volume}{86}}, \bibinfo{eid}{063519}
  (\bibinfo{year}{2012}), \eprint{1207.3326}.

\bibitem[{\citenamefont{{Story} et~al.}(2015)\citenamefont{{Story}, {Hanson},
  {Ade}, {Aird}, {Austermann}, {Beall}, {Bender}, {Benson}, {Bleem},
  {Carlstrom} et~al.}}]{sptpol}
\bibinfo{author}{\bibfnamefont{K.~T.} \bibnamefont{{Story}}},
  \bibinfo{author}{\bibfnamefont{D.}~\bibnamefont{{Hanson}}},
  \bibinfo{author}{\bibfnamefont{P.~A.~R.} \bibnamefont{{Ade}}},
  \bibinfo{author}{\bibfnamefont{K.~A.} \bibnamefont{{Aird}}},
  \bibinfo{author}{\bibfnamefont{J.~E.} \bibnamefont{{Austermann}}},
  \bibinfo{author}{\bibfnamefont{J.~A.} \bibnamefont{{Beall}}},
  \bibinfo{author}{\bibfnamefont{A.~N.} \bibnamefont{{Bender}}},
  \bibinfo{author}{\bibfnamefont{B.~A.} \bibnamefont{{Benson}}},
  \bibinfo{author}{\bibfnamefont{L.~E.} \bibnamefont{{Bleem}}},
  \bibinfo{author}{\bibfnamefont{J.~E.} \bibnamefont{{Carlstrom}}},
  \bibnamefont{et~al.}, \bibinfo{journal}{\apj} \textbf{\bibinfo{volume}{810}},
  \bibinfo{eid}{50} (\bibinfo{year}{2015}), \eprint{1412.4760}.

\bibitem[{\citenamefont{{Das} et~al.}(2011)\citenamefont{{Das}, {Sherwin},
  {Aguirre}, {Appel}, {Bond}, {Carvalho}, {Devlin}, {Dunkley}, {D{\"u}nner},
  {Essinger-Hileman} et~al.}}]{act}
\bibinfo{author}{\bibfnamefont{S.}~\bibnamefont{{Das}}},
  \bibinfo{author}{\bibfnamefont{B.~D.} \bibnamefont{{Sherwin}}},
  \bibinfo{author}{\bibfnamefont{P.}~\bibnamefont{{Aguirre}}},
  \bibinfo{author}{\bibfnamefont{J.~W.} \bibnamefont{{Appel}}},
  \bibinfo{author}{\bibfnamefont{J.~R.} \bibnamefont{{Bond}}},
  \bibinfo{author}{\bibfnamefont{C.~S.} \bibnamefont{{Carvalho}}},
  \bibinfo{author}{\bibfnamefont{M.~J.} \bibnamefont{{Devlin}}},
  \bibinfo{author}{\bibfnamefont{J.}~\bibnamefont{{Dunkley}}},
  \bibinfo{author}{\bibfnamefont{R.}~\bibnamefont{{D{\"u}nner}}},
  \bibinfo{author}{\bibfnamefont{T.}~\bibnamefont{{Essinger-Hileman}}},
  \bibnamefont{et~al.}, \bibinfo{journal}{Physical Review Letters}
  \textbf{\bibinfo{volume}{107}}, \bibinfo{eid}{021301} (\bibinfo{year}{2011}),
  \eprint{1103.2124}.

\bibitem[{\citenamefont{{Ade} et~al.}(2014)\citenamefont{{Ade}, {Akiba},
  {Anthony}, {Arnold}, {Atlas}, {Barron}, {Boettger}, {Borrill}, {Chapman},
  {Chinone} et~al.}}]{pb}
\bibinfo{author}{\bibfnamefont{P.~A.~R.} \bibnamefont{{Ade}}},
  \bibinfo{author}{\bibfnamefont{Y.}~\bibnamefont{{Akiba}}},
  \bibinfo{author}{\bibfnamefont{A.~E.} \bibnamefont{{Anthony}}},
  \bibinfo{author}{\bibfnamefont{K.}~\bibnamefont{{Arnold}}},
  \bibinfo{author}{\bibfnamefont{M.}~\bibnamefont{{Atlas}}},
  \bibinfo{author}{\bibfnamefont{D.}~\bibnamefont{{Barron}}},
  \bibinfo{author}{\bibfnamefont{D.}~\bibnamefont{{Boettger}}},
  \bibinfo{author}{\bibfnamefont{J.}~\bibnamefont{{Borrill}}},
  \bibinfo{author}{\bibfnamefont{S.}~\bibnamefont{{Chapman}}},
  \bibinfo{author}{\bibfnamefont{Y.}~\bibnamefont{{Chinone}}},
  \bibnamefont{et~al.}, \bibinfo{journal}{Physical Review Letters}
  \textbf{\bibinfo{volume}{113}}, \bibinfo{eid}{021301} (\bibinfo{year}{2014}),
  \eprint{1312.6646}.

\bibitem[{\citenamefont{{Keck Array} et~al.}(2016)\citenamefont{{Keck Array},
  {BICEP2 Collaborations}, {:}, {Ade}, {Ahmed}, {Aikin}, {Alexander},
  {Barkats}, {Benton}, {Bischoff} et~al.}}]{bicep}
\bibinfo{author}{\bibfnamefont{T.}~\bibnamefont{{Keck Array}}},
  \bibinfo{author}{\bibnamefont{{BICEP2 Collaborations}}},
  \bibinfo{author}{\bibnamefont{{:}}}, \bibinfo{author}{\bibfnamefont{P.~A.~R.}
  \bibnamefont{{Ade}}},
  \bibinfo{author}{\bibfnamefont{Z.}~\bibnamefont{{Ahmed}}},
  \bibinfo{author}{\bibfnamefont{R.~W.} \bibnamefont{{Aikin}}},
  \bibinfo{author}{\bibfnamefont{K.~D.} \bibnamefont{{Alexander}}},
  \bibinfo{author}{\bibfnamefont{D.}~\bibnamefont{{Barkats}}},
  \bibinfo{author}{\bibfnamefont{S.~J.} \bibnamefont{{Benton}}},
  \bibinfo{author}{\bibfnamefont{C.~A.} \bibnamefont{{Bischoff}}},
  \bibnamefont{et~al.}, \bibinfo{journal}{ArXiv e-prints}
  (\bibinfo{year}{2016}), \eprint{1606.01968}.

\bibitem[{\citenamefont{{Ackermann}
  et~al.}(2012{\natexlab{a}})\citenamefont{{Ackermann}, {Ajello}, {Albert},
  {Baldini}, {Ballet}, {Barbiellini}, {Bastieri}, {Bechtol}, {Bellazzini},
  {Bloom} et~al.}}]{fermi_auto_ps}
\bibinfo{author}{\bibfnamefont{M.}~\bibnamefont{{Ackermann}}},
  \bibinfo{author}{\bibfnamefont{M.}~\bibnamefont{{Ajello}}},
  \bibinfo{author}{\bibfnamefont{A.}~\bibnamefont{{Albert}}},
  \bibinfo{author}{\bibfnamefont{L.}~\bibnamefont{{Baldini}}},
  \bibinfo{author}{\bibfnamefont{J.}~\bibnamefont{{Ballet}}},
  \bibinfo{author}{\bibfnamefont{G.}~\bibnamefont{{Barbiellini}}},
  \bibinfo{author}{\bibfnamefont{D.}~\bibnamefont{{Bastieri}}},
  \bibinfo{author}{\bibfnamefont{K.}~\bibnamefont{{Bechtol}}},
  \bibinfo{author}{\bibfnamefont{R.}~\bibnamefont{{Bellazzini}}},
  \bibinfo{author}{\bibfnamefont{E.~D.} \bibnamefont{{Bloom}}},
  \bibnamefont{et~al.}, \bibinfo{journal}{\prd} \textbf{\bibinfo{volume}{85}},
  \bibinfo{eid}{083007} (\bibinfo{year}{2012}{\natexlab{a}}),
  \eprint{1202.2856}.

\bibitem[{\citenamefont{{Camera} et~al.}(2013)\citenamefont{{Camera},
  {Fornasa}, {Fornengo}, and {Regis}}}]{cross_letter}
\bibinfo{author}{\bibfnamefont{S.}~\bibnamefont{{Camera}}},
  \bibinfo{author}{\bibfnamefont{M.}~\bibnamefont{{Fornasa}}},
  \bibinfo{author}{\bibfnamefont{N.}~\bibnamefont{{Fornengo}}},
  \bibnamefont{and} \bibinfo{author}{\bibfnamefont{M.}~\bibnamefont{{Regis}}},
  \bibinfo{journal}{\apjl} \textbf{\bibinfo{volume}{771}}, \bibinfo{eid}{L5}
  (\bibinfo{year}{2013}), \eprint{1212.5018}.

\bibitem[{\citenamefont{{Camera} et~al.}(2015)\citenamefont{{Camera},
  {Fornasa}, {Fornengo}, and {Regis}}}]{cross_long}
\bibinfo{author}{\bibfnamefont{S.}~\bibnamefont{{Camera}}},
  \bibinfo{author}{\bibfnamefont{M.}~\bibnamefont{{Fornasa}}},
  \bibinfo{author}{\bibfnamefont{N.}~\bibnamefont{{Fornengo}}},
  \bibnamefont{and} \bibinfo{author}{\bibfnamefont{M.}~\bibnamefont{{Regis}}},
  \bibinfo{journal}{\jcap} \textbf{\bibinfo{volume}{6}}, \bibinfo{eid}{029}
  (\bibinfo{year}{2015}), \eprint{1411.4651}.

\bibitem[{\citenamefont{{Regis} et~al.}(2015)\citenamefont{{Regis}, {Xia},
  {Cuoco}, {Branchini}, {Fornengo}, and {Viel}}}]{pdm}
\bibinfo{author}{\bibfnamefont{M.}~\bibnamefont{{Regis}}},
  \bibinfo{author}{\bibfnamefont{J.-Q.} \bibnamefont{{Xia}}},
  \bibinfo{author}{\bibfnamefont{A.}~\bibnamefont{{Cuoco}}},
  \bibinfo{author}{\bibfnamefont{E.}~\bibnamefont{{Branchini}}},
  \bibinfo{author}{\bibfnamefont{N.}~\bibnamefont{{Fornengo}}},
  \bibnamefont{and} \bibinfo{author}{\bibfnamefont{M.}~\bibnamefont{{Viel}}},
  \bibinfo{journal}{Physical Review Letters} \textbf{\bibinfo{volume}{114}},
  \bibinfo{eid}{241301} (\bibinfo{year}{2015}), \eprint{1503.05922}.

\bibitem[{\citenamefont{{Cuoco} et~al.}(2015)\citenamefont{{Cuoco}, {Xia},
  {Regis}, {Branchini}, {Fornengo}, and {Viel}}}]{cross_rg}
\bibinfo{author}{\bibfnamefont{A.}~\bibnamefont{{Cuoco}}},
  \bibinfo{author}{\bibfnamefont{J.-Q.} \bibnamefont{{Xia}}},
  \bibinfo{author}{\bibfnamefont{M.}~\bibnamefont{{Regis}}},
  \bibinfo{author}{\bibfnamefont{E.}~\bibnamefont{{Branchini}}},
  \bibinfo{author}{\bibfnamefont{N.}~\bibnamefont{{Fornengo}}},
  \bibnamefont{and} \bibinfo{author}{\bibfnamefont{M.}~\bibnamefont{{Viel}}},
  \bibinfo{journal}{\apjs} \textbf{\bibinfo{volume}{221}}, \bibinfo{eid}{29}
  (\bibinfo{year}{2015}), \eprint{1506.01030}.

\bibitem[{\citenamefont{{Shirasaki}
  et~al.}(2014{\natexlab{a}})\citenamefont{{Shirasaki}, {Horiuchi}, and
  {Yoshida}}}]{rs}
\bibinfo{author}{\bibfnamefont{M.}~\bibnamefont{{Shirasaki}}},
  \bibinfo{author}{\bibfnamefont{S.}~\bibnamefont{{Horiuchi}}},
  \bibnamefont{and}
  \bibinfo{author}{\bibfnamefont{N.}~\bibnamefont{{Yoshida}}},
  \bibinfo{journal}{\prd} \textbf{\bibinfo{volume}{90}}, \bibinfo{eid}{063502}
  (\bibinfo{year}{2014}{\natexlab{a}}), \eprint{1404.5503}.

\bibitem[{\citenamefont{{Fornengo} et~al.}(2015)\citenamefont{{Fornengo},
  {Perotto}, {Regis}, and {Camera}}}]{kr}
\bibinfo{author}{\bibfnamefont{N.}~\bibnamefont{{Fornengo}}},
  \bibinfo{author}{\bibfnamefont{L.}~\bibnamefont{{Perotto}}},
  \bibinfo{author}{\bibfnamefont{M.}~\bibnamefont{{Regis}}}, \bibnamefont{and}
  \bibinfo{author}{\bibfnamefont{S.}~\bibnamefont{{Camera}}},
  \bibinfo{journal}{\apjl} \textbf{\bibinfo{volume}{802}}, \bibinfo{eid}{L1}
  (\bibinfo{year}{2015}), \eprint{1410.4997}.

\bibitem[{\citenamefont{{Meerburg} et~al.}(2013)\citenamefont{{Meerburg},
  {Dvorkin}, and {Spergel}}}]{tau21_2013}
\bibinfo{author}{\bibfnamefont{P.~D.} \bibnamefont{{Meerburg}}},
  \bibinfo{author}{\bibfnamefont{C.}~\bibnamefont{{Dvorkin}}},
  \bibnamefont{and} \bibinfo{author}{\bibfnamefont{D.~N.}
  \bibnamefont{{Spergel}}}, \bibinfo{journal}{\apj}
  \textbf{\bibinfo{volume}{779}}, \bibinfo{eid}{124} (\bibinfo{year}{2013}),
  \eprint{1303.3887}.

\bibitem[{\citenamefont{{Bullock} et~al.}(2001)\citenamefont{{Bullock},
  {Kolatt}, {Sigad}, {Somerville}, {Kravtsov}, {Klypin}, {Primack}, and
  {Dekel}}}]{concentration}
\bibinfo{author}{\bibfnamefont{J.~S.} \bibnamefont{{Bullock}}},
  \bibinfo{author}{\bibfnamefont{T.~S.} \bibnamefont{{Kolatt}}},
  \bibinfo{author}{\bibfnamefont{Y.}~\bibnamefont{{Sigad}}},
  \bibinfo{author}{\bibfnamefont{R.~S.} \bibnamefont{{Somerville}}},
  \bibinfo{author}{\bibfnamefont{A.~V.} \bibnamefont{{Kravtsov}}},
  \bibinfo{author}{\bibfnamefont{A.~A.} \bibnamefont{{Klypin}}},
  \bibinfo{author}{\bibfnamefont{J.~R.} \bibnamefont{{Primack}}},
  \bibnamefont{and} \bibinfo{author}{\bibfnamefont{A.}~\bibnamefont{{Dekel}}},
  \bibinfo{journal}{\mnras} \textbf{\bibinfo{volume}{321}},
  \bibinfo{pages}{559} (\bibinfo{year}{2001}), \eprint{astro-ph/9908159}.

\bibitem[{\citenamefont{{Ando}}(2014)}]{jfac}
\bibinfo{author}{\bibfnamefont{S.}~\bibnamefont{{Ando}}},
  \bibinfo{journal}{\jcap} \textbf{\bibinfo{volume}{10}}, \bibinfo{eid}{061}
  (\bibinfo{year}{2014}), \eprint{1407.8502}.

\bibitem[{\citenamefont{{Sheth} and {Tormen}}(1999)}]{halomodel}
\bibinfo{author}{\bibfnamefont{R.~K.} \bibnamefont{{Sheth}}} \bibnamefont{and}
  \bibinfo{author}{\bibfnamefont{G.}~\bibnamefont{{Tormen}}},
  \bibinfo{journal}{\mnras} \textbf{\bibinfo{volume}{308}},
  \bibinfo{pages}{119} (\bibinfo{year}{1999}), \eprint{astro-ph/9901122}.

\bibitem[{\citenamefont{{Planck Collaboration}
  et~al.}(2014{\natexlab{b}})\citenamefont{{Planck Collaboration}, {Ade},
  {Aghanim}, {Armitage-Caplan}, {Arnaud}, {Ashdown}, {Atrio-Barandela},
  {Aumont}, {Baccigalupi}, {Banday} et~al.}}]{planckCIB2013}
\bibinfo{author}{\bibnamefont{{Planck Collaboration}}},
  \bibinfo{author}{\bibfnamefont{P.~A.~R.} \bibnamefont{{Ade}}},
  \bibinfo{author}{\bibfnamefont{N.}~\bibnamefont{{Aghanim}}},
  \bibinfo{author}{\bibfnamefont{C.}~\bibnamefont{{Armitage-Caplan}}},
  \bibinfo{author}{\bibfnamefont{M.}~\bibnamefont{{Arnaud}}},
  \bibinfo{author}{\bibfnamefont{M.}~\bibnamefont{{Ashdown}}},
  \bibinfo{author}{\bibfnamefont{F.}~\bibnamefont{{Atrio-Barandela}}},
  \bibinfo{author}{\bibfnamefont{J.}~\bibnamefont{{Aumont}}},
  \bibinfo{author}{\bibfnamefont{C.}~\bibnamefont{{Baccigalupi}}},
  \bibinfo{author}{\bibfnamefont{A.~J.} \bibnamefont{{Banday}}},
  \bibnamefont{et~al.}, \bibinfo{journal}{\aap} \textbf{\bibinfo{volume}{571}},
  \bibinfo{eid}{A30} (\bibinfo{year}{2014}{\natexlab{b}}), \eprint{1309.0382}.

\bibitem[{\citenamefont{{Serra}
  et~al.}(2014{\natexlab{a}})\citenamefont{{Serra}, {Lagache}, {Dor{\'e}},
  {Pullen}, and {White}}}]{cibg}
\bibinfo{author}{\bibfnamefont{P.}~\bibnamefont{{Serra}}},
  \bibinfo{author}{\bibfnamefont{G.}~\bibnamefont{{Lagache}}},
  \bibinfo{author}{\bibfnamefont{O.}~\bibnamefont{{Dor{\'e}}}},
  \bibinfo{author}{\bibfnamefont{A.}~\bibnamefont{{Pullen}}}, \bibnamefont{and}
  \bibinfo{author}{\bibfnamefont{M.}~\bibnamefont{{White}}},
  \bibinfo{journal}{\aap} \textbf{\bibinfo{volume}{570}}, \bibinfo{eid}{A98}
  (\bibinfo{year}{2014}{\natexlab{a}}), \eprint{1404.1933}.

\bibitem[{\citenamefont{{Serra}
  et~al.}(2014{\natexlab{b}})\citenamefont{{Serra}, {Lagache}, {Dor{\'e}},
  {Pullen}, and {White}}}]{xCIBg}
\bibinfo{author}{\bibfnamefont{P.}~\bibnamefont{{Serra}}},
  \bibinfo{author}{\bibfnamefont{G.}~\bibnamefont{{Lagache}}},
  \bibinfo{author}{\bibfnamefont{O.}~\bibnamefont{{Dor{\'e}}}},
  \bibinfo{author}{\bibfnamefont{A.}~\bibnamefont{{Pullen}}}, \bibnamefont{and}
  \bibinfo{author}{\bibfnamefont{M.}~\bibnamefont{{White}}},
  \bibinfo{journal}{\aap} \textbf{\bibinfo{volume}{570}}, \bibinfo{eid}{A98}
  (\bibinfo{year}{2014}{\natexlab{b}}), \eprint{1404.1933}.

\bibitem[{\citenamefont{{De Bernardis} and {Cooray}}(2012)}]{subhalo}
\bibinfo{author}{\bibfnamefont{F.}~\bibnamefont{{De Bernardis}}}
  \bibnamefont{and} \bibinfo{author}{\bibfnamefont{A.}~\bibnamefont{{Cooray}}},
  \bibinfo{journal}{\apj} \textbf{\bibinfo{volume}{760}}, \bibinfo{eid}{14}
  (\bibinfo{year}{2012}), \eprint{1206.1324}.

\bibitem[{\citenamefont{{van den Bosch} et~al.}(2005)\citenamefont{{van den
  Bosch}, {Tormen}, and {Giocoli}}}]{subhalo1}
\bibinfo{author}{\bibfnamefont{F.~C.} \bibnamefont{{van den Bosch}}},
  \bibinfo{author}{\bibfnamefont{G.}~\bibnamefont{{Tormen}}}, \bibnamefont{and}
  \bibinfo{author}{\bibfnamefont{C.}~\bibnamefont{{Giocoli}}},
  \bibinfo{journal}{\mnras} \textbf{\bibinfo{volume}{359}},
  \bibinfo{pages}{1029} (\bibinfo{year}{2005}), \eprint{astro-ph/0409201}.

\bibitem[{\citenamefont{{Cirelli} et~al.}(2011)\citenamefont{{Cirelli},
  {Corcella}, {Hektor}, {H{\"u}tsi}, {Kadastik}, {Panci}, {Raidal}, {Sala}, and
  {Strumia}}}]{PPPC4DMID}
\bibinfo{author}{\bibfnamefont{M.}~\bibnamefont{{Cirelli}}},
  \bibinfo{author}{\bibfnamefont{G.}~\bibnamefont{{Corcella}}},
  \bibinfo{author}{\bibfnamefont{A.}~\bibnamefont{{Hektor}}},
  \bibinfo{author}{\bibfnamefont{G.}~\bibnamefont{{H{\"u}tsi}}},
  \bibinfo{author}{\bibfnamefont{M.}~\bibnamefont{{Kadastik}}},
  \bibinfo{author}{\bibfnamefont{P.}~\bibnamefont{{Panci}}},
  \bibinfo{author}{\bibfnamefont{M.}~\bibnamefont{{Raidal}}},
  \bibinfo{author}{\bibfnamefont{F.}~\bibnamefont{{Sala}}}, \bibnamefont{and}
  \bibinfo{author}{\bibfnamefont{A.}~\bibnamefont{{Strumia}}},
  \bibinfo{journal}{\jcap} \textbf{\bibinfo{volume}{3}}, \bibinfo{eid}{051}
  (\bibinfo{year}{2011}), \eprint{1012.4515}.

\bibitem[{\citenamefont{{Gilmore} et~al.}(2012)\citenamefont{{Gilmore},
  {Somerville}, {Primack}, and {Dom{\'{\i}}nguez}}}]{tau_table}
\bibinfo{author}{\bibfnamefont{R.~C.} \bibnamefont{{Gilmore}}},
  \bibinfo{author}{\bibfnamefont{R.~S.} \bibnamefont{{Somerville}}},
  \bibinfo{author}{\bibfnamefont{J.~R.} \bibnamefont{{Primack}}},
  \bibnamefont{and}
  \bibinfo{author}{\bibfnamefont{A.}~\bibnamefont{{Dom{\'{\i}}nguez}}},
  \bibinfo{journal}{\mnras} \textbf{\bibinfo{volume}{422}},
  \bibinfo{pages}{3189} (\bibinfo{year}{2012}), \eprint{1104.0671}.

\bibitem[{\citenamefont{{Ando} and {Komatsu}}(2013)}]{boost_sub}
\bibinfo{author}{\bibfnamefont{S.}~\bibnamefont{{Ando}}} \bibnamefont{and}
  \bibinfo{author}{\bibfnamefont{E.}~\bibnamefont{{Komatsu}}},
  \bibinfo{journal}{\prd} \textbf{\bibinfo{volume}{87}}, \bibinfo{eid}{123539}
  (\bibinfo{year}{2013}), \eprint{1301.5901}.

\bibitem[{\citenamefont{{Gao} et~al.}(2012)\citenamefont{{Gao}, {Frenk},
  {Jenkins}, {Springel}, and {White}}}]{gao_sim}
\bibinfo{author}{\bibfnamefont{L.}~\bibnamefont{{Gao}}},
  \bibinfo{author}{\bibfnamefont{C.~S.} \bibnamefont{{Frenk}}},
  \bibinfo{author}{\bibfnamefont{A.}~\bibnamefont{{Jenkins}}},
  \bibinfo{author}{\bibfnamefont{V.}~\bibnamefont{{Springel}}},
  \bibnamefont{and} \bibinfo{author}{\bibfnamefont{S.~D.~M.}
  \bibnamefont{{White}}}, \bibinfo{journal}{\mnras}
  \textbf{\bibinfo{volume}{419}}, \bibinfo{pages}{1721} (\bibinfo{year}{2012}),
  \eprint{1107.1916}.

\bibitem[{\citenamefont{{Xia} et~al.}(2015)\citenamefont{{Xia}, {Cuoco},
  {Branchini}, and {Viel}}}]{XiaGamma}
\bibinfo{author}{\bibfnamefont{J.-Q.} \bibnamefont{{Xia}}},
  \bibinfo{author}{\bibfnamefont{A.}~\bibnamefont{{Cuoco}}},
  \bibinfo{author}{\bibfnamefont{E.}~\bibnamefont{{Branchini}}},
  \bibnamefont{and} \bibinfo{author}{\bibfnamefont{M.}~\bibnamefont{{Viel}}},
  \bibinfo{journal}{\apjs} \textbf{\bibinfo{volume}{217}}, \bibinfo{eid}{15}
  (\bibinfo{year}{2015}), \eprint{1503.05918}.

\bibitem[{\citenamefont{{Ajello} et~al.}(2014)\citenamefont{{Ajello}, {Romani},
  {Gasparrini}, {Shaw}, {Bolmer}, {Cotter}, {Finke}, {Greiner}, {Healey},
  {King} et~al.}}]{blazarLF}
\bibinfo{author}{\bibfnamefont{M.}~\bibnamefont{{Ajello}}},
  \bibinfo{author}{\bibfnamefont{R.~W.} \bibnamefont{{Romani}}},
  \bibinfo{author}{\bibfnamefont{D.}~\bibnamefont{{Gasparrini}}},
  \bibinfo{author}{\bibfnamefont{M.~S.} \bibnamefont{{Shaw}}},
  \bibinfo{author}{\bibfnamefont{J.}~\bibnamefont{{Bolmer}}},
  \bibinfo{author}{\bibfnamefont{G.}~\bibnamefont{{Cotter}}},
  \bibinfo{author}{\bibfnamefont{J.}~\bibnamefont{{Finke}}},
  \bibinfo{author}{\bibfnamefont{J.}~\bibnamefont{{Greiner}}},
  \bibinfo{author}{\bibfnamefont{S.~E.} \bibnamefont{{Healey}}},
  \bibinfo{author}{\bibfnamefont{O.}~\bibnamefont{{King}}},
  \bibnamefont{et~al.}, \bibinfo{journal}{\apj} \textbf{\bibinfo{volume}{780}},
  \bibinfo{eid}{73} (\bibinfo{year}{2014}), \eprint{1310.0006}.

\bibitem[{\citenamefont{{Ajello} et~al.}(2012)\citenamefont{{Ajello}, {Shaw},
  {Romani}, {Dermer}, {Costamante}, {King}, {Max-Moerbeck}, {Readhead},
  {Reimer}, {Richards} et~al.}}]{fsrq}
\bibinfo{author}{\bibfnamefont{M.}~\bibnamefont{{Ajello}}},
  \bibinfo{author}{\bibfnamefont{M.~S.} \bibnamefont{{Shaw}}},
  \bibinfo{author}{\bibfnamefont{R.~W.} \bibnamefont{{Romani}}},
  \bibinfo{author}{\bibfnamefont{C.~D.} \bibnamefont{{Dermer}}},
  \bibinfo{author}{\bibfnamefont{L.}~\bibnamefont{{Costamante}}},
  \bibinfo{author}{\bibfnamefont{O.~G.} \bibnamefont{{King}}},
  \bibinfo{author}{\bibfnamefont{W.}~\bibnamefont{{Max-Moerbeck}}},
  \bibinfo{author}{\bibfnamefont{A.}~\bibnamefont{{Readhead}}},
  \bibinfo{author}{\bibfnamefont{A.}~\bibnamefont{{Reimer}}},
  \bibinfo{author}{\bibfnamefont{J.~L.} \bibnamefont{{Richards}}},
  \bibnamefont{et~al.}, \bibinfo{journal}{\apj} \textbf{\bibinfo{volume}{751}},
  \bibinfo{eid}{108} (\bibinfo{year}{2012}), \eprint{1110.3787}.

\bibitem[{\citenamefont{{Charles} et~al.}(2016)\citenamefont{{Charles},
  {S{\'a}nchez-Conde}, {Anderson}, {Caputo}, {Cuoco}, {Di Mauro},
  {Drlica-Wagner}, {Gomez-Vargas}, {Meyer}, {Tibaldo}
  et~al.}}]{2016PhR...636....1C}
\bibinfo{author}{\bibfnamefont{E.}~\bibnamefont{{Charles}}},
  \bibinfo{author}{\bibfnamefont{M.}~\bibnamefont{{S{\'a}nchez-Conde}}},
  \bibinfo{author}{\bibfnamefont{B.}~\bibnamefont{{Anderson}}},
  \bibinfo{author}{\bibfnamefont{R.}~\bibnamefont{{Caputo}}},
  \bibinfo{author}{\bibfnamefont{A.}~\bibnamefont{{Cuoco}}},
  \bibinfo{author}{\bibfnamefont{M.}~\bibnamefont{{Di Mauro}}},
  \bibinfo{author}{\bibfnamefont{A.}~\bibnamefont{{Drlica-Wagner}}},
  \bibinfo{author}{\bibfnamefont{G.~A.} \bibnamefont{{Gomez-Vargas}}},
  \bibinfo{author}{\bibfnamefont{M.}~\bibnamefont{{Meyer}}},
  \bibinfo{author}{\bibfnamefont{L.}~\bibnamefont{{Tibaldo}}},
  \bibnamefont{et~al.}, \bibinfo{journal}{\physrep}
  \textbf{\bibinfo{volume}{636}}, \bibinfo{pages}{1} (\bibinfo{year}{2016}),
  \eprint{1605.02016}.

\bibitem[{\citenamefont{{Abdo} et~al.}(2009)\citenamefont{{Abdo}, {Ackermann},
  {Ajello}, {Atwood}, {Axelsson}, {Baldini}, {Ballet}, {Barbiellini},
  {Bastieri}, {Baughman} et~al.}}]{2009ApJ...700..597A}
\bibinfo{author}{\bibfnamefont{A.~A.} \bibnamefont{{Abdo}}},
  \bibinfo{author}{\bibfnamefont{M.}~\bibnamefont{{Ackermann}}},
  \bibinfo{author}{\bibfnamefont{M.}~\bibnamefont{{Ajello}}},
  \bibinfo{author}{\bibfnamefont{W.~B.} \bibnamefont{{Atwood}}},
  \bibinfo{author}{\bibfnamefont{M.}~\bibnamefont{{Axelsson}}},
  \bibinfo{author}{\bibfnamefont{L.}~\bibnamefont{{Baldini}}},
  \bibinfo{author}{\bibfnamefont{J.}~\bibnamefont{{Ballet}}},
  \bibinfo{author}{\bibfnamefont{G.}~\bibnamefont{{Barbiellini}}},
  \bibinfo{author}{\bibfnamefont{D.}~\bibnamefont{{Bastieri}}},
  \bibinfo{author}{\bibfnamefont{B.~M.} \bibnamefont{{Baughman}}},
  \bibnamefont{et~al.}, \bibinfo{journal}{\apj} \textbf{\bibinfo{volume}{700}},
  \bibinfo{pages}{597} (\bibinfo{year}{2009}), \eprint{0902.1559}.

\bibitem[{\citenamefont{{Ando} et~al.}(2007)\citenamefont{{Ando}, {Komatsu},
  {Narumoto}, and {Totani}}}]{sando2007}
\bibinfo{author}{\bibfnamefont{S.}~\bibnamefont{{Ando}}},
  \bibinfo{author}{\bibfnamefont{E.}~\bibnamefont{{Komatsu}}},
  \bibinfo{author}{\bibfnamefont{T.}~\bibnamefont{{Narumoto}}},
  \bibnamefont{and} \bibinfo{author}{\bibfnamefont{T.}~\bibnamefont{{Totani}}},
  \bibinfo{journal}{\mnras} \textbf{\bibinfo{volume}{376}},
  \bibinfo{pages}{1635} (\bibinfo{year}{2007}), \eprint{astro-ph/0610155}.

\bibitem[{\citenamefont{{Gruppioni} et~al.}(2013)\citenamefont{{Gruppioni},
  {Pozzi}, {Rodighiero}, {Delvecchio}, {Berta}, {Pozzetti}, {Zamorani},
  {Andreani}, {Cimatti}, {Ilbert} et~al.}}]{sfg}
\bibinfo{author}{\bibfnamefont{C.}~\bibnamefont{{Gruppioni}}},
  \bibinfo{author}{\bibfnamefont{F.}~\bibnamefont{{Pozzi}}},
  \bibinfo{author}{\bibfnamefont{G.}~\bibnamefont{{Rodighiero}}},
  \bibinfo{author}{\bibfnamefont{I.}~\bibnamefont{{Delvecchio}}},
  \bibinfo{author}{\bibfnamefont{S.}~\bibnamefont{{Berta}}},
  \bibinfo{author}{\bibfnamefont{L.}~\bibnamefont{{Pozzetti}}},
  \bibinfo{author}{\bibfnamefont{G.}~\bibnamefont{{Zamorani}}},
  \bibinfo{author}{\bibfnamefont{P.}~\bibnamefont{{Andreani}}},
  \bibinfo{author}{\bibfnamefont{A.}~\bibnamefont{{Cimatti}}},
  \bibinfo{author}{\bibfnamefont{O.}~\bibnamefont{{Ilbert}}},
  \bibnamefont{et~al.}, \bibinfo{journal}{\mnras}
  \textbf{\bibinfo{volume}{432}}, \bibinfo{pages}{23} (\bibinfo{year}{2013}),
  \eprint{1302.5209}.

\bibitem[{\citenamefont{{Fields} et~al.}(2010)\citenamefont{{Fields},
  {Pavlidou}, and {Prodanovi{\'c}}}}]{2010ApJ...722L.199F}
\bibinfo{author}{\bibfnamefont{B.~D.} \bibnamefont{{Fields}}},
  \bibinfo{author}{\bibfnamefont{V.}~\bibnamefont{{Pavlidou}}},
  \bibnamefont{and}
  \bibinfo{author}{\bibfnamefont{T.}~\bibnamefont{{Prodanovi{\'c}}}},
  \bibinfo{journal}{\apjl} \textbf{\bibinfo{volume}{722}},
  \bibinfo{pages}{L199} (\bibinfo{year}{2010}), \eprint{1003.3647}.

\bibitem[{\citenamefont{{Ackermann}
  et~al.}(2012{\natexlab{b}})\citenamefont{{Ackermann}, {Ajello}, {Allafort},
  {Baldini}, {Ballet}, {Bastieri}, {Bechtol}, {Bellazzini}, {Berenji}, {Bloom}
  et~al.}}]{Ackermann12}
\bibinfo{author}{\bibfnamefont{M.}~\bibnamefont{{Ackermann}}},
  \bibinfo{author}{\bibfnamefont{M.}~\bibnamefont{{Ajello}}},
  \bibinfo{author}{\bibfnamefont{A.}~\bibnamefont{{Allafort}}},
  \bibinfo{author}{\bibfnamefont{L.}~\bibnamefont{{Baldini}}},
  \bibinfo{author}{\bibfnamefont{J.}~\bibnamefont{{Ballet}}},
  \bibinfo{author}{\bibfnamefont{D.}~\bibnamefont{{Bastieri}}},
  \bibinfo{author}{\bibfnamefont{K.}~\bibnamefont{{Bechtol}}},
  \bibinfo{author}{\bibfnamefont{R.}~\bibnamefont{{Bellazzini}}},
  \bibinfo{author}{\bibfnamefont{B.}~\bibnamefont{{Berenji}}},
  \bibinfo{author}{\bibfnamefont{E.~D.} \bibnamefont{{Bloom}}},
  \bibnamefont{et~al.}, \bibinfo{journal}{\apj} \textbf{\bibinfo{volume}{755}},
  \bibinfo{eid}{164} (\bibinfo{year}{2012}{\natexlab{b}}), \eprint{1206.1346}.

\bibitem[{\citenamefont{{Willott} et~al.}(2001)\citenamefont{{Willott},
  {Rawlings}, {Blundell}, {Lacy}, and {Eales}}}]{radioGLF}
\bibinfo{author}{\bibfnamefont{C.~J.} \bibnamefont{{Willott}}},
  \bibinfo{author}{\bibfnamefont{S.}~\bibnamefont{{Rawlings}}},
  \bibinfo{author}{\bibfnamefont{K.~M.} \bibnamefont{{Blundell}}},
  \bibinfo{author}{\bibfnamefont{M.}~\bibnamefont{{Lacy}}}, \bibnamefont{and}
  \bibinfo{author}{\bibfnamefont{S.~A.} \bibnamefont{{Eales}}},
  \bibinfo{journal}{\mnras} \textbf{\bibinfo{volume}{322}},
  \bibinfo{pages}{536} (\bibinfo{year}{2001}), \eprint{astro-ph/0010419}.

\bibitem[{\citenamefont{{Di Mauro} et~al.}(2014)\citenamefont{{Di Mauro},
  {Calore}, {Donato}, {Ajello}, and {Latronico}}}]{Lr_L}
\bibinfo{author}{\bibfnamefont{M.}~\bibnamefont{{Di Mauro}}},
  \bibinfo{author}{\bibfnamefont{F.}~\bibnamefont{{Calore}}},
  \bibinfo{author}{\bibfnamefont{F.}~\bibnamefont{{Donato}}},
  \bibinfo{author}{\bibfnamefont{M.}~\bibnamefont{{Ajello}}}, \bibnamefont{and}
  \bibinfo{author}{\bibfnamefont{L.}~\bibnamefont{{Latronico}}},
  \bibinfo{journal}{\apj} \textbf{\bibinfo{volume}{780}}, \bibinfo{eid}{161}
  (\bibinfo{year}{2014}), \eprint{1304.0908}.

\bibitem[{\citenamefont{{Bandara} et~al.}(2009)\citenamefont{{Bandara},
  {Crampton}, and {Simard}}}]{Mbh_M}
\bibinfo{author}{\bibfnamefont{K.}~\bibnamefont{{Bandara}}},
  \bibinfo{author}{\bibfnamefont{D.}~\bibnamefont{{Crampton}}},
  \bibnamefont{and} \bibinfo{author}{\bibfnamefont{L.}~\bibnamefont{{Simard}}},
  \bibinfo{journal}{\apj} \textbf{\bibinfo{volume}{704}}, \bibinfo{pages}{1135}
  (\bibinfo{year}{2009}), \eprint{0909.0269}.

\bibitem[{\citenamefont{{Franceschini}
  et~al.}(1998)\citenamefont{{Franceschini}, {Vercellone}, and
  {Fabian}}}]{Mbh_Lr}
\bibinfo{author}{\bibfnamefont{A.}~\bibnamefont{{Franceschini}}},
  \bibinfo{author}{\bibfnamefont{S.}~\bibnamefont{{Vercellone}}},
  \bibnamefont{and} \bibinfo{author}{\bibfnamefont{A.~C.}
  \bibnamefont{{Fabian}}}, \bibinfo{journal}{\mnras}
  \textbf{\bibinfo{volume}{297}}, \bibinfo{pages}{817} (\bibinfo{year}{1998}),
  \eprint{astro-ph/9801129}.

\bibitem[{\citenamefont{{Shirasaki}
  et~al.}(2014{\natexlab{b}})\citenamefont{{Shirasaki}, {Horiuchi}, and
  {Yoshida}}}]{xCFHLS}
\bibinfo{author}{\bibfnamefont{M.}~\bibnamefont{{Shirasaki}}},
  \bibinfo{author}{\bibfnamefont{S.}~\bibnamefont{{Horiuchi}}},
  \bibnamefont{and}
  \bibinfo{author}{\bibfnamefont{N.}~\bibnamefont{{Yoshida}}},
  \bibinfo{journal}{\prd} \textbf{\bibinfo{volume}{90}}, \bibinfo{eid}{063502}
  (\bibinfo{year}{2014}{\natexlab{b}}), \eprint{1404.5503}.

\bibitem[{\citenamefont{{Hivon} et~al.}(2002)\citenamefont{{Hivon},
  {G{\'o}rski}, {Netterfield}, {Crill}, {Prunet}, and {Hansen}}}]{master}
\bibinfo{author}{\bibfnamefont{E.}~\bibnamefont{{Hivon}}},
  \bibinfo{author}{\bibfnamefont{K.~M.} \bibnamefont{{G{\'o}rski}}},
  \bibinfo{author}{\bibfnamefont{C.~B.} \bibnamefont{{Netterfield}}},
  \bibinfo{author}{\bibfnamefont{B.~P.} \bibnamefont{{Crill}}},
  \bibinfo{author}{\bibfnamefont{S.}~\bibnamefont{{Prunet}}}, \bibnamefont{and}
  \bibinfo{author}{\bibfnamefont{F.}~\bibnamefont{{Hansen}}},
  \bibinfo{journal}{\apj} \textbf{\bibinfo{volume}{567}}, \bibinfo{pages}{2}
  (\bibinfo{year}{2002}), \eprint{astro-ph/0105302}.

\bibitem[{\citenamefont{{Madhavacheril}
  et~al.}(2014)\citenamefont{{Madhavacheril}, {Sehgal}, and
  {Slatyer}}}]{cmbandothers}
\bibinfo{author}{\bibfnamefont{M.~S.} \bibnamefont{{Madhavacheril}}},
  \bibinfo{author}{\bibfnamefont{N.}~\bibnamefont{{Sehgal}}}, \bibnamefont{and}
  \bibinfo{author}{\bibfnamefont{T.~R.} \bibnamefont{{Slatyer}}},
  \bibinfo{journal}{\prd} \textbf{\bibinfo{volume}{89}}, \bibinfo{eid}{103508}
  (\bibinfo{year}{2014}), \eprint{1310.3815}.

\bibitem[{\citenamefont{{The Fermi LAT Collaboration}}(2015)}]{fermi2}
\bibinfo{author}{\bibnamefont{{The Fermi LAT Collaboration}}},
  \bibinfo{journal}{\jcap} \textbf{\bibinfo{volume}{9}}, \bibinfo{eid}{008}
  (\bibinfo{year}{2015}), \eprint{1501.05464}.

\bibitem[{\citenamefont{{Abramowski} et~al.}(2011)\citenamefont{{Abramowski},
  {Acero}, {Aharonian}, {Akhperjanian}, {Anton}, {Barnacka}, {Barres de
  Almeida}, {Bazer-Bachi}, {Becherini}, {Becker} et~al.}}]{gchalo}
\bibinfo{author}{\bibfnamefont{A.}~\bibnamefont{{Abramowski}}},
  \bibinfo{author}{\bibfnamefont{F.}~\bibnamefont{{Acero}}},
  \bibinfo{author}{\bibfnamefont{F.}~\bibnamefont{{Aharonian}}},
  \bibinfo{author}{\bibfnamefont{A.~G.} \bibnamefont{{Akhperjanian}}},
  \bibinfo{author}{\bibfnamefont{G.}~\bibnamefont{{Anton}}},
  \bibinfo{author}{\bibfnamefont{A.}~\bibnamefont{{Barnacka}}},
  \bibinfo{author}{\bibfnamefont{U.}~\bibnamefont{{Barres de Almeida}}},
  \bibinfo{author}{\bibfnamefont{A.~R.} \bibnamefont{{Bazer-Bachi}}},
  \bibinfo{author}{\bibfnamefont{Y.}~\bibnamefont{{Becherini}}},
  \bibinfo{author}{\bibfnamefont{J.}~\bibnamefont{{Becker}}},
  \bibnamefont{et~al.}, \bibinfo{journal}{Physical Review Letters}
  \textbf{\bibinfo{volume}{106}}, \bibinfo{eid}{161301} (\bibinfo{year}{2011}),
  \eprint{1103.3266}.

\bibitem[{\citenamefont{{Ackermann} et~al.}(2015)\citenamefont{{Ackermann},
  {Albert}, {Anderson}, {Atwood}, {Baldini}, {Barbiellini}, {Bastieri},
  {Bechtol}, {Bellazzini}, {Bissaldi} et~al.}}]{dsph}
\bibinfo{author}{\bibfnamefont{M.}~\bibnamefont{{Ackermann}}},
  \bibinfo{author}{\bibfnamefont{A.}~\bibnamefont{{Albert}}},
  \bibinfo{author}{\bibfnamefont{B.}~\bibnamefont{{Anderson}}},
  \bibinfo{author}{\bibfnamefont{W.~B.} \bibnamefont{{Atwood}}},
  \bibinfo{author}{\bibfnamefont{L.}~\bibnamefont{{Baldini}}},
  \bibinfo{author}{\bibfnamefont{G.}~\bibnamefont{{Barbiellini}}},
  \bibinfo{author}{\bibfnamefont{D.}~\bibnamefont{{Bastieri}}},
  \bibinfo{author}{\bibfnamefont{K.}~\bibnamefont{{Bechtol}}},
  \bibinfo{author}{\bibfnamefont{R.}~\bibnamefont{{Bellazzini}}},
  \bibinfo{author}{\bibfnamefont{E.}~\bibnamefont{{Bissaldi}}},
  \bibnamefont{et~al.}, \bibinfo{journal}{Physical Review Letters}
  \textbf{\bibinfo{volume}{115}}, \bibinfo{eid}{231301} (\bibinfo{year}{2015}),
  \eprint{1503.02641}.

\bibitem[{\citenamefont{{Aleksi{\'c}} et~al.}(2014)\citenamefont{{Aleksi{\'c}},
  {Ansoldi}, {Antonelli}, {Antoranz}, {Babic}, {Bangale}, {Barres de Almeida},
  {Barrio}, {Becerra Gonz{\'a}lez}, {Bednarek} et~al.}}]{magic}
\bibinfo{author}{\bibfnamefont{J.}~\bibnamefont{{Aleksi{\'c}}}},
  \bibinfo{author}{\bibfnamefont{S.}~\bibnamefont{{Ansoldi}}},
  \bibinfo{author}{\bibfnamefont{L.~A.} \bibnamefont{{Antonelli}}},
  \bibinfo{author}{\bibfnamefont{P.}~\bibnamefont{{Antoranz}}},
  \bibinfo{author}{\bibfnamefont{A.}~\bibnamefont{{Babic}}},
  \bibinfo{author}{\bibfnamefont{P.}~\bibnamefont{{Bangale}}},
  \bibinfo{author}{\bibfnamefont{U.}~\bibnamefont{{Barres de Almeida}}},
  \bibinfo{author}{\bibfnamefont{J.~A.} \bibnamefont{{Barrio}}},
  \bibinfo{author}{\bibfnamefont{J.}~\bibnamefont{{Becerra Gonz{\'a}lez}}},
  \bibinfo{author}{\bibfnamefont{W.}~\bibnamefont{{Bednarek}}},
  \bibnamefont{et~al.}, \bibinfo{journal}{\jcap} \textbf{\bibinfo{volume}{2}},
  \bibinfo{eid}{008} (\bibinfo{year}{2014}), \eprint{1312.1535}.

\bibitem[{\citenamefont{{G{\'o}rski} et~al.}(2005)\citenamefont{{G{\'o}rski},
  {Hivon}, {Banday}, {Wandelt}, {Hansen}, {Reinecke}, and {Bartelmann}}}]{hp}
\bibinfo{author}{\bibfnamefont{K.~M.} \bibnamefont{{G{\'o}rski}}},
  \bibinfo{author}{\bibfnamefont{E.}~\bibnamefont{{Hivon}}},
  \bibinfo{author}{\bibfnamefont{A.~J.} \bibnamefont{{Banday}}},
  \bibinfo{author}{\bibfnamefont{B.~D.} \bibnamefont{{Wandelt}}},
  \bibinfo{author}{\bibfnamefont{F.~K.} \bibnamefont{{Hansen}}},
  \bibinfo{author}{\bibfnamefont{M.}~\bibnamefont{{Reinecke}}},
  \bibnamefont{and}
  \bibinfo{author}{\bibfnamefont{M.}~\bibnamefont{{Bartelmann}}},
  \bibinfo{journal}{\apj} \textbf{\bibinfo{volume}{622}}, \bibinfo{pages}{759}
  (\bibinfo{year}{2005}), \eprint{astro-ph/0409513}.

\end{thebibliography}

\end{document}